%% file: GW_Interferometry_Maxwell.tex
\documentclass[11pt, fleqn]{article}

\usepackage[utf8]{inputenc}
\usepackage[T1]{fontenc}

\usepackage[
	a4paper,
	margin=3cm,
	]{geometry}
\usepackage{subcaption, float}
\usepackage[hidelinks]{hyperref}

\usepackage[backref=true,sorting=none,url=false,style=numeric-comp,backend=bibtex]{biblatex}
\bibliography{bibliography}

\usepackage{lmodern}
\usepackage{microtype}

\usepackage{mathtools,amssymb} 
\usepackage{tensor}
\usepackage{mathrsfs}
\allowdisplaybreaks

\usepackage{cleveref}
\numberwithin{equation}{subsection}
\renewcommand*{\theequation}{%
	\ifnum\value{subsection}=0 \thesection
	\else \thesubsection \fi
	.\arabic{equation}%
}

\usepackage{textgreek} 

\input{abbreviations.tex}

\newcommand{\checked}[2]{}
\newcommand{\checkedtogether}[1]{}
\newcommand{\ptc}[1]{}
\newcommand{\ptcheck}[1]{}

\newcommand{\red}[1]{#1}

\usepackage{authblk} 

\begin{document}
\title{The Electromagnetic Field\\ in Gravitational Wave Interferometers%
\thanks{Preprint UWThPh-2021-11}
}
\author[1]{Thomas B.~Mieling}
\author[1]{Piotr T.~Chruściel}
\author[2]{Stefan Palenta}
\affil[1]{University of Vienna, Faculty of Physics and TURIS Research Platform, Vienna, Austria}
\affil[2]{University of Vienna, Faculty of Physics, Vienna, Austria}
\date{\today}
\maketitle
\begin{abstract}
	We analyse the response of laser interferometric gravitational wave detectors using the full Maxwell equations in curved spacetime in the presence of weak gravitational waves.
	Existence and uniqueness of solutions is ensured by setting up a suitable boundary value problem. This puts on solid ground previous approximate calculations. We find consistency with previous results obtained from eikonal expansions at the level of accuracy accessible to current gravitational wave detectors.
\end{abstract}

\tableofcontents
\input{optics_subfiles/introduction.tex}
\input{optics_subfiles/phase_emission.tex}
\input{optics_subfiles/setup.tex}
\section{Scalar Waves}
\label{s:scalar waves}
\input{optics_subfiles/scalar_emission.tex}
\input{optics_subfiles/scalar_reflection.tex}
\section{Maxwell’s Equations}
\label{s:maxwell}
\input{optics_subfiles/maxwell_introduction.tex}

\input{optics_subfiles/maxwell_unperturbed.tex}
\input{optics_subfiles/maxwell_boundary.tex}
\input{optics_subfiles/maxwell_emission.tex}
\input{optics_subfiles/maxwell_reflection.tex}

\input{optics_subfiles/maxwell_interferometry.tex}

\input{optics_subfiles/discussion.tex}

\clearpage

\section*{Acknowledgements}
Research supported in part by the Austrian Science Fund (FWF), Project P34274, and by the Vienna University Research Platform TURIS.
TBM gratefully acknowledges funding via a fellowship of the Vienna Doctoral School in Physics (VDSP) and the Austrian Science Fund (FWF) through GRIPS (Grant No. P30817-N36).
PTC was further supported in part by the Polish National Center of Science (NCN) under grant 2016/21/B/ST1/00940.

\appendix

\input{optics_subfiles/maxwell_wave_eq.tex}
\input{optics_subfiles/appendix-emission-covariant.tex}

\input{optics_subfiles/appendix-constraint.tex}
\input{optics_subfiles/appendix-energy.tex}
\input{optics_subfiles/appendix-reflection-covariant.tex}
\input{optics_subfiles/appendix-polarisation.tex}
\clearpage
\printbibliography[heading=bibintoc]

\end{document}

%% file: abbreviations.tex
\def\const{\text{const}}

\def\half{\tfrac{1}{2}}
\def\ihalf{\tfrac{i}{2}}
\def\quarter{\tfrac{1}{4}}

\let\t\tensor
\let\p\partial
\DeclareMathOperator{\wop}{\square}

\def\gs{\mathrm g}	
\def\dd{\mathrm d}	

\DeclareMathOperator{\diag}{diag}
\DeclareMathOperator{\divergence}{div}

\def\o#1{{(#1)}}

\def\particular{\text{part.}}

\def\omegaG{\omega_g}

\def\Fb{{\bar{F}}}	
\def\permeability{\text{\textmu}}
\def\permittivity{\text{\textepsilon}}

\def\kh{{\hat \kappa}}
\def\mn{m.\kh}	

\def\Aa{{\mathscr A}}
\def\Zz{{\hat Z}}
\def\cI{{\textsc{i}}}
\def\cII{{\textsc{ii}}}

%% file: optics_subfiles/introduction.tex
\section{Introduction}

Interferometric gravitational wave detectors are the only instruments so far that have been able to directly detect gravitational waves (\red{GWs}). While the signals observed are certainly derivable by solving Maxwell’s equations in a GW\ metric, much to our surprise we have not found satisfactory calculations of this kind in the literature.
Instead, most approaches to the problem use various approximations which can be ordered in sequence of increasing detail as
\begin{center}
	\small
	Geodesic Deviation
	\,\textrightarrow\
	Geometric Optics
	\,\textrightarrow\,
	Eikonal Expansion
	\,\textrightarrow\,
	Maxwell’s Equations.
\end{center}
The perhaps simplest method consists in using an approximate solution of Jacobi’s equation for the \emph{geodesic deviation} of nearby world lines, which model the trajectories of the mirrors at the ends of the two interferometer arms \cite{Forward1978, Schutz1987}.
One key assumption here is that the arm lengths are sufficiently short, so that the geodesic deviation vector is a reasonable approximation of the arm lengths. Further, this implicitly assumes that laser interferometry directly measures such deviation vectors. As shown in \cite{Finn2009}, this can be understood as a limiting case of geometric optics methods.

The \emph{geometric optics approximation} consists in computing the difference in round-trip times of light rays in the interferometer arms, which is then related to the phase shift \cites{Weiss1972,Estabrook1975,Poplawski2006,Rakhmanov2008,Finn2009,Rakhmanov2009,Melissinos2010,Koop2014,Saulson2017,Blaut2019}.
While the results of this method are generally believed to be correct, they provide only very limited information about the precise form of the electromagnetic field producing the detected signal.

Indeed, the geometric optics equations arise at first order in \emph{eikonal expansions}, where one explicitly introduces a frequency parameter $\omega$ and an eikonal function $\psi$, seeking solutions possessing a series expansion of the amplitude in inverse powers of $\omega$ \cite{Ehlers1967}. The validity of such high-frequency approximations in the considered context has been questioned in \cite{Calura1999}, and some previous attempts to obtain formulae applicable for all incidence angles of the gravitational wave have led to expressions which are undefined if the GW\ propagates parallelly to one of the interferometer arms \cites{Lobo1992,Angelil2015}.

Finally, previous  important and significant  analyses starting with the \emph{full Maxwell equations} in the context of GW\ detection have used approximation schemes which raise concerns about well-posedness of the resulting equations, in particular about uniqueness of solutions \cites{Cooperstock1968,Lobo1992,Cooperstock1993,Montanari1998,Calura1999,Angelil2015}.
This significant question of uniqueness of the proposed solutions likewise plagues all the methods mentioned so far.

We resolve the  issues arising in these schemes by setting up a boundary value problem for Maxwell’s equations, where light sent into the interferometer arms is described by suitably prescribed values of the electromagnetic field on infinite planes, which we refer to as \emph{emission surfaces}. The key fact is that this is the only setup known to us which models reasonably the physical problem and which guarantees \emph{both existence and uniqueness} of solutions in terms of boundary data. \red{Here we use the word “data” in the PDE sense, namely fields that we prescribe in the problem at hand.}

The question then arises, what are the physically correct boundary data at the emission surface. There is freedom which ultimately needs to be tied to the experiment considered.

It turns out that there are several issues here.
First, one needs to model the apparatus in the absence of gravitational waves. One can then imagine plane waves, or Bessel waves, or Gauss waves, or else, streaming from the emission surface. Since the latter are a superposition of the former, we chose here to consider the simplest model of a plane wave emitted orthogonally to the emission surface, leaving the remaining important cases for future work. This is thus our model for the unperturbed wave.

Next, suppose that the gravitational wave is of order $\epsilon \ll 1$. In \Cref{s:boundary value problem} we show existence of solutions of our model with a convergent expansion in powers of $\epsilon$.
The fact that the expansion converges implies in particular that the first non-trivial coefficient in the expansion can be determined by solving a truncated equation. The resulting solutions of Maxwell’s equations are uniquely determined by certain boundary data, which can be prescribed as a convergent power series in $\epsilon$.
A careful choice of boundary terms of first order in $\epsilon$ is needed to eliminate the already-mentioned meaningless solutions when gravitational waves are parallel to the electromagnetic wave.

Some freedom still remains in the boundary data, and we calculate the first non-trivial term in an $\epsilon$-expansion of the associated solution for a specific choice of boundary data in \Cref{s:Maxwell wave equation emission}.
We show that there exist $\epsilon$-accurate solutions expressible in a form compatible with eikonal expansions, and that their amplitude prefactors are analytic in $\omegaG/\omega$, where $\omegaG$ and $\omega$ are the frequencies of the gravitational and electromagnetic waves, respectively.
We stress that the Maxwell field as a whole \emph{does not} have an expansion in terms of $\omegaG/\omega$, only the amplitude prefactor does; this clarifies the difficulties pointed out in~\cite{Calura1999,Montanari1998}.

In \Cref{s:overview eikonal} we show quite generally that the freedom in the boundary data does not affect the leading order, both in $\epsilon$ and in $\omegaG/\omega$, of the phase shift of the solutions, leading thus to an unambiguous interference pattern at this level of accuracy. Hence, we confirm that a calculation of the interference pattern based on eikonal expansions, and thus on geometric optics, is consistent with Maxwell equations within the model analysed here, and provides the key information needed for current experiments.

While our calculations are carried out in TT coordinates, which seem to be most convenient for the problem at hand, we show in \Cref{s25III21.1} that the leading-order interference pattern is independent of the coordinates in which the analysis is performed.

Our analysis applies not only to vacuum but also to linear dielectrics with refractive index $n \geq 1$, so that the results can be compared with those obtained for optical fibres~\cite{Mieling:2021b}, see \Cref{s25VI21.1}. Another motivation for this is to investigate whether new phenomena arise from different propagation speeds of the gravitational and electromagnetic waves. While there are explicit changes arising both in the phase and in the amplitude of the field, their net result in the interferometer response turn out to be a simple rescaling of the output signal, see \Cref{s:interferometry signals LFL}.

Our key findings are summarised in \Cref{s9VII21.1}, where explicit formulae are given for the electromagnetic field after its round-trip in one of the interferometer arms. The resulting interferometer response is computed in \Cref{s:interferometry signal general}, and the low-frequency limit is analysed in \Cref{s:interferometry signals LFL}.

\section{A Boundary Value Problem for Maxwell’s Equations}
\label{s:boundary value problem}

Consider Maxwell’s equations or the scalar wave equation in a metric of the form
\begin{equation}
	\t g{_\mu_\nu}
		= \t \eta{_\mu_\nu}
		+ \epsilon \t h{_\mu_\nu}(\epsilon, \t x{^\alpha})
\end{equation}
\red{with $ \epsilon \ll 1$,}
where $\t \eta{_\mu_\nu}$ is the Minkowski metric and the $\t h{_\mu_\nu}$’s are bounded real analytic functions of the coordinates and of $\epsilon$.
For example, the reader can think of the metric \eqref{16XII20.p1} below where the error terms are ignored:
\begin{equation}
\label{6I21.3}
	\t g{_\mu_\nu}
		= \t \eta{_\mu_\nu}
		+ \epsilon \t A{_\mu_\nu} \cos[\omegaG(x - t) + \chi]\,,
\end{equation}
where the $\t A{_\mu_\nu}$’s, $\omegaG$ and $\chi$ are constants;
it should be clear from the arguments below that this suffices for the calculations \red{relevant for nowadays interferometric GW detectors}.
The metric \eqref{6I21.3} is indeed real analytic, and depends analytically upon $\epsilon$ and $\omegaG$.

In the notation of \eqref{eq:Maxwell complex tensor}, \Cref{app:covariant emission equation}, Maxwell’s equations in the absence of sources are
\begin{equation}\label{16XII20.1}
	\t\nabla{_\mu} \t{\mathcal F}{^\mu^\nu} = 0\,.
\end{equation}
As explained in more detail shortly, we will seek solutions of \eqref{16XII20.1} with real-analytic data prescribed on a real-analytic timelike hypersurface $\Sigma$.
Detailed calculations are carried out for the case where $\Sigma$ is a timelike coordinate hyperplane $\Sigma =\{ \t m{_i} \t x{^i} = 0\}$ in TT coordinates.

\red{As  well-known}, one can extract from \eqref{16XII20.1} a system of equations which propagates the data in directions transverse to $\Sigma$, so that $\Sigma$ is \emph{not} characteristic for \eqref{16XII20.1}. The remaining equation is then a (complex) constraint equation on $\Sigma$ which, again by a standard argument
\red{which we reproduce in \Cref{ss9VIII21.1} for the convenience of the reader,} propagates away from $\Sigma$ if satisfied there.

Assuming that the free data on $\Sigma$ are analytic, the Cauchy-Kovalevskaya theorem shows existence of a neighbourhood of $\Sigma$ on which a solution exists. By Holmgren's uniqueness theorem, the solution is unique in the class of smooth solutions. Further standard results show that  the solution is real analytic in $\epsilon$ and any further parameters (such as $\omegaG$, $m_i$, etc.) if the boundary data on $\Sigma$ are. Hence, there exists a convergent series
\begin{equation}\label{16XII20.2}
	\t {\mathcal F}{_\mu_\nu}
		= \sum_{k=0}^\infty \t* {\mathcal F}{^{\o k}_\mu_\nu} \epsilon^k \,,
\end{equation}
where the (uniquely determined) expansion coefficients $\t* {\mathcal F}{^{\o k}_\mu_\nu}$ are independent of $\epsilon$.
\red{Given the accuracy of  the instruments that exist or are currently  planned, for applications in gravitational wave detection  it suffices to compute $\t* {\mathcal F}{^{\o 1}_\mu_\nu}$.}

Solving the equations explicitly (for suitable boundary data), we verify that the truncated series $\mathcal F^\o0 + \epsilon \mathcal F^\o1$ can be decomposed into amplitude and phase contributions as commonly assumed. Moreover, we show that, within our model, the amplitude is analytic in $\omegaG/\omega$ so that the eikonal expansion of the amplitude in inverse powers of $\omega$ is justified at this level of approximation:
we have not found an abstract argument for this last step of the analysis, we had to check the structure of $\t* {\mathcal F}{^{\o1}_\mu_\nu}$.
Analyticity in $\omegaG/\omega$ implies that our result  coincides with the one obtained by a suitably-truncated (cf.~\cite{Mieling:2021a}) eikonal approximation of our model. In other words, we have provided a justification that eikonal expansions give the correct result at the expected level of accuracy.

While we have concentrated on the Maxwell equations so far, identical arguments apply to the massless scalar wave equation, which is described in \Cref{s:scalar waves} below as a preparation for the Maxwell case.

It should be recognised that the assumption of real-analyticity of $\t g{_\mu_\nu}$ plays a key role in the justification-part of our analysis. While this is a common assumption in the setting considered, one does not expect “real life metrics” to be real-analytic, in which case our argument will break down. Indeed, the boundary-value problem considered here is well-posed in the analytic case, but is known to be ill-posed in general. Whatever their shortcomings, our arguments justify rigorously some results in the literature concerning the problem at hand, and allow us to pinpoint problems with some existing treatments.

\section{Coordinate-Independence}
\label{s25III21.1}

The question has been raised, to what extent the predictions for the interferometric phase shifts are coordinate invariant. Our approach to this question, as a boundary-value problem for Maxwell equations, allows to give a clear answer.

Indeed, let $\Sigma$ be a smooth timelike hypersurface, describing the emission surface, on which smooth data are given as described in \Cref{s:boundary value problem}, with a corresponding solution of the Maxwell equations which we denote by $\mathcal{F}^{\mu\nu}$. (As already hinted-to above, there are no general existence theorems in this setting except in the case of analytic data and of an analytic metric; this is however irrelevant for the problem of coordinate-independence, for if a solution does not exist, there is nothing to prove.)
By Holmgren's uniqueness theorem (which does not require analyticity of solutions, smoothness suffices) the field $\mathcal{F}^{\mu\nu}$ is unique, in particular it is independent of the coordinate system which is used to calculate the tensor field $\mathcal{F}^{\mu\nu} $.

Let further $\Sigma'$ be an analytic timelike hypersurface, describing the mirror. Our proposal for the boundary conditions at the mirror can be found in \Cref{s:Maxwell reflection data} below. In this proposal the boundary data at $\Sigma'$ are read off the field $\t{\mathcal F}{^\mu^\nu}$ and lead to a boundary value problem for the returning Maxwell field, say $\t{\check{\mathcal F}}{^\mu^\nu}$, essentially identical to the one in  \Cref{s:boundary value problem}. Thus, the resulting electromagnetic field between the mirrors, $\t{\mathcal F}{^\mu^\nu} + \t{\check{\mathcal F}}{^\mu^\nu}$ is defined uniquely in our setting.
This field, together with the corresponding field from the second arm, determines the interference pattern at the detector. If there exists a corresponding solution of the Maxwell equations (which is the case in the model considered here), then this solution is unique and independent of the coordinates by an identical argument.

The above makes it clear that both our solutions and the resulting signal at the detector are unique, and hence coordinate-independent.


%% file: optics_subfiles/phase_emission.tex
\section{Justification of Eikonal Expansions}
\label{s:overview eikonal}

As shown in detail below, both the scalar field equation and the wave equations implied by Maxwell’s equations are of the general form
\begin{equation}
	\label{eq:overview wave equation}
	\left(
		\t \delta{_I^J}
		\t \gamma{^\mu^\nu} \t\p{_\mu} \t\p{_\nu}
		+ \epsilon \omegaG \t a{_I^J^\mu} \t\p{_\mu}
		+ \epsilon \omegaG^2 \t b{_I^J}
	\right) \t \phi{_J}
	= 0\,,
\end{equation}
for some field $\phi$ with an index $J$.
Here the $\t a{_I^J^\mu}$’s and the $\t b{_I^J}$’s are linear combinations of $\sin(u)$ and $\cos(u)$, where $u$ is the phase of the gravitational wave
\begin{equation}
	\label{eq:u def}
	u = \t\kappa{_\mu} \t x{^\mu} + \chi\,,
\end{equation}
and $\gamma$ is the optical metric (see \eqref{eq:optical metric} below)
\begin{equation}
	\t \gamma{^\mu^\nu}
	= \t \gamma{^{\o0}^\mu^\nu}
	- \epsilon \t A{^\mu^\nu} \cos(u)
	+ O(\epsilon^2)\,.
\end{equation}
Writing the field $\phi$ as a perturbation of a plane wave
\begin{equation}
	\t \phi{_I} = \t\varphi{^{\o0}_I} e^{i \t k{_\mu} \t x{^\mu}} + \epsilon \t\phi{^{\o1}_I} + O(\epsilon^2)\,,
\end{equation}
where the $\t\varphi{^{\o0}_I}$’s are constants and $\t\phi{^{\o1}_I}$ does not depend upon $\epsilon$, one obtains, within our model for light emission from a surface $\Sigma$ (described in detail below), a boundary value problem of the form
\begin{align}
	\label{eq:overview wave eq}
	\wop^\o0 \t\phi{^{\o1}_I}
		&= \omega^2 \t f{_1_I}(\Omega, u) e^{i \t k{_\mu} \t x{^\mu}}\,,
		\\
		\t\phi{^{\o1}_I}|_\Sigma
		&= \t f{_2_I} (\Omega, u) e^{- i \omega t}\,,
		\\
	m(\t\phi{^{\o1}_I})|_\Sigma
		&= i n \omega \t f{_3_I} (\Omega, u) e^{- i \omega t}\,,
\end{align}
where $\t f{_i_J}$’s are linear combinations of $\sin(u)$ and $\cos(u)$, with coefficients which are analytic in $\Omega = \omegaG/\omega$, and $m$ is the unperturbed unit normal to $\Sigma$. In particular  the coefficients $\t f{_1_I}$ in \eqref{eq:overview wave eq} take the form
\begin{equation}
	\label{eq:overview f1 expansion}
	\t f{_1_I} = - n^2 A(m,m) \t \varphi{^{\o0}_I} \cos(u) + O(\Omega)\,.
\end{equation}
By explicit calculations we show that if the $\t f{_i_J}$’s satisfy a certain emission condition (which we verify for our model), then the solutions are of the form
\begin{equation}
	\label{eq:overview perturbation factoring}
	\t\phi{^{\o1}_I}
		= \left[
			i \t \varphi{^{\o0}_I} \psi^\o1
			+ \t \varphi{^{\o1}_I}(\Omega, u, \omegaG m.x)
		\right] e^{i \t k{_\mu} \t x{^\mu}}\,.
\end{equation}
Here, the real-valued function $\psi^\o1$ is of order $1/\Omega$ and satisfies
\begin{equation}
	\label{eq;overview eikonal eq perturbation}
	2 \t \gamma{^{\o0}^\mu^\nu} \t k{_\mu} \t \p{_\nu} \psi^\o1
		- \t A{^\mu^\nu} \t k{_\mu} \t k{_\nu} \cos(u)= 0\,,
\end{equation}
and the $\t \varphi{^{\o1}_I}$’s are analytic functions of $\Omega$, which are of order unity.
Consequently, the overall field can be written as
\begin{equation}
	\begin{split}
		\t\phi{_I}
			&=
			\left[
				\t\varphi{^{\o0}_I}
				+ i \epsilon \t \varphi{^{\o0}_I} \psi^\o1
				+ \epsilon \t \varphi{^{\o1}_I}
			\right] e^{i \t k{_\mu} \t x{^\mu}}
			+ O(\epsilon^2)\\
			&=
			\left[
				\t\varphi{^{\o0}_I}
				+ \epsilon \t \varphi{^{\o1}_I}
			\right]
			\left[
				1 + i \epsilon \psi^\o1
			\right]
			e^{i \t k{_\mu} \t x{^\mu}}
			+ O(\epsilon^2)
			\\
			&= \left[
				\t\varphi{^{\o0}_I} + \epsilon \t\varphi{^{\o1}_I}
			\right]e^{i(\t k{_\mu} \t x{^\mu} + \epsilon \psi^\o1)}
			+ O(\epsilon^2)
			= \t\varphi{_I}\, e^{i \psi} + O(\epsilon^2)
			\,,
	\end{split}
\end{equation}
where the function $\psi$ is
\begin{equation}
	\psi := \t k{_\mu} \t x{^\mu} + \epsilon \psi^\o1.
\end{equation}
Due to \eqref{eq;overview eikonal eq perturbation}, $\psi$ satisfies the eikonal equation to first order in $\epsilon$:
\begin{equation}
	\t \gamma{^\mu^\nu} (\t \p{_\mu} \psi) (\t \p{_\nu} \psi) = O(\epsilon^2)\,.
\end{equation}
Moreover, the amplitude $\t\varphi{_I} = \t\varphi{^{\o0}_I} + \epsilon \t\varphi{^{\o1}_I}$ is analytic in the frequency ratio $\Omega$ and thus analytic in the inverse frequency $1/\omega$. This justifies the use of eikonal expansions, at least at first order in $\epsilon$.

Our formulae show that the phase perturbation $\psi^\o1$ is independent of $f_2$, $f_3$, and also of the terms not written explicitly in \eqref{eq:overview f1 expansion}. Instead, the details of the boundary conditions and
the lower order terms in the wave equation only affect the amplitude perturbation $\t \varphi{^{\o1}_I}$. In this sense, the phase does not depend on the precise details of the boundary data. In particular  as $\psi^\o1$ is of order $1/\Omega$, it constitutes the dominant contribution to $\phi^\o1$. This shows that previous analyses using geometric optics correctly determine the leading-order formula for $\phi^\o1$ in an expansion in both parameters $\epsilon$ and $\Omega$.

%% file: optics_subfiles/setup.tex
\section{The Setup}
\label{s:setup}

We consider a linearised gravitational wave of amplitude $\epsilon$ in transverse-traceless gauge. Thus, neglecting non-linearities, the metric can be written as
\begin{equation}
\label{6I21.1}
	\t g{_\mu_\nu}
	= \t \eta{_\mu_\nu}
	+ \epsilon \t h{_\mu_\nu}\,,
\end{equation}
where $\epsilon$ is a small constant and $\t h{_\mu_\nu}$ is spatial, transverse, traceless, and satisfies the Minkowskian wave equation:
\begin{align}
	\label{6I21.2}
	\t h{_\mu_0} &= 0\,,
	&
	\t\eta{^\mu^\nu} \t\p{_\mu} \t h{_\nu_\rho} &= 0\,,
	&
	\t\eta{^\mu^\nu} \t h{_\mu_\nu} &= 0\,,
	&
	\t\eta{^\mu^\nu}  \t\p{_\mu} \t\p{_\nu} \t h{_\rho_\sigma} &= 0\,.
\end{align}
\red{Here $\epsilon$ should be small enough so that the existence arguments above apply. A sharp estimate for this number is beyond the existing mathematical techniques, but we expect that  values typical for current GW detectors are compatible with our analysis.
}

As explained in \Cref{s:boundary value problem}, we are concerned with solutions of the scalar wave equation and of the Maxwell equations which are convergent power series in $\epsilon$. Since only the first-order term is of experimental relevance, it suffices to solve a linear equation for the first coefficient in this expansion.
By linearity, it suffices to consider the case where $\t h{_\mu_\nu}$ has a single frequency $\omegaG$ and propagates in a fixed direction.
We are then led to consider a metric tensor
\begin{equation}
	\label{16XII20.p1}
	\t g{_\mu_\nu}
		= \t\eta{_\mu_\nu}
		+ \epsilon \t A{_\mu_\nu} \cos(\kappa.x + \chi)
		+ O(\epsilon^2)\,,
\end{equation}
where $\chi$ is a constant phase offset, $\kappa$ is the GW\ wave-covector
\begin{equation}
	\kappa = \omegaG( \t\kh{_i} \dd \t x{^i} - \dd t)\,,
\end{equation}
$\t\kh{_i}$ being a constant unit vector in the background geometry, and where for any covector $k$ we write
\begin{equation}
	k.x \equiv k_\mu x^\mu\,,
\end{equation}
so that $\kappa.x = \omegaG( \t\kh{_i} \t x{^i} - t)$.
The signature used here is
\begin{equation}
	(\t\eta{_\mu_\nu}) = \diag(-1,+1,+1,+1)\,.
\end{equation}
The gauge conditions now imply
\begin{align}
	\t A{_\mu_0} &= 0\,,
	&
	\t \delta{^i^j} \t A{_i_j} &= 0\,,
	&
	\t A{_\mu_\nu} \t\kappa{^\nu} &= 0\,,
\end{align}
i.e.\ $A$ is purely spatial, traceless, and orthogonal to the GW\ wave-vector $\kappa$ (whose index is raised with the Minkowski metric).

We  consider a region filled by an isotropic linear dielectric of constant refractive index $n$. The dielectric is at rest in the coordinate system above, so that its four-velocity is $u = \p/\p x^0$.
This is a very crude approximation which ignores the elastic properties of the dielectric medium; we plan to return to this  in future work.
The associated optical metric is given by
\begin{equation}
	\label{eq:optical metric}
	\t\gamma{^\mu^\nu} = \t g{^\mu^\nu} + (1-n^2) \t u{^\mu} \t u{^\nu}\,.
\end{equation}
Here, the inverse metric is given by
\begin{equation}
	\t g{^\mu^\nu}
	= \t\eta{^\mu^\nu}
	- \epsilon \t A{^\mu^\nu} \cos(\kappa.x + \chi)
	+ O(\epsilon^2)\,,
\end{equation}
where $\t A{^\mu^\nu} \equiv \t\eta{^\mu^\rho} \t\eta{^\nu^\sigma} \t A{_\rho_\sigma}$ numerically coincides with $\t A{_\mu_\nu}$ (in the used coordinate system).
Hence, the optical metric at first order in $\epsilon$ is
\begin{align}
	\t \gamma{^{\o0}^\mu^\nu} &= \diag(-n^2, 1, 1, 1)\,,
	&
	\t \gamma{^{\o1}^\mu^\nu} &= - \t h{^\mu^\nu}\,.
\end{align}

%% file: optics_subfiles/scalar_emission.tex
Before considering Maxwell’s equations and wave solutions thereof in \Cref{s:maxwell}, we first study a simplified model based on the scalar wave equation. It is folklore knowledge, and confirmed by our analysis (cf.~\Cref{s:Maxwell wave equation emission,s:Maxwell reflection data} below), that this suffices to determine the phase shift in interferometric experiments.

Since Maxwell’s equations imply a wave equation for each component of the electromagnetic field, we consider a scalar field $\phi$ satisfying the scalar wave equation
\begin{equation}
	\label{eq:scalar wave equation}
	\square_\gamma \phi
	\equiv \t \gamma{^\mu^\nu} \t \p{_\mu} \t \p{_\nu} \phi
	= 0\,,
\end{equation}
where $\square_\gamma$ is the wave operator associated to the optical metric~\eqref{eq:optical metric}.
 The first equality here holds because we are working with harmonic coordinates.

\subsection{Emission From a Laser}

To model the emission of light by a laser, we impose the boundary conditions
\begin{align}
	\label{eq:boundary conditions laser}
	\phi|_\Sigma
		&= e^{-i \omega t} \,,
	&
	\nu(\phi)|_\Sigma
		&= i n \omega e^{-i \omega t} \,,
\end{align}
where $\Sigma$ is the timelike hypersurface from which the wave is emitted,
\red{$\nu$ denotes its unit normal, and $\nu(\phi) = \t\nu{^\mu} \t\p{_\mu} \phi$ is the usual notation for the action of a vector field on a function.}

One motivation for these boundary conditions comes from geometric optics with which we want to compare.
There, one writes the field as $\mathscr A e^{i \psi}$, where the gradient of the eikonal $\psi$ is required to be null.
Seeking solutions which resemble plane waves in flat space as closely as possible, we assume the field to take the same value on the emission surface as one would have for a plane wave in Minkowski space.
In the language of geometric optics, this corresponds to a constant amplitude $\mathscr A$ equal to $1$ on $\Sigma$, and the eikonal $\psi$ equal to $-\omega t$ there.
Having prescribed $\psi|_\Sigma = -\omega t$, it follows that $\dd \psi = -\omega \dd t + \alpha \nu$, where $\nu$ is the geometric normal to $\Sigma$ (normalized with respect to the full metric $g$). The null condition then entails $\alpha^2 = n^2 \omega^2$, and the sign of $\alpha$ determines the direction into which light is emitted. Choosing a positive sign of $\alpha$  (emitting along $\nu$), and neglecting the derivative  of $\mathscr A$ in the normal direction,%
\footnote{The contribution of the normal derivatives of $\mathscr A$ within the eikonal approximation approach (cf.~e.g.~\cite[Equation~(4.20)]{Mieling:2021a}) is of order of $\omegaG/\omega$; one can check that this changes the final solution $\phi$ by terms of order $O(\omegaG/\omega)$ which can be determined explicitly within our scheme, and which is therefore of the same order as the (physically irrelevant) error terms in our final approximate solution.}
yields the boundary values in \eqref{eq:boundary conditions laser}.

More significantly, as follows from our analysis below, these boundary values can also be motivated from Maxwell’s equations.
There, we prescribe the values of the fields on $\Sigma$, and Maxwell’s equations then, in turn, determine the normal derivatives. It turns out that the boundary conditions arising there differ from those here only by polarisation corrections, so that the toy model discussed here is sufficient to determine the phase.

Summarizing: the first condition encodes that every point of the emission surface $\Sigma$ emits light of the same proper frequency $\omega$ ($t$ measures proper time of clocks at rest), with the same phase, and the second equation means that the emission is normal to $\Sigma$.
While there are certainly more accurate models for the light emission of lasers (e.g.\ using Gaussian waves), we only consider this simple case.

For the emission surface, we take the coordinate-plane
\begin{equation}
	\Sigma = \{\t m{_i} \t x{^i} = 0 \}
 \,,
\end{equation}
for some (constant) coefficients $\t m{_i}$, normalized to $m.m = 1$., 
Here and in the following, for two spatial vectors $v, w$, we denote by $v.w$ their Euclidean scalar product
\begin{equation}
	v.w \equiv \t \delta{^i^j} \t v{_i} \t w{_j}\,.
\end{equation}
This does not apply to $\t\kappa{_\mu}$ and $\t x{^\mu}$, which are \emph{not} spatial vectors: instead, $\kappa.x$ denotes the contraction $\t \kappa{_\mu}\t x{^\mu}$.

Choosing the orientation of $\nu$ to be the same as the one for the conormal
\begin{equation}
	m := \t m{_i}\, \dd \t x{^i}\,,
\end{equation}
one has
\begin{equation}
	\label{eq:conormal normalised}
	\t \nu{_i}
		= \t m{_i} \left( 1 + \half \epsilon A(m,m) \cos(\kappa.x + \chi) \right)\,,
\end{equation}
where we have used the notation
\begin{equation}
	A(m,m) = \t A{^i^j} \t m{_i} \t m{_j}\,.
\end{equation}
The associated normal vector field  $\t \nu{^\alpha} = \t g{^\alpha^\beta} \t \nu{_\beta}$
is thus
\begin{equation}
	\label{eq:perturbed normal}
	\t \nu{^i}
		= \t \delta{^i^j} \t m{_j} \left( 1 + \half \epsilon A(m,m) \cos(\kappa.x + \chi) \right)
		- \epsilon \t A{^i^j} \t m{_j} \cos(\kappa.x + \chi) \,.
\end{equation}

Analogously to the discussion of Maxwell fields in \Cref{s:boundary value problem}, our boundary conditions \eqref{eq:boundary conditions laser} fit well into both the Cauchy-Kovalevskaya theorem, which provides a solution in our setting, and the Holmgren uniqueness theorem, which asserts that solutions of such boundary value problems are unique within the class of smooth solutions.
In any case this provides a toy model for the problem at hand, and we provide the unique solution below up to, experimentally irrelevant, error terms $O(\epsilon^2)$.

For the perturbative setting, consider first $\epsilon = 0$ (unperturbed problem) where the solution is evidently given by
\begin{align}
	\phi^\o0
		&= e^{i k.x}\,,
\shortintertext{where}
	\label{eq:wave vector unperturbed}
	\t k{_\mu} &= \omega(-1, n m)\,.
\end{align}
In the perturbed case, we thus write
\begin{equation}
	\phi
		= \phi^\o0 + \epsilon \phi^\o1 + O(\epsilon^2)\,.
\end{equation}
Inserting this into the scalar wave equation \eqref{eq:scalar wave equation} and the boundary conditions \eqref{eq:boundary conditions laser} leads, at first order in $\epsilon$, to the boundary value problem
\begin{equation}
	\label{eq:scalar wave problem 1}
	\begin{aligned}
	\wop_{\gamma^\o0} \phi^\o1
		&= - n^2 \omega^2 A(m,m) \cos(\kappa.x + \chi) e^{i k.x}\,,
	\\
	\phi^\o1|_\Sigma
		&= 0\,,
	\\
	m(\phi^\o1)|_\Sigma
		&= \ihalf n \omega A(m,m) \cos(\kappa.x + \chi) e^{- i \omega t}\,.
	\end{aligned}
\end{equation}
Of course, the right-hand sides are also evaluated at $\Sigma$ whenever the left-hand sides are.

\subsubsection{General Expression for Emitted Waves}
\label{s:emission general}

Since we will later consider light emission modelled by Maxwell fields, we will now solve a slightly more general problem (suppressing the superscript “$\o1$” momentarily):
\begin{align}
	\wop_{\gamma^\o0} \phi
		&= - \omega^2 f_1 e^{i k.x}\,,
	\\
	\phi|_\Sigma
		&= f_2 e^{- i \omega t}\,,
	\\
	m(\phi)|_\Sigma
		&= i n \omega f_3 e^{- i \omega t}\,,
\end{align}
where $f_1$ is a trigonometric function (by which we mean  a linear combination of $\sin$ and $\cos$) of $\kappa.x + \chi$, and $f_2, f_3$ are linear combinations of such functions of  $\kappa.x|_\Sigma + \chi$,  $\chi - \omegaG t$, and possibly involve additive constants.
This form covers all the boundary values problems, arising in the description of light emission,  considered in this work.

We find it useful to expand the trigonometric functions in terms of complex exponentials as follows:
\begin{align}
	\label{eq:scalar problem split 1}
	f_1
		&= \alpha^+ e^{+i (\kappa.x + \chi)} + \alpha^- e^{-i (\kappa.x + \chi)}\,,
	\\
	\label{eq:scalar problem split 2}
	f_2
		&= \beta_1^+ e^{+i (\kappa.x + \chi)} + \beta_1^- e^{-i (\kappa.x + \chi)}
			+ \gamma_1^+ e^{+i(\chi - \omegaG t)} + \gamma_1^- e^{-i(\chi - \omegaG t)}
			+ 2 \delta_1\,,
	\\
	\label{eq:scalar problem split 3}
	f_3
		&= \beta_2^+ e^{+i (\kappa.x + \chi)} +  \beta_2^- e^{-i (\kappa.x + \chi)}
			+ \gamma_2^+ e^{+i(\chi - \omegaG t)} + \gamma_2^- e^{-i(\chi - \omegaG t)}
			+ 2 \delta_2\,,
\end{align}
with the right-hand sides of
\eqref{eq:scalar problem split 2} and \eqref{eq:scalar problem split 3} implicitly restricted to $\Sigma$.
The factor $2$ in front of $\delta_1$ and $\delta_2$ is introduced for notational simplicity in the calculations which follow.
The solution to this problem can then be written as
\begin{equation}
	\label{eq:field decomposition pm}
	\phi = \phi^+ + \phi^-\,,
\end{equation}
where $\phi^+ \equiv \phi^+(\alpha^+, \beta_1^+, \beta_2^+, \gamma_1^+, \gamma_2^+)$ is the solution to the problem
\begin{align}
	\wop_{\gamma^\o0} \phi^+
		&= - \alpha^+ \omega^2 e^{i(k + \kappa).x + i \chi}\,,
		\\
	\phi^+|_\Sigma
		&= (\beta_1^+ e^{i (\kappa.x + \chi)}
			+ \gamma_1^+ e^{i (\chi - \omegaG t)}
			+ \delta_1 )e^{- i \omega t}\,,
		\\
	m(\phi^+)|_\Sigma
		&= i n \omega (\beta_2 e^{i (\kappa.x + \chi)}
			+ \gamma_2^+ e^{i (\chi - \omegaG t)}
			+ \delta_2 )e^{- i \omega t}\,,
\end{align}
which we will construct shortly, and $\phi^-$ is obtained from $\phi^+$ by replacing the various coefficients $\alpha^+, \beta_1^+, \ldots$ by the corresponding quantities $\alpha^-, \beta_1^-, \ldots$, and reversing the sign of both $\omegaG$ and $\chi$.

In fact, it is not necessary to write down the phase offset $\chi$ explicitly.
Indeed, given a solution for $\chi = 0$, the general case is obtained by the substitution
\begin{equation}
	\label{eq:GW phase substitution rule}
	(\alpha^\pm, \beta_{1,2}^\pm, \gamma_{1,2}^\pm)
		\to e^{\pm i \chi} (\alpha^\pm, \beta_{1,2}^\pm, \gamma_{1,2}^\pm)\,,
\end{equation}
while keeping $\delta_{1,2}$ unchanged.

Hence, we consider the boundary value problem
\begin{align}
	\label{eq:problem wave equation}
	\wop_{\gamma^\o0} \phi
		&= -\omega^2 \alpha e^{i(k + \kappa).x}\,,
		\\
	\label{eq:problem value}
	\phi|_\Sigma
		&= (\beta_1 e^{i \kappa.x}
			+ \gamma_1 e^{- i \omegaG t}
			+ \delta_1 )e^{- i \omega t}\,,
		\\
	\label{eq:problem derivative}
	m(\phi)|_\Sigma
		&= i n \omega (\beta_2 e^{i \kappa.x}
			+ \gamma_2 e^{- i \omegaG t}
			+ \delta_2 )e^{- i \omega t}\,,
\end{align}
for generic parameters $\alpha, \beta_{1,2}, \gamma_{1,2}, \delta_{1,2}$.
The wave-vectors occurring here are
\begin{align}
	\t k{_\mu}
		&= \omega(-1, n m)\,,
	&
	\t \kappa{_\mu}
		&= \omegaG(-1, \kh)\,.
\end{align}
We constrict explicitly a solution $\phi = \phi(\alpha^+, \beta^+, \gamma^+, \delta)$, which is the unique solution by Holmgren’s theorem.

\paragraph{Construction of the Solution}

A particular solution to the inhomogeneous wave equation \eqref{eq:problem wave equation} is
\begin{align}
	\label{eq:scalar wave particular solution}
	\phi_\particular
		&= - \frac{\alpha}{\Omega \sigma}e^{i(k + \kappa).x}\,,
	\\
\shortintertext{where}
	\label{eq:scalar wave sigma}
	\sigma
		&= - \frac{\t \gamma{^{\o0}^\mu^\nu} \t {(k+\kappa)}{_\mu} \t {(k+\kappa)}{_\nu}}{\omega \omegaG}
		= 2 n (n - \mn) + \Omega (n^2 -1)\,,
\end{align}
provided that $\Omega$ and $\sigma$ do not vanish. We will assume this in all calculations that follow, and consider the low frequency limit $\Omega \to 0$ and the “collinear vacuum limit” (where $n = 1$ and $\mn \to 1$ so that $\sigma\to 0$) at the end of our calculations.

To satisfy the boundary conditions, we add plane waves which restrict to the desired behaviour on $\Sigma$.
The functions $\exp(i(k + \kappa - \omega \zeta m).x)$ and $\exp(i(k - \omega \xi m).x - i \omegaG t)$ satisfy the homogeneous wave equation for the following values of $\zeta$ and $\xi$:
\begin{align}
	\label{eq:scalar wave zeta}
	\zeta_{1,2}
		&= n + \Omega \mn
		\mp \sqrt{(n + \Omega \mn)^2 + \Omega \sigma}\,,
	\\
	\xi_{1,2}
		&= n \mp n( 1 + \Omega)\,,
\end{align}
where the first (second) subscript index refers to the upper (lower) sign.
Defining also the reflected wave vector
\begin{equation}
	\t {\check k}{_\mu}
		= \omega(-1, - n m)\,,
\end{equation}
we make the ansatz
\begin{align}
	\label{eq:phi decomposition}
	\phi
		&= \phi_S + \phi_R\,,
	\\
\shortintertext{with}
	\label{eq:phi part s}
	\phi_S
		&= - \frac{\alpha}{\Omega \sigma}e^{i(k + \kappa).x} \left(
			1 - e^{-i \omega \zeta_1 m.x}
		\right)\,,
	\\
	\label{eq:phi part r}
	\begin{split}
	\phi_R
		&= \left( \lambda_1 e^{-i \omega \zeta_1 m.x} + \lambda_2 e^{-i \omega \zeta_2 m.x} \right)e^{i(k + \kappa).x}\\
		&+ \left( \mu_1 e^{-i \omega \xi_1 m.x} + \mu_2 e^{-i \omega \xi_2 m.x} \right)e^{i(k.x - \omegaG t)}
		+ \nu_1 e^{i k.x} + \nu_2 e^{i \check k.x}
		\,.
	\end{split}
\end{align}
Here we have split the function $\phi$ into a “seemingly singular part” $\phi_S$ which is undefined for $\Omega \sigma = 0$, and a remainder $\phi_R$. As we will see, both $\phi_S$ ad $\phi_R$ have well-behaved low-frequency and collinear vacuum limits.
This ansatz satisfies the boundary conditions \eqref{eq:problem value} and \eqref{eq:problem derivative} if
\begin{align}
	\label{eq:scalar boundary system 1}
	\lambda_1 + \lambda_2
		&= \beta_1\,,
	&
	\zeta_1 \lambda_1 + \zeta_2 \lambda_2
		&= \beta_1 (n + \Omega \mn)
		- n \beta_2 - \frac{\alpha \zeta_1}{\Omega \sigma}\,,
	\\
	\mu_1 + \mu_2
		&= \gamma_1\,,
	&
	\xi_1 \mu_1 + \xi_2 \mu_2
		&= n(\gamma_1 - \gamma_2)\,,
	\\
	\nu_1 + \nu_2
		&= \delta_1\,,
	&
	\nu_1 - \nu_2
		&= \delta_2\,,
\end{align}
with the solution
\begin{align}
	\lambda_{1,2}
		&= \frac{1}{2} \left(
			\beta_1
			\pm \frac{
				n \beta_2 + \frac{\alpha \zeta_1}{\Omega \sigma}
			}{\sqrt{(n + \Omega \mn)^2 + \Omega \sigma}}
		\right)\,,
	\\
	\mu_{1,2}
		&= \frac{1}{2} \left(
			\gamma_1 \pm \frac{\gamma_2}{1 + \Omega}
		\right)\,,
	\\
	\label{eq:scalar boundary solution 3}
	\nu_{1,2}
		&= \half (\delta_1 \pm \delta_2)\,.
\end{align}
Recall that for computational reasons $\Omega$ is allowed to be negative, but since we are interested in the low-frequency regime $|\Omega| < 1$, the case $1 + \Omega = 0$ is irrelevant for our purposes and, in fact, we will assume $1 + \Omega > 0$.

\paragraph{Emission Condition}
Note that $\zeta_1$ and $\xi_2$ are of order $\Omega$. The plane waves multiplied by $\lambda_1$, $\mu_1$ and $\nu_1$ thus have wave vectors which are close to the unperturbed wave vector $k$. In contrast, since $\zeta_2$ and $\xi_2$ have an expansion of the form $2 n + O(\Omega)$, the plane waves multiplied by $\lambda_2$, $\mu_2$ and $\nu_2$ correspond to “counter-propagating waves” whose wave vectors are close to the \emph{reflected} wave vector $\check k = k - 2 n \omega m$.

The condition for absence of such counter-propagating components is $\lambda_2 = \mu_2 = \nu_2 = 0$, which translates to the following conditions on the boundary data:
\begin{align}
	\label{eq:emission condition full}
	\beta_1 \sqrt{(n + \Omega \mn)^2 + \Omega \sigma}
		- n \beta_2 - \frac{\alpha \zeta_1}{\Omega \sigma} &= 0\,,
	&
	(1 + \Omega) \gamma_1 - \gamma_2 &= 0\,,
	&
	\delta_1 - \delta_2 &= 0\,.
\end{align}
 These conditions could have been imposed from the beginning, but the analysis without assuming \eqref{eq:emission condition full}  makes it clear that only a limited subclass of solutions are of the eikonal form.
If these conditions are met, the solution simplifies to
\begin{equation}
	\label{eq:scalar simplified 1}
	\phi/e^{i k.x}
		= -\frac{\alpha}{\Omega \sigma} \left( e^{i \kappa.x} - e^{i (\kappa.x - \omega \zeta_1 m.x)} \right)
		+ \beta_1 e^{i (\kappa.x - \omega \zeta_1 m.x)}
		+ \gamma_1 e^{-i (\omegaG t + \omega \xi_1 m.x) }
		+ \delta_1\,.
\end{equation}

We now consider the following two limiting cases which have been excluded so far:
(i) the collinear vacuum limit where $n = 1$ and $\mn \to 1$, so that $\sigma \to 0$ according to \eqref{eq:scalar wave sigma}, and (ii) the low frequency limit $\Omega \to 0$.
Both singularities turn out to be removable.

\paragraph{Collinear Vacuum Limit}

Consider first the collinear vacuum limit of \eqref{eq:scalar simplified 1}, where $n = 1$ and $\mn \to 1$.
The last two terms in \eqref{eq:scalar simplified 1} remain unchanged in this limit, and the second to last term tends to $\beta_1 e^{i(\kappa.x)}$ since $\zeta_1 \to 0$ in this limit.
Finally, for the remaining term, one has
\begin{equation}
	\lim_{\mn \to 1}
		\frac{1}{\Omega \sigma} \left( e^{-i \omega \zeta_1 m.x} - 1\right)
		= \frac{i \omega m.x}{2( 1+ \Omega)}\,,
\end{equation}
so all terms in \eqref{eq:scalar simplified 1} have a finite collinear vacuum limit.
This shows that the singularity at $\mn = 1$ and $n = 1$ is removable.

\paragraph{Low-Frequency Limit}

Next, consider the limit $\Omega \to 0$ of \eqref{eq:scalar simplified 1}.
There, one has
\begin{equation}
	\lim_{\Omega \to 0}
		\frac{1}{\Omega \sigma} \left( e^{-i \omega \zeta_1 m.x} - 1\right)
		= \frac{i \omega m.x}{2 n}\,.
\end{equation}
Hence, the limit $\Omega \to 0$ is also well-behaved and the singularity at $\Omega = 0$ is removable.

\paragraph{Eikonal Expansion}

Using the explicit result \eqref{eq:scalar simplified 1}, one can now prove the claims made in \Cref{s:overview eikonal}.
The formula just obtained is already reminiscent of \eqref{eq:overview perturbation factoring}, but the exact details require further analysis.

Since $\sigma$ is analytic near $\Omega = 0$ with the expansion $\sigma = 2 n (n - \mn) + O(\Omega)$, one has
\begin{equation}
	\frac{1}{\Omega \sigma}
	= \frac{1}{2 n \Omega (n - \mn)} + f(\Omega)\,;
\end{equation}
here and in what follows $f(\Omega)$ denotes a generic function which may change from line to line and which is analytic in $\Omega$ in a neighbourhood of zero.
Similarly, $\zeta_1$ is analytic in $\Omega$ in a neighbourhood of zero and has the expansion $\zeta_1 = \Omega (\mn - n) + O(\Omega^2)$. Consequently, $\kappa.x - \omega \zeta_1 m.x = \kappa_0.x + f(\Omega) \omegaG m.x$,
where
\begin{equation}
	\label{eq:kappa.x_0}
	\kappa_0.x := \kappa.x + (n - \mn) \omegaG m.x\,.
\end{equation}
It then follows that
\begin{equation}
	-\frac{\alpha}{\Omega \sigma} \left( e^{i \kappa.x} - e^{i (\kappa.x - \omega \zeta_1 m.x)} \right)
	= - \frac{\alpha}{2 n \Omega} \frac{e^{i \kappa.x} - e^{i \kappa_0.x}}{n - \mn}
	+ f(\Omega, \kappa.x, \omegaG m.x)\,.
\end{equation}
Similarly, the remaining terms in \eqref{eq:scalar simplified 1} are of the form $f(\Omega, \kappa.x, \omegaG t, \omegaG m.x)$: the $\beta_1$-term is of the form just considered, and the $\gamma_1$-term can be written as $\gamma_1 e^{i \omegaG (t - n m.x)}$.
Putting all this together, one arrives at the result
\begin{equation}
	\phi/e^{i k.x} = - \frac{\alpha}{2 n \Omega} \frac{e^{i \kappa.x} - e^{i \kappa_0.x}}{n - \mn} e^{i \kappa.x}
		+ f(\Omega, \kappa.x, \omegaG t, \omegaG m.x)\,.
\end{equation}

We are primarily concerned with perturbations of plane waves solutions, with the plane wave taking the form $\phi^\o0 = \varphi^\o0 e^{i k.x}$ for some constant $\varphi^\o0$.
For $\varphi^\o0 \neq 0$ it suffices to consider the case $\varphi^\o0 = 1$ by linearity.
In this case, the structure of the wave equation \eqref{eq:overview wave equation} implies that $\alpha = \half n^2 A(m,m) + O(\Omega)$ so that the overall correction is found to be
\begin{equation}
	\phi^\o1
		= i \psi^\o1 + f(\Omega, \kappa.x, \omegaG t, \omegaG m.x)\,,
\end{equation}
where
\begin{equation}
	\psi^\o1 = - \half n A(m,m) \frac{\sin(\kappa.x + \chi) - \sin(\kappa_0.x + \chi)}{\Omega (n - \mn)}\,.
\end{equation}
By a direct calculation one readily verifies that this function satisfies
\begin{equation}
	\t\gamma{^{\o0}^\mu^\nu} \t k{_\mu} \t\p{_\nu} \psi^\o1 = \half n^2 \omega^2 h(m,m)\,,
\end{equation}
which is the perturbed eikonal equation \eqref{eq;overview eikonal eq perturbation}.

Alternatively, if the unperturbed field has $\varphi^\o0 = 0$ (this occurs, for example, for polarized EM\ plane waves, where two components of each of the unperturbed electric and magnetic fields are zero), $\alpha$ is of order $\Omega$ and one obtains instead a formula of the kind $\phi^\o1 = f(\Omega, \kappa.x, \omegaG t, \omegaG m.x)$ without a leading $1/\Omega$ term.

All in all, this proves the claims made in \Cref{s:overview eikonal}.

\paragraph{Next-To-Leading Order Expansion}
For later applications, we give an approximate form of \eqref{eq:scalar simplified 1} which takes into account both the perturbation of the phase, and also the first correction of the amplitude. Since phase perturbations arise at order $\Omega^{-1}$ and the first amplitude corrections arise at order $\Omega^0$, the desired approximation is obtained by neglecting terms of order $\Omega$ and higher.

At this level of accuracy, the emission conditions  \eqref{eq:emission condition full} simplify to yield
\begin{align}
	\label{eq:scalar emission condition}
	\frac{\alpha}{2n^2} + \beta_1 - \beta_2 &= O(\Omega)\,,
	&
	\gamma_1 - \gamma_2 &= O(\Omega)\,,
	&
	\delta_1 - \delta_2 &= O(\Omega)\,.
\end{align}
If these conditions are satisfied, the field perturbation is given by \eqref{eq:scalar simplified 1}, in which case one may approximate the phases as
\begin{align}
	\label{eq:emitted phase expansion}
	\kappa.x - \omega \zeta_1 m.x
		&= \kappa_0.x - \frac{1 - (\mn)^2}{2 n} \Omega \omegaG m.x
		+ O(\Omega^2 \omegaG m.x)\,,
		\\
	- \omegaG t - \omega \xi_1 m.x
		&= \Omega k.x\,,
\end{align}
where $\kappa_0.x$ is defined in \eqref{eq:kappa.x_0}.
Expanding the exponentials in this way and using
\begin{equation}
	\label{eq:emitted amplitude expansion}
	\frac{1}{\Omega \sigma}
		= \frac{1}{2 \Omega n (n - \mn)} - \frac{n^2 - 1}{4 n^2 (n - \mn)^2} + O(\Omega)\,,
\end{equation}
one arrives at
\begin{equation}
	\label{eq:scalar template emission}
	\begin{split}
		\phi^\pm/e^{i k.x} ={}&
			\mp \alpha^\pm\, \frac{e^{\pm i \kappa.x} - e^{\pm i \kappa_0.x}}{2 \Omega n (n - \mn)}
			+ \alpha^\pm (n^2 - 1) \frac{e^{\pm i \kappa.x} - e^{\pm i \kappa_0.x}}{4 n^2 (n - \mn)^2}
			\\&
			+ \left(
				\beta_1^\pm
				\mp i \alpha^\pm \frac{1 - (\mn)^2}{4 n^2(n - \mn)} \omegaG m.x
			\right) e^{\pm i \kappa_0.x}
			\\&
			+ \gamma_1^\pm e^{\pm i \omegaG (n m.x - t)}
			+ \delta_1
			+ O(\Omega \omegaG m.x) + O(\Omega)
			\,,
	\end{split}
\end{equation}
where the first error term grows with the distance $m.x$, but the second is uniform in the distance.

\subsubsection{Emitted Scalar Wave}

Let us now specialize the general solution just constructed to the concrete problem \eqref{eq:scalar wave problem 1}.
Then the coefficients in
\eqref{eq:scalar problem split 1}—\eqref{eq:scalar problem split 3} read
\begin{align}
	\label{eq:scalar emitted parameters}
	\alpha^\pm &= \half n^2 A(m,m)\,,
	&
	\beta_1^\pm &= 0\,,
	&
	\beta_2^\pm &= \quarter A(m,m)\,,
\end{align}
and the other parameters $\gamma_{1,2}^\pm$, $\delta_{1,2}$ vanish.
These parameters do satisfy the conditions \eqref{eq:scalar emission condition}, so that one may use \eqref{eq:scalar template emission} to obtain
\begin{equation}
\begin{split}
	\phi^\pm
	= - \half A(m,m) e^{i k.x}\bigg[
	&
		\pm \frac{n}{2 \Omega}
		\frac{e^{\pm i \kappa.x} - e^{\pm i \kappa_0.x}}{n - \mn}
	- (n^2 - 1)\frac{e^{\pm i \kappa.x} - e^{\pm i \kappa_0.x}}{4 (n - \mn)^2}
	\\
	&
		\pm i \frac{1 - (\mn)^2}{4(n - \mn)} \omegaG m.x e^{\pm  i \kappa_0.x}
	\bigg]
	+ O(\Omega \omegaG m.x) + O(\Omega)\,,
\end{split}
\end{equation}
Computing $\phi^\o1 = \phi^+ + \phi^-$ and restoring the gravitational wave phase shift $\chi$, the first order perturbation is found to be
\begin{equation}
	\label{eq:scalar emission result}
	\begin{split}
			\phi^\o1 = - \half A(m,m) e^{i k.x}\bigg[
			&
			\frac{i n}{\Omega}
			\frac{\sin(\kappa.x + \chi) - \sin(\kappa_0.x + \chi)}{n - \mn}
			\\&
			- (n^2 - 1)\frac{\cos(\kappa.x + \chi) - \cos(\kappa_0.x + \chi)}{2 (n - \mn)^2}
			\\&
			- \frac{1 - (\mn)^2}{2(n - \mn)} \omegaG m.x \sin(\kappa_0.x + \chi)
		\bigg]
		+ O(\Omega \omegaG m.x) + O(\Omega) \,.
	\end{split}
\end{equation}
Since derivatives of the first term in brackets are of order $\omega$, while those of the remaining terms are of order $\omegaG$, we interpret the purely imaginary first term as a phase correction (recall that in geometric optics, the eikonal is assumed to be rapidly varying, while the amplitude is slowly varying), and write the resulting overall field as
\begin{align}
	\label{eq:scalar field emitted}
	\phi
		&= \mathscr A e^{i \psi} + O(\epsilon^2) + O(\epsilon \Omega) + O(\epsilon \Omega \omegaG m.x)\,,
	\\
\shortintertext{where}
	\begin{split}
		\mathscr A
			&= 1
			+ \quarter \epsilon A(m,m) \bigg[
				\frac{1 - (\mn)^2}{n - \mn} \sin(\kappa_0.x + \chi) \omegaG m.x\\
				&\hspace{3cm}
				+ (n^2 - 1) \frac{\cos(\kappa.x + \chi) - \cos(\kappa_0.x + \chi)}{(n - \mn)^2}
			\bigg]
			\,,
	\end{split}
		\\
	\psi
		&= k.x - \half \epsilon n A(m,m) \frac{\sin(\kappa.x + \chi) - \sin(\kappa_0.x + \chi)}{\Omega (n - \mn)}\,.
\end{align}
%
\checkedtogether{29I21} 

%% file: optics_subfiles/scalar_reflection.tex
\subsection{Reflection At a Mirror}

We model the mirror by a timelike hypersurface $\Sigma'$ with the following reflection properties.
An impinging wave $\phi$ gives rise to a reflected wave $\check \phi$ such that
\begin{enumerate}
	\item[(a)] the overall field $\phi + \check \phi$ vanishes on $\Sigma'$, and
	\item[(b)] the normal derivative of $\check \phi$ is the same as that of $\phi$.
\end{enumerate}
This is made precise by the equations
\begin{align}
\label{eq:boundary conditions mirror}
	(\check \phi + \phi)|_{\Sigma'} &= 0\,,
	&
	\nu(\check \phi - \phi)|_{\Sigma'} &= 0\,.
\end{align}

To illustrate this in flat space-time, consider a plane wave $\phi = \exp(i k.x)$ with $\t{k}{_\mu} = (-\omega, k_1, k_2, k_3)$ impinging on a mirror $\Sigma'$, which is described by $x = 0$.
Choosing the normal $\nu $ to be $ \p/\p x^1$ one has $\nu(\phi)|_{\Sigma'} = i k_1 \phi|_{\Sigma'}$.
The reflected wave $\check \phi = - \exp(i \check k.x)$ with $\t{\check k}{_\mu} = (-\omega, -k_1, k_2, k_3)$ then restricts to $-\phi$ on $\Sigma'$ and its normal derivative is
\begin{equation}
	\nu(\check \phi)|_{\Sigma'}
		= i \t {\check k}{_1} \check \phi|_{\Sigma'}
		= - i \t k{_1} \check \phi|_{\Sigma'}
		= i \t k{_1} \phi|_{\Sigma'}
		= \nu(\phi)|_{\Sigma}\,.
\end{equation}
Since generic waves can be decomposed into plane waves and this argument applies to every Fourier component individually, these conditions are not restricted to plane waves.

In the case of normal incidence in flat spacetime we have $k_\mu = (-\omega, n \omega, 0, 0)$ and $\check k_\mu = (- \omega, - n\omega, 0, 0)$. However, we shall work in coordinates which are not necessarily adapted to the reflection surface, with both $\Sigma$ and $\Sigma'$ given by $m.x = \const$.

Here it should be kept in mind that different components of the electromagnetic field behave differently at mirrors. We will use the current model for scalar waves to get a first insight, before considering the full problem using Maxwell’s equations below.

As a simple model for the mirror, we consider the hypersurface
\begin{equation}
	\Sigma' = \{ \t m{_i} \t x{^i} = \ell \} \,,
\end{equation}
where the coefficients $m_i$ are the same as for the emission surface $\Sigma$.
The emission and reflection surfaces are thus parallel as coordinate-planes and separated by a coordinate distance $\ell$.
\checkedtogether{1II21}

\subsubsection{General Expression for Reflected Waves}
\label{s:reflection general}

Before computing the reflected field from the incident field \eqref{eq:scalar field emitted}, we consider a more general problem (much like \Cref{s:emission general}) and specialize to the concrete problem later. This will be particularly useful for the discussion of Maxwell fields in \Cref{s:Maxwell wave equation reflection}.

Proceeding as in \eqref{eq:field decomposition pm} and following, we thus consider the boundary value problem
\begin{align}
	\label{eq:reflection problem 1}
	\wop_{\gamma^\o0} \check \phi
		&= - \check \alpha \omega^2 e^{i (\check k + \kappa).x + i n \omega \ell}\,,
	\\
	\label{eq:reflection problem 2}
	\check \phi|_{\Sigma'}
		&= \left(
			\check \beta_1 e^{i \kappa.x}
			+ \check \gamma_1 e^{-i \omegaG t}
			+ \check \delta_1
		\right) e^{-i \omega t }\,,
	\\
	\label{eq:reflection problem 3}
	m(\check \phi)|_{\Sigma'}
		&= - i n \omega \left(
			\check \beta_2 e^{i \kappa.x}
			+ \check \gamma_2 e^{-i \omegaG t}
			+ \check \delta_2
		\right) e^{-i \omega t}
	\,,
\end{align}
where the relevant wave vectors are
\begin{align}
	\t k{_\mu}
		&= \omega(-1, n m)\,,
	&
	\t {\check k}{_\mu}
		&= \omega(-1, - n m)\,,
	&
	\t \kappa{_\mu}
		&= \omegaG(-1, \kh)\,.
\label{24V21.51}
\end{align}
Here the phase of the source term in the right-hand side was chosen for convenience ($\check k.x + n \omega \ell$ restricts to $-\omega t$ on $\Sigma'$). This is no restriction, as additional phase shifts can always be absorbed in the coefficients $\check \alpha$, $\check \beta_{1,2}$ etc.

As in the previous discussion, we work with a vanishing gravitational phase offset $\chi$.
The general case of non-zero $\chi$ is obtained by a substitution similar to \eqref{eq:GW phase substitution rule}.

The analysis proceeds in almost the same way as in \Cref{s:emission general}.
In fact, the only modification is the sign change of $m$ and the addition of phases of the form $e^{i n \omega \ell}$. In a way similar to \eqref{eq:phi decomposition}, we write
\begin{align}
	\check \phi
		&= \check \phi_S + \check \phi_R\,,
	\\
\shortintertext{where}
	\label{eq:reflected general S}
	\check \phi_S
		&= - \frac{\check \alpha}{\Omega \check \sigma} e^{i(\check k + \kappa).x + i n \omega \ell} \left( 1 - e^{+i \omega \check\zeta_1 (m.x-\ell) } \right)\,,
	\\
	\label{eq:reflected general R}
	\begin{split}
		\check \phi_R
		&= \left(
				\check \lambda_1 e^{i \omega \check \zeta_1 (m.x - \ell)}
				+ \check \lambda_2 e^{i \omega \check \zeta_2 (m.x - \ell)}
			\right) e^{i(\check k + \kappa).x + i n \omega \ell}
		\\
		&+ \left(
			\check \mu_1 e^{i \omega \check \xi_1 (m.x - \ell)}
			+ \check \mu_2 e^{i \omega \check \xi_2 (m.x - \ell)}
		\right) e^{i(\check k.x - \omegaG t + n \omega \ell)}
		\\
		&+ \check \nu_1 e^{i(\check k.x + n \omega \ell)}
		+ \check \nu_2 e^{i( k.x - n \omega \ell)}.
		\end{split}
\end{align}
Here the coefficients $\check \sigma$, $\check \zeta_{1,2}$ and $\check \xi_{1,2}$ are
\begin{align}
	\check \sigma
		&= 2 n (n + \mn) + \Omega (n^2 - 1)\,,
	\\
	\check \zeta_{1,2}
		&= n - \Omega \mn \mp \sqrt{( n - \Omega \mn)^2 + \Omega \check \sigma}\,,
	\\
	\check \xi_{1,2}
		&= n \mp n ( 1 + \Omega)\,,
\end{align}
which differ from the previously defined quantities $\sigma$, $\zeta_{1,2}$ and $\xi_{1,2}$ merely by change of sign of $m$ (in particular, $\check \xi_{1,2} = \xi_{1,2}$).

To implement the boundary conditions, we now impose
\begin{align}
	\check \lambda_1 + \check \lambda_2
		&= \check \beta_1\,,
	&
	\check \zeta_1 \check \lambda_1 + \check \zeta_2 \check \lambda_2
		&= \check \beta_1 (n - \Omega \mn) - n \check \beta_2 - \frac{\check \alpha \check \zeta_1}{\Omega \check \sigma} \,,
	\\
	\check \mu_1 + \check \mu_2
		&= \check \gamma_1\,,
	&
	\check \xi_1 \check \mu_1 + \check \xi_2 \check \mu_2
		&= n (\check \gamma_1 - \check \gamma_2)\,,
	\\
	\check \nu_1 + \check \nu_2
		&= \check \delta_1\,,
	&
	\check \nu_2 - \check \nu_2
		&= \check \delta_2\,,
\end{align}
with the solution
\begin{align}
	\label{eq:reflected lambda}
	\check \lambda_{1,2}
		&= \frac{1}{2} \left(
			\check \beta_1
			\pm \frac{
				n \check \beta_2 + \frac{\check \alpha \check \zeta_1}{\Omega \check \sigma}
			}{\sqrt{(n - \Omega \mn)^2 + \Omega \check \sigma}}
		\right)\,,
	\\
	\check \mu_{1,2}
		&= \frac{1}{2} \left(
			\check \gamma_1 \pm \frac{\check \gamma_2}{1 + \Omega}
		\right)\,,
	\\
	\check \nu_{1,2}
		&= \half (\check \delta_1 \pm \check \delta_2)\,,
\end{align}
cf.\ \eqref{eq:scalar boundary system 1} — \eqref{eq:scalar boundary solution 3}.
\checkedtogether{1II21}

\paragraph{Reflection Condition}
As for the emitted wave, we are primarily interested in the case where no counter-propagating terms arise, which means $\check \lambda_2 = \check \mu_2 = \check \nu_2 = 0$, or equivalently
\begin{align}
	\label{eq:reflection condition full}
	\sqrt{(n - \Omega \mn)^2 + \Omega \check \sigma} \check \beta_1
		- n \check \beta_2 - \frac{\check \alpha \check \zeta_1}{\Omega \check \sigma} &= 0\,,
	&
	(1 + \Omega) \check\gamma_1 - \check\gamma_2 &= 0\,,
	&
	\check\delta_1 - \check\delta_2 &= 0\,.
\end{align}
If these conditions are met, the field simplifies to
\begin{equation}
	\label{eq:scalar simplified 2}
	\begin{split}
		\check \phi/e^{i(k.x + n \omega \ell)}
			={}& - \frac{\check\alpha }{\Omega \check\sigma} \left(
				e^{i \kappa.x} - e^{i \kappa.x + i \omega \check\zeta_1 (m.x - \ell)}
			\right)
			\\&
			+ \check\beta_1 e^{i(\kappa.x + \omega \check\zeta_1 (m.x-\ell))}
			+ \check\gamma_1 e^{-i\omegaG(t + n(m.x-\ell))}
			+ \check\delta_1\,.
	\end{split}
\end{equation}

\paragraph{Next-To-Leading Order Expansion}
Expanding \eqref{eq:reflection condition full} in powers of $\Omega$, and considering that the boundary values for the reflected wave may contain $1/\Omega$ terms, the condition for the absence of counter-propagating terms at first order in $\Omega$ is found to be
\begin{align}
	\label{eq:counter-propagation reflection}
	\frac{\check \alpha}{2 n^2} + \check \beta_1 - \check \beta_2 (1 - \Omega) &= O(\Omega)\,,
	&
	\check \gamma_1 - \check \gamma_2(1 - \Omega) &= O(\Omega)\,,
	&
	\check \delta_1 - \check \delta_2 &= O(\Omega)\,.
\end{align}
Expanding then the first coefficient in \eqref{eq:scalar simplified 2} as
\begin{equation}
	\frac{1}{\Omega \check \sigma}
		= \frac{1}{2 \Omega n(n + \mn)} - \frac{n^2 - 1}{4 n^2 (n + \mn)^2} + O(\Omega)\,,
\end{equation}
cf.\ \eqref{eq:emitted amplitude expansion}, and expanding the phase as
\begin{equation}
	\label{eq:reflected phase expansion}
	\kappa.x + \check \zeta_1 (m.x - \ell)
		= \kappa_1.x - \frac{1 - (\mn)^2}{2 n} \Omega \omegaG(\ell - m.x)
		+ O(\Omega^2 \omegaG(\ell - m.x))\,,
\end{equation}
where
\begin{equation}
	\kappa_1.x
		= \kappa.x + (n + \mn) \omegaG (\ell - m.x)\,,
\end{equation}
cf.\ \eqref{eq:emitted phase expansion}, one arrives at
\begin{equation}
	\label{eq:scalar template reflection}
	\begin{split}
		\check \phi^\pm/e^{i (\check k.x + n \omega \ell)}
			=&{} \mp \check \alpha^\pm \frac{e^{\pm i \kappa.x} - e^{\pm i \kappa_1.x}}{2 \Omega n (n + \mn)}
			+ \check \alpha^\pm (n^2 - 1) \frac{e^{\pm i \kappa.x} - e^{\pm i \kappa_1.x}}{4 n^2 (n + \mn)^2}
			+ \check \beta_1^\pm e^{\pm i \kappa_1.x}
			\\&
			\mp i \frac{1 - (\mn)^2}{2 n} \left[
				\frac{\check \alpha^\pm}{2 n (n + \mn)} \pm \Omega \check \beta_1^\pm
				\right] \omegaG( \ell - m.x) e^{\pm i \kappa_1.x}
			\\&
			+ \check\gamma_1^\pm e^{\mp i\omegaG(t + n(m.x-\ell))}
			+ \check\delta_1^\pm
			+ O(\Omega \omegaG(\ell - m.x))
			+ O(\Omega)	\,.
	\end{split}
\end{equation}
As in \eqref{eq:scalar template emission}, the first error term grows with the distance from $\Sigma'$, while the second one is uniform in the distance.

\subsubsection{Reflected Scalar Wave}

In the unperturbed problem  $\epsilon = 0$, the reflected wave is given by
\begin{align}
	\label{eq:scalar reflected unperturbed}
	\check \phi^\o0
		&= - e^{i \check k.x + 2 i n \omega \ell}\,,
\shortintertext{where}
	\t{\check k}{_\mu}
		&= \omega(-1, - n m)\,.
\end{align}
In the perturbed case, we thus write
\begin{equation}
	\check \phi = \check \phi^\o0 + \epsilon \check \phi^\o1 + O(\epsilon^2)\,.
\end{equation}
From the perturbed wave equation \eqref{eq:scalar wave equation} one then finds that $\check \phi^\o1$ satisfies the inhomogeneous wave equation
\begin{equation}
	\wop_{\gamma^\o0} \check \phi^\o1
		= - n^2 \omega^2 A(m,m) \cos(\kappa.x  + \chi) \check \phi^\o0\,,
\end{equation}
cf. \eqref{eq:scalar wave problem 1}.
According to \eqref{eq:boundary conditions mirror}, the boundary conditions for $\phi^\o1$ are
\begin{align}
	\check \phi^\o1 |_{\Sigma'}
		&= - \phi^\o1 |_{\Sigma'}\,,
	&
	m(\check \phi^\o1)|_{\Sigma'}
		&= m(\phi^\o1)|_{\Sigma'}
		+ \nu^\o1(\phi^\o0 - \check \phi^\o0) |_{\Sigma'} \,.
\end{align}
Using the explicit formula \eqref{eq:perturbed normal}, one finds that the terms arising from the perturbation of the normal $\nu$ cancel, so that it suffices to implement the conditions
\begin{align}
	\check \phi^\o1 |_{\Sigma'}
		&= - \phi^\o1 |_{\Sigma'}\,,
	&
	m(\check \phi^\o1)|_{\Sigma'}
		&= m(\phi^\o1)|_{\Sigma'} \,.
\end{align}
Using the previously derived expression \eqref{eq:scalar emission result} for the emitted wave and setting
\begin{equation}
	\label{eq:varpi def}
	\varpi = (n - \mn) \omegaG \ell\,
\end{equation}
such that $\kappa_0.x = \kappa.x + \varpi$ on $\Sigma'$, one finds the coefficients $\check \alpha^\pm$ and $\check \beta^\pm_{1,2}$ in \eqref{eq:reflection problem 1} — \eqref{eq:reflection problem 3} to be
\begin{align}
\label{19V21.1}
	\check \alpha^\pm
		&= - \half n^2 A(m,m) e^{i n \omega \ell \pm i \chi}\,,
	\\
	\begin{split}
		\check \beta_1^\pm
			&= \quarter A(m,m) e^{i n \omega \ell \pm i \chi} \bigg[
				\pm \frac{n}{\Omega} \frac{1 - e^{\pm i \varpi}}{n - \mn}
				- (n^2 -1) \frac{1 - e^{\pm i \varpi}}{2(n - \mn)^2}
				\\&\hspace{4cm}
				\pm i \omegaG \ell \frac{1 - (\mn)^2}{2(n - \mn)} e^{\pm i \varpi}
			\bigg]
			+ O(\Omega) + O(\Omega \omegaG \ell)
			\,,
	\end{split}
	\\
	\begin{split}
		\check \beta_2^\pm
			&= \check \beta_1^\pm (1 \pm \Omega \mn/n)
			- \quarter A(m,m) e^{i n \omega \ell \pm i \chi \pm i \varpi}
			+ O(\Omega) + O(\Omega \omegaG \ell)
			\,,
	\end{split}
\label{19V21.3}
\end{align}
and the coefficients $\check \gamma_{1,2}^\pm$, $\check \delta_{1,2}$ vanish.
These parameters satisfy \eqref{eq:counter-propagation reflection},  so that no counter-propagating terms arise (at the considered level of accuracy) and hence \eqref{eq:scalar template reflection} applies. This yields the following result for the reflected wave:
\begin{equation}
\begin{split}
	\check \phi^\pm
		&= \quarter A(m,m) \check \phi^\o0 e^{\pm i \chi} \bigg[
			\mp \frac{n}{\Omega} \left(
				\frac{e^{\pm i \kappa.x} - e^{\pm i \kappa_1.x}}{n + \mn}
				+ \frac{e^{\pm i \kappa_1.x} - e^{\pm i \kappa_1.x \pm i \varpi}}{n - \mn}
			\right)
			\\&
			\pm \frac{i}{2} \frac{1 - (\mn)^2}{n^2 - (\mn)^2} \left(
				2 \mn\, \omegaG (\ell-m.x) e^{\pm i \kappa_1.x}
				- (n + \mn) \omegaG(2 \ell - m.x)e^{\pm i (\kappa_1.x + \varpi)}
			\right)
			\\&
			+ \half (n^2 - 1) \left(
				\frac{e^{\pm i \kappa.x} - e^{\pm i \kappa_1.x}}{(n + \mn)^2}
				+ \frac{e^{\pm i \kappa_1.x} - e^{\pm i (\kappa_1.x + \varpi)}}{(n - \mn)^2}
			\right)
		\bigg]\,.
\end{split}
\end{equation}
Computing the overall field as $\check \phi = \check \phi^\o0 + \varepsilon (\check \phi^+ + \check \phi^-)$, we find
\begin{align}
	\label{eq:scalar field reflected}
	\check\phi
		&= - \check{\mathscr A} e^{i \check \psi} + O(\epsilon^2) + O(\epsilon \Omega) + O(\epsilon \Omega \omegaG \ell)\,,
	\\
\shortintertext{where}
	\begin{split}
		\check{\mathscr A}
			&= 1 - \quarter \epsilon A(m,m) \bigg[
				\frac{1 - (\mn)^2}{n^2 - (\mn)^2} \bigg(
					2 \mn\, \omegaG (\ell- m.x) \sin(\kappa_1.x + \chi)
					\\&\hspace{3.5cm}
					- (n + \mn) \omegaG (2 \ell - m.x) \sin(\kappa_1.x + \chi + \varpi)
				\bigg)
				\\&\qquad
				- (n^2 - 1) \bigg(
					\frac{\cos(\kappa.x + \chi) - \cos(\kappa_1.x + \chi)}{(n + \mn)^2}
					\\&\hspace{3.5cm}
					+ \frac{\cos(\kappa_1.x + \chi) - \cos(\kappa_1.x + \chi + \varpi)}{(n - \mn)^2}
				\bigg)
			\bigg]
			\,,
	\end{split}
	\\
	\label{eq:scalar phase result}
	\begin{split}
		\check\psi
		&= \check k.x + 2 n \omega \ell
		- \half \epsilon n A(m,m) \bigg(
			\frac{\sin(\kappa.x + \chi) - \sin(\kappa_1.x + \chi)}{\Omega(n + \mn)}
			\\&\hspace{3.5cm}
			+\frac{\sin(\kappa_1.x + \chi) - \sin(\kappa_1.x + \chi + \varpi)}{\Omega(n - \mn)}
			\bigg)\,.
	\end{split}
\end{align}
%
\checkedtogether{9II21}

To express the returning field in a concise notation and to compare with the final results of Ref.~\cite{Mieling:2021a} (see \Cref{s:interferometry}), we define, for any wave-vector $\tilde k$
\begin{equation}
	\label{eq:def integrated waveform}
	H(\tilde k, u_2, u_1)
		= \frac{\t {\tilde k}{_\mu} \t {\tilde k}{_\nu}}{2 \gamma^\o0(k, \kappa)}
		\int_{u_1}^{u_2} \t h{^\mu^\nu}(u) \,\dd u\,,
\end{equation}
so that for the considered metric perturbation $\t h{^\mu^\nu}(u) = \t A{^\mu^\nu} \cos(u)$ one has
\begin{equation}
	H(k, u_2, u_1)
		= - \half n A(m, m) \frac{\sin u_2 - \sin u_1}{\Omega (n - m.\kh)}\,.
\end{equation}
The analogous expression with $k$ replaced by $\check k$ is obtained by reversing the sign of $m$.
With this notation, the eikonal perturbation can be written concisely as
\begin{equation}
	\label{eq:scalar result phase perturbation}
	\psi^\o1
		= H(\check k, \kappa.x + \chi, \kappa_1.x + \chi )
		+ H(k, \kappa_1.x + \chi, \kappa_1.x + \chi + \varpi )\,.
\end{equation}
Similarly, we set
\begin{equation}
	\label{eq:def difference in waveform}
	\mathring H(\tilde k, u_2, u_1)
		= \frac{\t {\tilde k}{_\mu} \t {\tilde k}{_\nu}}{2 \gamma^\o0(\tilde k, \kappa)}
		\left[ \t h{^\mu^\nu}(u_2) - \t h{^\mu^\nu}(u_1)\right] \,,
\end{equation}
so that the amplitude perturbation can be written as
\begin{equation}
	\label{eq:scalar result amplitude perturbation}
	\begin{split}
		\Aa^\o1
			=&
			- \quarter A(m,m) \bigg[
				\frac{1 - (\mn)^2}{n^2 - (\mn)^2} \bigg(
					2 \mn\, \omegaG (\ell- m.x) \sin(\kappa_1.x + \chi)
					\\&\hspace{3.5cm}
					- (n + \mn) \omegaG (2 \ell - m.x) \sin(\kappa_1.x + \chi + \varpi)
				\bigg)
				\bigg]
				\\&
				+ \half (n^2 - 1) \omegaG^2 \left(
					\frac{\mathring H(\check k,\kappa.x + \chi,\kappa_1.x + \chi)}{\gamma^\o0(\kappa, \check k)}
					+ \frac{\mathring H(k,\kappa_1.x + \chi,\kappa_1.x + \chi + \varpi)}{\gamma^\o0(\kappa, k)}
				\right)\,.
	\end{split}
\end{equation}

%% file: optics_subfiles/maxwell_introduction.tex
Consider the source-free Maxwell equations in the form
\begin{align}
	\dd F &= 0\,,
	&
	\divergence \Fb &= 0 \,,
\end{align}
where the field strength $F$ (a two-form) comprises the electric and magnetic fields $E$ and $B$, and the excitation tensor $\Fb$ (a bivector) comprises the fields $D, H$;  see \eqref{eq:decomposition DEBH} below.
Here “$\dd$” denotes the exterior derivative and “$\divergence$” denotes the divergence with respect to the spacetime metric $g$.
In this work, we consider linear isotropic dielectrics only, for which the relationship between $F$ and $\Fb$ takes the form
\begin{equation}
	\label{eq:optical metric maxwell}
	\permeability \t\Fb{^\alpha^\beta} = \t \gamma{^\alpha^\rho} \t \gamma{^\beta^\sigma} \t F{_\rho_\sigma}\,,
\end{equation}
where $\permeability$ is the permeability, and $\gamma$ is the optical metric as defined in \eqref{eq:optical metric}, cf.\ Ref.~\cite{Gordon1923}.

We apply the 3+1 decomposition given in Ref.~\cite{Beig2018}, where we use the expansion
\begin{align}
	\t g{_0_0}
		&= -1 + O(\epsilon^2)\,,
		&
	\t g{_0_i}
		&= \t g{_i_0}
		= 0 + O(\epsilon^2)\,,
		\\
	\label{eq:spatial metric}
	\t\gs{_i_j}
		&= \t g{_i_j}
		= \t \delta{_i_j} + \epsilon \t h{_i_j} + O(\epsilon^2) \,,
		&
	\t h{_i_j}
		&= \t A{_i_j} \cos(\kappa.x + \chi) \,,
\end{align}
where $\gs$ denotes the \emph{spatial metric}, i.e.\ the Riemannian metric induced by $g$ on slices of constant $t \equiv \t x{^0}$.
 For a linearised gravitational field in TT-gauge the error terms above are zero, but our calculations allow for the above.
Since we expand both the metric tensor and the electromagnetic field to first order in $\epsilon$, all subsequent equations are understood to be correct up to $O(\epsilon^2)$, where the error term is not always written explicitly. Since $\t g{_0_0} = -1 + O(\epsilon^2)$ and $\det \gs = 1 + O(\epsilon^2)$, the definitions in Ref.~\cite{Beig2018} reduce  to
\begin{align}
	\label{eq:decomposition DEBH}
	D^i &= \t \Fb{^0^i}\,,
	&
	E_i &= \t F{_i_0}\,,
	&
	B^i &= \half \t\varepsilon{^i^j^k} \t F{_j_k}\,,
	&
	H_i &= \half \t\varepsilon{_i_j_k} \t\Fb{^j^k}\,,
\end{align}
where $\t\varepsilon{_i_j_k} = \t\varepsilon{^i^j^k}$ is the three-dimensional Levi-Civita symbol.
Maxwell’s equations then take the form
\begin{align}
	\t\p{_0} \t B{^i} + \t\varepsilon{^i^j^k} \t\p{_j} \t E{_k} &= 0\,,
	&
	\t\p{_i} \t B{^i} &= 0\,,
	\\
	\t\p{_0} \t D{^i} - \t\varepsilon{^i^j^k} \t\p{_j} \t H{_k} &= 0\,,
	&
	\t\p{_i} \t D{^i} &= 0\,.
\end{align}
These look exactly as in Minkowski spacetime, but the dependence upon the gravitational field enters through the constitutive equation \eqref{eq:optical metric maxwell}.
Indeed, using \eqref{eq:optical metric maxwell} as well as equation \eqref{eq:optical metric} for the optical metric $\gamma$, one finds
\begin{align}
	\t D{^i}
		&= \permittivity\, \t\gs{^i^j} \t E{_j}\,,
	&
	\t B{^i}
		&= \permeability\, \t\gs{^i^j} \t H{_j}\,,
\end{align}
where $\t\gs{^i^j}$ is the contravariant spatial metric (i.e.\ $\t\gs{^i^j}$ is the matrix inverse to $\t\gs{_i_j}$), $\permittivity$ the permittivity of the medium and $\permeability$ its permeability.

In order to exploit the electric-magnetic symmetry in the absence of external charges and currents, it is useful to define the complex vector field
\begin{equation}
	\label{eq:Maxwell complex vector}
	\t Z{^i}
		= \permeability \t D{^i}
		+ j n \t B{^i}\,,
\end{equation}
where $n = \sqrt{\permittivity \permeability}$ and $j$ is a second imaginary unit, independent of $i$ (in particular commuting with it), while we reserve the usage of $i$ for the usual complex description of waves.
The field equations then reduce to
\begin{align}
	\label{eq:maxwell 3+1}
	n \t\p{_0} \t Z{^i}
	+ j \t \varepsilon{^i^j^k} \p_j( \t \gs{_k_l} \t Z{^l}) &= 0\,,
	&
	\t\p{_i} \t Z{^i} &= 0\,,
\end{align}
provided that both $\permittivity$ and $\permeability$ are constant.

As shown in \Cref{s:Maxwell wave equation}, these equations imply the following wave equation:
\begin{equation}
	\label{eq:maxwell wave equation}
	n^2   \t {\ddot Z}{^i} - \Delta(\t Z{^i})
	+ 2 \epsilon \t R{^{\o1}^i_j} \t Z{^j}
	- 2 \epsilon \t\delta{^j^k} \t\Gamma{^{\o1}^i_j_l} \t\p{_k} \t Z{^l}
	+ j \epsilon \t\varepsilon{^i^j^k}[ n \t{\dot \Gamma}{^{\o1}_k_j_l} \t Z{^l} + n \t{\dot h}{_k_l} \t\p{_j} \t Z{^l} ]
	= 0\,,
\end{equation}
where $\Delta (\cdot)$ is the scalar Laplacian defined with respect to the perturbed spatial metric $\t\gs{_i_j}$, $\t\Gamma{^{\o1}^i_j_l}$ are the spatial Christoffel symbols, and $\t R{^{\o1}_i_j}$ is the spatial Ricci tensor, both truncated to first order in $\epsilon$.
Explicitly, one has
\begin{equation}
	\t R{^{\o1}_i_j} = \half \omegaG^2 \t h{_i_j}\,,
\end{equation}
and
\begin{align}
	\label{eq:Christoffel symbols spatial}
	\t \Gamma{^{\o1}_i_j_k}
		&= - \omegaG \t \gamma{_i_j_k} \sin(\kappa.x + \chi)\,,
\shortintertext{where}
	\label{eq:Christoffel amplitude}
	\t \gamma{_i_j_k}
		&= \half \left(
			\t\kh{_j} \t A{_k_i}
			+ \t\kh{_k} \t A{_j_i}
			- \t\kh{_i} \t A{_j_k}
		\right)\,.
\end{align}

In the following \Cref{s:Maxwell unperturbed}, we review the description of plane EM\ waves in the \emph{absence} of \red{GWs} using the complex notation presented here. Boundary data describing the emission of plane EM\ waves in the \emph{presence} of \red{GWs}   are constructed in \Cref{s:Maxwell emission data}, from which the perturbed emitted field is computed in \Cref{s:Maxwell wave equation emission}. The boundary data for reflection at perfect mirrors are then considered in \Cref{s:Maxwell reflection data}, and the reflected EM\ wave is computed in \Cref{s:Maxwell wave equation reflection}.

%% file: optics_subfiles/maxwell_unperturbed.tex
\subsection{The Unperturbed Field}
\label{s:Maxwell unperturbed}

In the unperturbed case (flat space), monochromatic plane waves can be written as
\begin{equation}
    \t Z{^{\o0}^l}
        = \t \zeta{^l} e^{i k.x}\,,
\end{equation}
where the wave vector is
\begin{equation}
    \t k{_\mu} = \omega(-1, n m)\,,
\end{equation}
for some unit vector $m$, and $\t \zeta{^i}$ is a constant $i$-real but $j$-complex vector (assuming linear polarisation). Maxwell’s equations further imply
\begin{equation}
    \label{eq:polarisation relation unperturbed}
    \t \zeta{^i} - j \t\varepsilon{^i^j^k} \t m{_j} \t\zeta{_k} = 0\,,
\end{equation}
cf.\ \eqref{eq:maxwell 3+1}. Contracting with the $j$-complex conjugate $\t*\zeta{^*_i}$, $\t \zeta{_i}$ and $\t m{_i}$, one obtains
\begin{align}
	\label{eq:polarisation conditions unperturbed}
	\zeta_i \zeta^i &= 0\,,
	&
	m_i \zeta^i &= 0\,,
	&
	j \varepsilon^{ijk} \zeta^*_i m_j \zeta_k &\geq 0\,,
\end{align}
which, together, are equivalent to \eqref{eq:polarisation relation unperturbed}.

Decomposing the complex field $Z$ into the $j$-real fields $D$ and $B$ according to \eqref{eq:Maxwell complex vector}, \eqref{eq:polarisation conditions unperturbed} one finds that $(E, B, m)$ forms a right-handed orthogonal system and
\begin{equation}
    \t E{_i} \t D{^i} = \t B{_i} \t H{^i}\,.
\end{equation}

We choose to normalise $\zeta$ according to
\begin{equation}
    \t \zeta{^*_i} \t \zeta{^i} = 1\,.
\end{equation}
This leaves a $j$-phase degree of freedom, which corresponds to the freedom of choosing the polarisation of the electromagnetic field.

Note that \eqref{eq:polarisation relation unperturbed} is invariant under
\begin{align}
    m &\to - m\,,
    &
    \zeta &\to \zeta^*\,,
\end{align}
so that counter-propagating waves (as arising from normal reflection at a mirror) are given by $\t\zeta{^*^l} e^{i \check k.x}$, where the reflected wave vector is $\t {\check k}{_\mu} = \omega(-1, - n m)$.

In the perturbed case, we find it useful to decompose the field as
\begin{equation}
	\label{eq:Maxwell complex basis expansion}
    \t Z{^i} = a \t\zeta{^i} + b \t \zeta{^*^i} + c \t m{^i}\,,
\end{equation}
where, in general, all three functions $a,b,c$ are non-zero.
For later reference, we note the useful identities
\begin{align}
	\label{eq:polarisation cross products}
	j \varepsilon^{ijk} m_j \zeta_k &= + \zeta^i\,,
	&
	j \varepsilon^{ijk} m_j \zeta^*_k &= - \zeta^{*i}\,,
	&
	j \varepsilon^{ijk} \zeta_j \zeta^*_k &= + m^i\,,
\end{align}
where the first equation is due to \eqref{eq:polarisation relation unperturbed}, the second equation is obtained from the first by complex conjugation, and the third one is obtained by expanding the left-hand side in the basis $(m, \zeta, \zeta^*)$ and determining the coefficients from suitable contractions.
Moreover, we have
\begin{equation}
	\label{eq:complex basis delta}
	\t \zeta{^*^i} \t \zeta{^j} + \t \zeta{^i} \t \zeta{^*^j}
		= \t\delta{^i^j} - \t m{^i} \t m{^j}\,,
\end{equation}
and since the metric perturbation $h$ is traceless, one finds
\begin{equation}
	\label{eq:complex basis h traceless}
	h(\zeta^*, \zeta) + \half h(m,m) = 0\,.
\end{equation}
\checkedtogether{30IV21}

%% file: optics_subfiles/maxwell_boundary.tex
\subsection{Boundary Values}
\label{s:Maxwell emission data}

In this section, we construct boundary data for Maxwell’s equations on the emission surface $\Sigma = \{m.x = 0\}$, which model  the radiation sent out by a laser.

To describe the emission of plane waves with a given frequency $\omega$, we require the field on $\Sigma$ to be of the form
\begin{equation}
	\label{eq:Maxwell emission form}
	\t Z{^k} = \t\Zz{^k} e^{-i \omega t}\,,
\end{equation}
where $\t\Zz{^k}$ is $i$-real and normalized according to
\begin{equation}
	\label{eq:Maxwell emission normalisation}
	\gs(\Zz^*, \Zz) = 1
	\qquad\text{at the spatial origin}\,,
\end{equation}
where $\Zz^*$ denotes the $j$-complex conjugate of $\Zz$; as we will see below, the normalisation $\gs(\Zz^*, \Zz) = 1$ \emph{cannot} be imposed everywhere on $\Sigma$.
Moreover, we demand
\begin{align}
	\label{eq:Maxwell emission assumption decomposed}
	\t\Zz{^i} \t\nu{_i} &= 0\,,
	&
	\t \gs{_i_j} \t\Zz{^i} \t\Zz{^j} &= 0\,,
	&
	j \t\varepsilon{^i^j^k} \t*\Zz{^*_i} \t\nu{_j} \t\Zz{_k} &> 0\,,
\end{align}
i.e.\ the fields $(\nu, D, B)$ form a right-handed orthogonal basis ($D$ and $B$ are thus tangent to $\Sigma$) and satisfy $\t E{_i} \t D{^i} = \t B{_i} \t H{^i}$ on $\Sigma$.
As in the unperturbed case, \eqref{eq:Maxwell emission assumption decomposed} is equivalent to
\begin{equation}
	\label{eq:Maxwell emission assumption}
	\t\Zz{^i} - j \t\varepsilon{^i^j^k} \t \nu{_j} \t\gs{_k_l} \t\Zz{^l} = 0\,.
\end{equation}
A four-dimensional-covariant formulation of this equation is given in \Cref{app:covariant emission equation}.

Using the expansion \eqref{eq:Maxwell complex basis expansion}, the requirement that $Z$ be tangent to the emission surface is seen to be equivalent to $c = 0$ on $\Sigma$, so it remains to prescribe the functions $a$ and $b$ there.

\paragraph{The Unperturbed Case}

Before considering the general case, let us briefly return to the unperturbed problem where the spatial metric reduces to the flat Euclidean metric and the surface conormal is given by $\t\nu{_i} = \t m{_i}$.

Decomposing the electromagnetic field as in \eqref{eq:Maxwell complex basis expansion} and projecting the field equations \eqref{eq:maxwell 3+1} with $\epsilon = 0$ onto the basis $\zeta, \zeta^*, m$, one obtains Maxwell’s equations in the form
\begin{align}
	\label{eq:maxwell unperturbed m-deriv c}
	\zeta(a) + \zeta^*(b) + m(c) = 0\,,
	\\
	\label{eq:maxwell unperturbed m-deriv a}
	n \p_0 a + m(a) - \zeta^*(c) = 0\,,
	\\
	\label{eq:maxwell unperturbed m-deriv b}
	n \p_0 b - m(b) + \zeta(c) = 0\,,
	\\
	\label{eq:maxwell constraint unperturbed}
	n \p_0 c - \zeta(a) + \zeta^*(b) = 0\,.
\end{align}
The last equation here does not contain any $m$-derivatives and is thus a “constraint equation” for boundary data prescribed on $\Sigma$, while the remaining equations determine the normal derivatives of $a$, $b$ and $c$, once their values on $\Sigma$ are specified.

In flat space, \eqref{eq:Maxwell emission assumption} is equivalent to $b = c = 0$ on $\Sigma$, so that the “constraint equation” \eqref{eq:maxwell constraint unperturbed} on $\Sigma$ reduces to $\zeta(a) = 0$. Choosing adapted coordinates $\tilde x, \tilde y$ on $\Sigma$ such that $\sqrt 2 \zeta = \p_{\tilde x} + j \p_{\tilde y}$ (which is always possible by choosing $\p_{\tilde x}$, $\p_{\tilde y}$ and $m$ to be a right-handed orthonormal basis), this reduces to the Cauchy-Riemann equation
\begin{equation}
	\label{eq:Cauchy Riemann}
	\frac{\p a}{\p \tilde x} + j \frac{\p a}{\p \tilde y} = 0\,.
\end{equation}
Requiring $a$ to be bounded, one finds that it must be spatially constant (by Liouville's theorem) and thus a function of time $t$ alone.
Hence, one also has $\zeta^*(a) = 0$, so that \eqref{eq:maxwell unperturbed m-deriv c} yields $m(c) = 0$.

Thus, assuming \eqref{eq:Maxwell emission assumption} in flat space and requiring the field on $\Sigma$ to be bounded, one obtains the following boundary values on the emission surface $\Sigma$
\begin{align}
	n \p_0 a + m(a) &= 0\,,
	&
	b = m(b) &= 0\,,
	&
	c = m(c) &= 0\,,
\end{align}
where $a$ is an arbitrary function of time $t$ alone.
In particular  to describe the emission of monochromatic plane waves of frequency $\omega$, we shall take $a  = e^{-i \omega t}$, which entails $m(a) = - i n \omega a$.

Note that since all field components satisfy the wave equation, $b$ and $c$ vanish identically everywhere and $a$ is the only remaining degree of freedom.
\ptcheck{28II20 up to here}

\paragraph{The Perturbed Case}

Let us now consider the perturbed case, where we still consider the fields on $\Sigma$ only.
Since $b$ and $c$ vanish identically in the unperturbed case, we write
\begin{equation}
	\label{eq:Maxwell complex basis expansion perturbed}
	Z
	= a \zeta + \epsilon (b^\o1 \zeta^* + c^\o1 m)
	\equiv a^\o0 \zeta + \epsilon(a^\o1 \zeta + b^\o1 \zeta^* + c^\o1 m)\,.
\end{equation}
Inserting this into the field equations \eqref{eq:maxwell 3+1},
one obtains Maxwell’s equations at first order in $\epsilon$ in the form
\begin{align}
	\label{eq:maxwell perturbed m-deriv c general}
	\zeta(a^\o1)
		+ \zeta^*(b^\o1)
		+ m(c^\o1)
		&= 0\,,
		\\
	\label{eq:maxwell perturbed m-deriv a general}
	n\t\p{_0} a
		+ \nu(a)
		- \epsilon \zeta^*(c^\o1)
		+ \epsilon a^\o0 j \t\varepsilon{^i^j^k} \t*\zeta{^*_i} (\t\p{_j} \t h{_k_l}) \t\zeta{^l}
		&= 0\,,
		\\
	\label{eq:maxwell perturbed m-deriv b general}
	n\t\p{_0} b^\o1
		- m(b^\o1)
		+ \zeta(c^\o1)
		+ j \t\varepsilon{^i^j^k} \t\zeta{_i} \t\p{_j} (a^\o0 \t h{_k_l}) \t\zeta{^l}
		&= 0\,,
		\\
	\label{eq:maxwell constraint perturbed general}
	n\t\p{_0} c^\o1
		- \zeta(a^\o1)
		+ \zeta^*(b^\o1)
		+ j \t\varepsilon{^i^j^k} \t m{_i} \t\p{_j} (a^\o0 \t h{_k_l}) \t\zeta{^l}
		&= 0\,.
\end{align}
Equation \eqref{eq:maxwell perturbed m-deriv c general} is obtained by inserting \eqref{eq:Maxwell complex basis expansion perturbed} into the second part of \eqref{eq:maxwell 3+1}, and the equations \eqref{eq:maxwell perturbed m-deriv b general} and \eqref{eq:maxwell constraint perturbed general} are obtained by contracting the first part of \eqref{eq:maxwell 3+1} with either $\t\zeta{_i}$ or $\t m{_i}$. To arrive at \eqref{eq:maxwell perturbed m-deriv a general}, one can contract the first equation in \eqref{eq:maxwell 3+1} with $\t*\zeta{^*_i}$ to obtain
\begin{equation}
	\label{eq:maxwell perturbed m-deriv a intermediate}
	n\t\p{_0} a
		+ \t\gs{^i^j} \t\theta{_i} \t\p{_j} a
		- \epsilon \zeta^*(c^\o1)
		+ \epsilon a^\o0 j \t\varepsilon{^i^j^k} \t*\zeta{^*_i} (\t\p{_j} \t h{_k_l}) \t\zeta{^l}
		= 0\,,
\end{equation}
where $\theta$ is defined as
\begin{equation}
	\label{eq:theta def}
	\t\theta{_i} = - j \t\gs{_i_j} \t\varepsilon{^j^k^l} \t*\zeta{^*_k} \t\gs{_l_m} \t\zeta{^m}\,.
\end{equation}
Expanding $\theta$ in powers of $\epsilon$, one has $\t\theta{_i} = \t m{_i} + \epsilon \t\theta{^{\o1}_i} + O(\epsilon^2)$, by virtue of \eqref{eq:polarisation cross products}.
Using the contractions
\begin{align}
	\t*\theta{^{\o1}_i} \zeta^{i}
		&= 0 \,,
	&
	\t*\theta{^{\o1}_i} \zeta^{*i}
		&= h(m, \zeta^*) \,,
	&
	\t*\theta{^{\o1}_i} m^i
		&= h(m,m) + h(\zeta^*, \zeta) \,,
\end{align}
as well as \eqref{eq:complex basis h traceless}, one finds
\begin{equation}
	\label{eq:maxwell boundary vector decomposition}
	\t\theta{_i} = \t\nu{_i} + \epsilon \t\zeta{_i} h(m, \zeta^*) + O(\epsilon^2)\,.
\end{equation}
Here $\nu$ is the unit conormal to $\Sigma$ as defined in \eqref{eq:conormal normalised}. Inserting this into \eqref{eq:maxwell perturbed m-deriv a intermediate} and using $\zeta(a^\o0)=0$ then leads to \eqref{eq:maxwell perturbed m-deriv a general}, as claimed.
\checkedtogether{30IV21}

As in the unperturbed case, \eqref{eq:maxwell constraint perturbed general} is a constraint equation for boundary data on $\Sigma$, while the remaining equations determine the normal derivatives of $a^\o1$, $b^\o1$ and $c^\o1$, once their values on $\Sigma$ are specified.

Let us now implement the assumptions \eqref{eq:Maxwell emission form}, \eqref{eq:Maxwell emission normalisation} and \eqref{eq:Maxwell emission assumption}, together with $\Zz = \zeta + O(\epsilon)$. The general solution for $\Zz$ satisfying these conditions is
\begin{equation}
	\label{eq:Maxwell boundary z}
	\t \Zz{^i}
		= \t \zeta{^i} (1 + j \epsilon \tilde a)
		- \half \epsilon [ \t \delta{^i^j} - \t m{^i} \t m{^j} ]
		\t h{_j_k} \t \zeta{^k}\,,
\end{equation}
where $\tilde a$ is an arbitrary function.
In components, this is the same as
\begin{align}
	\label{eq:maxwell boundary abc}
	a &= [1 + \epsilon j \tilde a - \half \epsilon h(\zeta^*, \zeta)] e^{- i \omega t}\,,
	&
	b^\o1 &= - \half h(\zeta, \zeta) e^{- i \omega t}\,,
	&
	c^\o1 &= 0\,.
\end{align}
Inserting this into \eqref{eq:maxwell constraint perturbed general}, the constraint equation reduces to
\begin{equation}
	\label{eq:maxwell emission constraint}
	\zeta(a^\o1)
	+ a^\o0 \t\Gamma{^{\o1}_i_j_k} \t\zeta{^*^i} \t\zeta{^j} \t\zeta{^k}
	= 0\,,
\end{equation}
which is equivalent to
\begin{equation}
	\label{eq:maxwell emission constraint2}
	j \zeta(\tilde a)
	+ \half \zeta (h(\zeta^*, \zeta))
	- \half \zeta^*(h(\zeta, \zeta )) = 0\,.
\end{equation}
\checkedtogether{30IV21}

To understand the implications of this equation, suppose we have two solutions $\tilde a_1$ and $\tilde a_2$.
Then their difference satisfies the Cauchy-Riemann equation \eqref{eq:Cauchy Riemann}, so that if both functions are bounded, their difference is spatially constant. Hence, the bounded solutions to \eqref{eq:maxwell emission constraint2} are parameterised by an arbitrary complex function of time alone.

A simple choice would be to prescribe $\tilde a(t, \t x{^i} = 0) = 0$.
However, to allow for alternative models of laser emission, which might differ in their prescriptions for the time evolution of the emitted polarisation (e.g.\ one could demand the polarisation vector to be parallel transported in time in the sense of the perturbed metric), we allow $\tilde a$ to satisfy
\begin{equation}
	\label{eq;maxwell emitted polarisation perturbation}
	\tilde a|_{\t x{^i} = 0}
	\equiv \alpha(t)
	:= \alpha_0 + \alpha_c \cos(-\omegaG t + \chi) + \alpha_s \sin(-\omegaG t + \chi)\,,
\end{equation}
where $\alpha_0$, $\alpha_c$ and $\alpha_s$ (as well as $\chi$) are $j$-real constants.
This means that the emitted polarization vector at the coordinate origin is normalized (corresponding to an energy density at the coordinate origin which remains constant in time), but we allow for perturbations of the plane of polarization which oscillate with the same frequency as the metric perturbation.

As will be shown below, the parameters $\alpha_0$, $\alpha_c$ and $\alpha_s$ have no impact on the signal detected in the interferometer considered and can thus be set to zero for this concrete application. However, as they describe the perturbation of the emitted polarisation, they would be observable in potential experiments sensitive to light polarisation.
We note also that the precise form of $\alpha(t)$ in \eqref{eq;maxwell emitted polarisation perturbation} fits into our calculations without further due, while more general functions would require further analysis.

Defining
\begin{equation}
	\label{eq: varsigma def}
	\varsigma = \frac{1}{2 \kh.\zeta} \left[
		(\kh.\zeta) A(\zeta^*, \zeta)
		- (\kh.\zeta^*) A(\zeta, \zeta)
	\right]\,,
\end{equation}
the unique solution to the constraint equation \eqref{eq:maxwell emission constraint2} which remains bounded on $\Sigma$ is found to be
\begin{equation}
	\label{eq: tilde a def}
	\tilde a
		= j \varsigma [ \cos(\kappa.x) - \cos(-\omegaG t + \chi) ]
		+ \alpha_0 + \alpha_c \cos(-\omegaG t + \chi) + \alpha_s \sin(-\omegaG t + \chi)\,.
\end{equation}
The first term here has a non-zero $j$-imaginary   part and thus describes $\epsilon$-oscillations in the emitted amplitude. Such a behaviour is unavoidable, as it is a consequence of the constraint equation \eqref{eq:maxwell emission constraint2}.

Having prescribed the field on $\Sigma$, the normal derivatives of $a$, $b^\o1$ and $c^\o1$ can now be obtained from the remaining Maxwell equations
\begin{align}
	\label{eq:maxwell boundary normal a}
	n\t\p{_0} a
		+ \t\gs{^i^j} \t\nu{_i} \t\p{_j} a
		+ \epsilon a^\o0 j \t\varepsilon{^i^j^k} \t*\zeta{^*_i} (\t\p{_j} \t h{_k_l}) \t\zeta{^l}
		&= 0\,,
		\\
	\label{eq:maxwell boundary normal b}
	n\t\p{_0} b^\o1
		- m(b^\o1)
		+ j \t\varepsilon{^i^j^k} \t\zeta{_i} \t\p{_j} (a^\o0 \t h{_k_l}) \t\zeta{^l}
		&= 0\,,
		\\
	\label{eq:maxwell boundary normal c}
	\zeta(a^\o1) + \zeta^*(b^\o1) + m(c^\o1) &= 0\,.
\end{align}
Similarly to the scalar wave case, from now on we assume $\omegaG \ll \omega$.

From \eqref{eq:maxwell boundary normal a} it follows  that the normal derivative of $a$ is
\begin{equation}
	\nu(a)/\omega = i n a + O(\Omega)\,,
\end{equation}
since $n \p_0 a = - i n \omega a(1 + \epsilon O(\omegaG/\omega))$ by virtue of \eqref{eq:maxwell boundary abc}.

Using the explicit form of $b^\o1$ in \eqref{eq:maxwell boundary abc}, we have 
\begin{equation}
	\tfrac{1}{\omega} n \p_0 b^{(1)}
		= - i n b^{(1)}
		 + O(\omegaG/\omega)\,,
\end{equation}
since $h$ varies with frequency $\omegaG$ only.
Furthermore, since $a^\o0$ is constant on $\Sigma$ but varies rapidly in the direction of $m$ (again by a factor $\omega/\omegaG$ faster than the metric $h$), we have
\begin{equation}
\begin{split}
	\tfrac{1}{\omega} j \varepsilon^{ijk} \zeta_i \p_j ( a^{(0)} \t h{_k_l} ) \zeta^l
		&= j m(a^\o0/\omega) \varepsilon^{ijk} \zeta_i m_j \t h{_k_l} \zeta^l
		 + O(\omegaG/\omega)
		 \\
		&= - m(a^\o0/\omega) h(\zeta, \zeta)
		 + O(\omegaG/\omega)
		 \\
		&= - i n a^\o0 h(\zeta, \zeta)
		 + O(\omegaG/\omega)
		 \\
		&= 2 i n b^\o1
		 + O(\omegaG/\omega)
		 \,,
\end{split}
\end{equation}
where we have used \eqref{eq:polarisation cross products}, $m(a^\o0) = i n \omega a^\o0$, and again the explicit form of $b^\o1$ given in \eqref{eq:maxwell boundary abc} to arrive at the last line.
Plugging all this into \eqref{eq:maxwell boundary normal b}, one obtains
\begin{equation}
	m(b^\o1)/\omega = i n b^\o1 + O(\Omega)\,.
\end{equation}
Finally, a direct calculation shows that
\begin{equation}
	\zeta^*(b^\o1) = - a^\o0 \t*\Gamma{^{\o1}_i_j_k} \zeta^i \zeta^j \zeta^{*k}\,,
\end{equation}
\checkedtogether{3.5.21}
so that equation~\eqref{eq:maxwell boundary normal c} leads to
\begin{equation}
	m(c) = a^\o0 \left(
		\t*\Gamma{^{\o1}_i_j_k} \zeta^{*i} \zeta^j \zeta^k
		+ \t*\Gamma{^{\o1}_i_j_k} \zeta^i \zeta^j \zeta^{*k}
	\right)\,.
\end{equation}
Inserting \eqref{eq:complex basis delta} as well as the explicit expression for the spatial Christoffel symbols \eqref{eq:Christoffel symbols spatial}, this yields
\begin{equation}
	m(c) = \half \omegaG a^\o0 (\kh.\zeta) A(m,m) \sin(\kappa.x + \chi)\,.
\end{equation}
To summarize, we have obtained the following boundary data for Maxwell equations on the emission surface $\Sigma =\{m.x = 0\}$.
\begin{align}
	\label{eq:Maxwell a emitted value}
	a
		&=
			e^{-i\omega t} \left[
			1 + \epsilon j \tilde a - \half \epsilon h(\zeta^*, \zeta)
			\right] \,,
		\\
	\label{eq:Maxwell b emitted value}
	b^\o1
		&= - \half e^{-i \omega t} h(\zeta, \zeta) \,,
		\\
	\label{eq:Maxwell c emitted value}
	c^\o1
		&= 0 \,,
		\\
	\label{eq:Maxwell a emitted deriv}
	\nu(a)
		&= i n \omega a + O(\omegaG)\,,
		\\
	\label{eq:Maxwell b emitted deriv}
	\nu^\o0(b^\o1)
		&= i n \omega b^\o1 + O(\omegaG) \,,
		\\
	\label{eq:Maxwell c emitted deriv}
	\nu^\o0(c^\o1)
		&= \half \omegaG a^\o0 (\kh.\zeta) A(m,m) \sin(\kappa.x + \chi) \,.
\end{align}
\checkedtogether{3.5.21} 

%% file: optics_subfiles/maxwell_emission.tex
\subsection{The Emitted Wave}
\label{s:Maxwell wave equation emission}

The wave equation for the perturbation of the emitted field is now obtained by inserting the unperturbed expression
\begin{equation}
	\t Z{^{\o0}^i}
		= \t \zeta{^i} e^{i k.x} \,,
\end{equation}
with $\t k{_\mu} = \omega(-1, n m)$, into the general wave equation \eqref{eq:maxwell wave equation}.
This leads to
\begin{equation}
\begin{split}
	\Delta(\t Z{^i}) - n^2 \t{\ddot Z}{^i}
	={}& - i \epsilon n \omega \left[
		2 \t \Gamma{^{\o1}^i_j_k} \t m{^j} \t \zeta{^k}
		- n j \t \varepsilon{^i^j^k} \t m{_j} \t {\dot h}{_k_l} \t \zeta{^l}
	\right]e^{i k.x}
	\\
	&+ \epsilon \left[
		2 \t R{^{\o1}^i_j} \t \zeta{^j}
		+ n j \t\varepsilon{^i^j^k} \t{\dot \Gamma}{^{\o1}_k_j_l} \t \zeta{^l}
	\right]e^{i k.x}\,,
\end{split}
\end{equation}
where $\Delta$ is the scalar Laplacian defined with respect to the \emph{perturbed} spatial metric ${\t \delta{_i_j} + \epsilon \t h{_i_j}}$.
Projecting this equation onto the basis $\zeta, \zeta^*, m$, one finds the wave equations for the components $a, b, c$ to be
\begin{align}
	\label{eq:Maxwell a emitted wave eq}
	\begin{split}
		\wop_{\gamma^\o0} a^\o1
			={}& - n^2 \omega^2 A(m,m) \cos(\kappa.x + \chi) e^{i k.x}\\
			&+ i n \omegaG \omega \left[
				2 \t \gamma{_i_j_k} \t \zeta{^*^i} \t m{^j} \t \zeta{^k}
				+ n A(\zeta^*, \zeta)
			\right] \sin(\kappa.x + \chi) e^{i k.x}\\
			&+ \omegaG^2 \left[
				A(\zeta^*, \zeta)
				+ n j \t \varepsilon{^i^j^k} \t*\zeta{^*_i} \t \gamma{_k_j_l} \t \zeta{^l}
			\right] \cos(\kappa.x + \chi) e^{i k.x}
			\,,
	\end{split}
	\\
	\label{eq:Maxwell b emitted wave eq}
	\begin{split}
		\wop_{\gamma^\o0} b^\o1
			={}& i n \omegaG \omega \left[
				2 \t \gamma{_i_j_k} \t \zeta{^i} \t m{^j} \t \zeta{^k}
				- n A(\zeta, \zeta)
			\right] \sin(\kappa.x + \chi) e^{i k.x}\\
			&+ \omegaG^2 \left[
				A(\zeta, \zeta)
				+ n j \t \varepsilon{^i^j^k} \t \zeta{_i} \t \gamma{_k_j_l} \t \zeta{^l}
			\right] \cos(\kappa.x + \chi) e^{i k.x}\,,
	\end{split}
	\\
	\label{eq:Maxwell c emitted wave eq}
	\begin{split}
		\wop_{\gamma^\o0} c^\o1
			={}& 2 i n \omegaG \omega
				\t \gamma{_i_j_k} \t m{^i} \t m{^j} \t \zeta{^k}
			\sin(\kappa.x + \chi) e^{i k.x}\\
			&+ \omegaG^2 \left[
				A(m, \zeta)
				+ n j \t \varepsilon{^i^j^k} \t m{_i} \t \gamma{_k_j_l} \t \zeta{^l}
			\right] \cos(\kappa.x + \chi) e^{i k.x}\,,
	\end{split}
\end{align}
The leading order term in \eqref{eq:Maxwell a emitted wave eq} is the same as for the scalar wave equation \eqref{eq:scalar wave problem 1}, so that the sub-leading terms can be interpreted as polarisation terms.

Having determined the wave equations and the boundary values for the functions $a$, $b$ and $c$, the emitted electromagnetic wave can now be obtained using the formulae derived in Section \ref{s:emission general}.

\paragraph{The $a$ component}

For the $a$ component, the value of $\alpha^\pm$ can be read from \eqref{eq:Maxwell a emitted wave eq}, $\beta^\pm_1$ is determined by \eqref{eq:Maxwell a emitted value}, and $\beta^\pm_2$ is obtained from \eqref{eq:Maxwell a emitted deriv}. These parameters $\alpha^\pm$, $\beta^\pm_1$ and $\beta^\pm_2$ are almost identical to those of the emitted scalar wave, as given in \eqref{eq:scalar emitted parameters}, so that we merely state the correction terms to be added to the values given there. We denote them by $\delta \alpha, \delta \beta$ etc. Finally, due to \eqref{eq;maxwell emitted polarisation perturbation}, the coefficients $\gamma_{1,2}$ and $\delta$ are no longer zero in general:
\begin{align}
	\delta\alpha^\pm
		&= \mp n \Omega \left(
			\t \gamma{_i_j_k} \t\zeta{^*^i} \t m{^j} \t \zeta{^k}
			+ \half n A(\zeta^*, \zeta)
		\right)
		+ O(\Omega^2)\,,
	\\
	\delta\beta_{1,2}^\pm
		&= - \quarter A(\zeta^*, \zeta) - \half \varsigma
		+ O(\Omega)\,,
		\\
	\gamma_{1,2}^\pm
		&= \half \varsigma + j \half [\alpha_c \mp i \alpha_s]
		+ O(\Omega)\,,
		\\
	\delta_{1,2}
		&= \half j \alpha_0
		+ O(\Omega)\,.
\end{align}

\paragraph{The $b$ component}

For the $b$ component, the parameters $\alpha^\pm$ are determined by \eqref{eq:Maxwell b emitted wave eq}, where we obtain at leading order, using the explicit form of the Christoffel symbols \eqref{eq:Christoffel amplitude},%
\begin{equation}
	2 \t \gamma{_i_j_k} \t \zeta{^i} \t m{^j} \t \zeta{^k} - n A(\zeta, \zeta)
	= (\mn - n) A(\zeta, \zeta)\,,
\end{equation}
Next, the parameters $\beta^\pm_1$ can be read from \eqref{eq:Maxwell a emitted value}, and the values of $\beta^\pm_2$ are determined by \eqref{eq:Maxwell a emitted deriv}. The non-vanishing components are then
\begin{align}
	\alpha^\pm
		&= \pm \half n \Omega (n - \mn) A(\zeta, \zeta)
		+ O(\Omega^2)\,,
	\\
	\beta_{1,2}^\pm
		&= - \quarter A(\zeta, \zeta)
		+ O(\Omega)\,.
\end{align}

\paragraph{The $c$ component}

For the remaining $c$ component, the value of $\alpha^\pm$ are determined from \eqref{eq:Maxwell c emitted wave eq}.
Using again the explicit form of the Christoffel symbols, one obtains
\begin{equation}
	2 \t \gamma{_i_j_k} \t m{^i} \t m{^j} \t \zeta{^k}
		= (\kh.\zeta) A(m,m)\,,
\end{equation}
which determines $\alpha^\pm$ to leading order in $\Omega$:
\begin{align}
	\alpha^\pm
		&= \mp \half n \Omega (\kh.\zeta) A(m,m)
		+ O(\Omega^2)\,,
\end{align}
while all other parameters vanish to the considered order, as follows from \eqref{eq:Maxwell c emitted value} and \eqref{eq:Maxwell c emitted deriv}.

\paragraph{Result}

All the above parameter sets satisfy the emission condition \eqref{eq:scalar emission condition}, so that the fields are given by the general formula \eqref{eq:scalar template emission}. This yields the following result for the emitted field:
\begin{align}
	\begin{split}
		\label{eq:maxwell emitted a}
		a^\o1
		={}& \phi^\o1
		+ \left(
			\t \gamma{_i_j_k} \t\zeta{^*^i} \t m{^j} \t \zeta{^k}
			+ \half n A(\zeta^*, \zeta)
		\right) \frac{\cos(u) - \cos(u_0)}{n - \mn} e^{i k.x}
		\\&
		- \half A(\zeta^*, \zeta) \cos(u_0) e^{i k.x}
		+ j \alpha_0 e^{i k.x}
		\\&
		- \varsigma [ \cos(u_0) - \cos(\omegaG (n m.x - t) + \chi) ] e^{i k.x}
		\\&
		+ j [\alpha_c \cos(\omegaG(n m.x - t) + \chi) + \alpha_s \sin(\omegaG(n m.x - t) + \chi)] e^{i k.x}
		\\&
		+ O(\Omega) + O(\Omega \omegaG m.x)\,,
	\end{split}
	\\
	\label{eq:maxwell emitted b}
	b^\o1
	=& - \half h(\zeta, \zeta) e^{i k.x}
	+ O(\Omega) + O(\Omega \omegaG m.x)\,,
	\\
	\label{eq:maxwell emitted c}
	c^\o1
		=& \half A(m,m) (\kh.\zeta) \frac{\cos(u) - \cos(u_0)}{n - \mn} e^{i k.x}
		+ O(\Omega) + O(\Omega \omegaG m.x)\,,
\end{align}
where $\phi^\o1$ is the perturbation of the emitted scalar field given in \eqref{eq:scalar emission result}, and where we have used the abbreviations
\begin{align}
	u &= \kappa.x + \chi\,,
	&
	u_0 &= \kappa_0.x + \chi\,.
\end{align}
Note that \eqref{eq:maxwell emitted b} entails that
\begin{equation}
	g(Z, Z) = O(\epsilon \Omega) + O(\epsilon \Omega \omegaG m.x)\,,
\end{equation}
so all scalar invariants of the electromagnetic field vanish to the considered accuracy, cf.~\eqref{eq:maxwell invariants complex notation}.
\checkedtogether{14.5.21} 

%% file: optics_subfiles/maxwell_reflection.tex
\subsection{Reflection at a Mirror}
\label{s:Maxwell reflection data}

Having found the electromagnetic wave emitted from a laser, we now consider its reflection off a perfectly reflecting mirror.

Recall that the jumps (denoted by $\varDelta D, \varDelta B$ etc.) of the fields at an interface with unit normal $\nu$ satisfy
\begin{align}
	\nu \cdot \varDelta D &= \mathfrak s\,,&
	\nu \cdot \varDelta B &= 0\,,&
	\nu \times \varDelta E &= 0\,,&
	\nu \times \varDelta H &= \mathfrak j\,,
\end{align}
where $\mathfrak s$ is the surface charge density and $\mathfrak j$ is the surface current density.

Since all fields vanish inside (or behind) the mirror, the overall field at the reflecting side of the mirror (being a sum of the incident and the reflected fields) satisfies
\begin{align}
	\nu \cdot D &= \mathfrak s\,,&
	\nu \cdot B &= 0\,,&
	\nu \times E &= 0\,,&
	\nu \times H &= \mathfrak j\,.
\end{align}

Let $Z$ denote the incident Maxwell field, with $Z^*$ its $j$-complex conjugate,  and let $\check Z$ be the reflected field.
Using our notation, the standard conditions for reflection on a perfect mirror can be written as
\begin{align}
	\label{eq:Maxwell mirror conditions}
	\nu \cdot (\check Z - Z^*) &= 0\,,
	&
	\nu \times (\check Z + Z^*) &= 0\,.
\end{align}
Indeed, the imaginary part of the first equation is equivalent to $\nu \cdot B = 0$, and the real part of the latter is equivalent to $\nu \times E = 0$.
The remaining parts (i.e.\ the real part of the first and the imaginary part of the second equation) say that the surface charges and currents have equal contributions from the incident and the reflected waves, cf.~equations~(56) and (58) in Ref.~\cite{Arnoldus2006}.
This can be understood as follows.
The reflected wave is caused by the movement of charge carriers which respond to the impinging wave as to cancel the field inside the conductor.
But the true sources must be computed from the total field (the incident \emph{and} the reflected wave) and a linear response of the mirror demands that the contributions of both parts are proportional to each other.
Equality then follows from the condition that the amplitudes of the incident and reflected waves coincide.

Equivalently, the equation can be interpreted to say that the normal part of the electric field and the tangential parts of the magnetic field are unchanged by the reflection process, while the tangential part of the electric field and the normal part of the magnetic field change sign.

A covariant formulation of \eqref{eq:Maxwell mirror conditions} is given in \Cref{app:reflection constraint consistent}.
There, it is also shown that this condition is consistent in the sense that if the incident field $Z$ satisfies the constraint equation arising from Maxwell’s equations, then so does the reflected field $\check Z$.

The unique solution to \eqref{eq:Maxwell mirror conditions} is given by
\begin{equation}
	\label{eq:Maxwell mirror solution}
	\check Z = - Z^* + 2 \gs(\nu, Z^*) \nu\,,
\end{equation}
i.e.\ $\check Z$ is obtained by reflecting $Z$ along the normal $\nu$, taking the $j$-complex conjugate and reversing the sign of the field. In other words, equation~\eqref{eq:Maxwell mirror conditions} is equivalent to~\eqref{eq:Maxwell mirror solution}.

Inserting the incident field of the form
\begin{equation}
	Z = a \zeta + \epsilon ( b^\o1 \zeta^* + c^\o1 m )\,,
\end{equation}
a direct calculation shows that the reflected field is given by
\begin{equation}
	\check Z = - a^* \zeta^* - \epsilon( b^{\o1*} \zeta - c^{\o1*} m)\,.
\end{equation}
Writing (everywhere) the field as
\begin{equation}
	\check Z =
	\check b \zeta^*
	+ \epsilon ( \check a^\o1 \zeta + \check c^\o1 m )
	\equiv
	(\check b^\o0 + \epsilon \check b^\o1) \zeta^*
	+ \epsilon ( \check a^\o1 \zeta + \check c^\o1 m )
	\,,
\end{equation}
the components of the reflected field on the mirror surface are seen to be
\begin{align}
	\label{eq:Maxwell reflection abc}
	\check a^\o1
		&= - b^\o1{}^*
	\,,
	&
	\check b
		&= - a^*\,,
	&
	\check c^\o1
		&= c^\o1{}^* \,.
\end{align}

Note that by virtue of equation \eqref{eq:maxwell emitted b}, we have
\begin{equation}
	\label{eq:maxwell check a1}
	\check a^\o1 = - \half \check b^\o0 h(\zeta^*, \zeta^*)\,.
\end{equation}

Having found the values of the functions $\check a, \check b$ and $\check c$ on the mirror surface $\Sigma'$, we must compute their normal derivatives from Maxwell’s equations.
One of the conditions stems from the divergence equation
\begin{equation}
	\nabla_i \check Z^i = 0\,,
\end{equation}
and the other ones are obtained from the evolution equation
\begin{equation}
	n \p_0 \check Z^i
	+ j \varepsilon^{ijk} \gs_{kl} \p_j \check Z^l
	+ j \varepsilon^{ijk} (\p_j \gs_{kl}) \check Z^l = 0\,.
\end{equation}
\checkedtogether{14.5.21}
We obtain, using the vector $\theta$ defined in \eqref{eq:theta def}:
\begin{align}
	\label{eq:reflected b boundary intermediate}
	n \p_0 \check b - \t\gs{^i^j} \theta^*_i \p_j \check b
		+ \epsilon \zeta(\check c^\o1)
		+ j \epsilon \check b^\o0 \varepsilon^{ijk} \zeta_i (\p_j \t h{_k_l}) \zeta^{*l}
		= 0\,,
	\\
	n \p_0 \check a^\o1
		+ m(\check a^\o1)
		- \zeta^*(\check c^\o1)
		+ j \varepsilon^{ijk} \zeta_i^* \p_j( \check b^\o0 \t h{_k_l} ) \zeta^{*l}
		= 0\,,
 \label{10.VII21.5}
	\\
	n \p_0 \check c^\o1
		- \zeta(\check a^\o1)
		+ \zeta^*(\check b^\o1)
		+ j \varepsilon^{ijk} m_i \p_j( \check b^\o0 \t h{_k_l}) \zeta^{*l}
		= 0\,,
 \label{10.VII21.4}
	\\
	\zeta(\check a^\o1) + \zeta^*(\check b^\o1) + m(\check c^\o1)
		= 0\,.
\end{align}
\checkedtogether{3.3.20}
Here we have used the fact that $\check b^\o0$ coincides with the unperturbed reflected scalar wave, defined in \eqref{eq:scalar reflected unperturbed}:
\begin{equation}
	\check b^\o0
		\equiv \check \phi^\o0
		= - e^{i (\check k.x + 2 n \omega \ell)}\,,
\end{equation}
and thus $\zeta^*(\check b^\o0) = 0$.

Equation \eqref{10.VII21.4}, obtained by contracting the evolution equation with $m_i$, is the “constraint equation” which, as shown in \Cref{app:reflection constraint consistent}, is automatically satisfied.

Using the explicit expression for $\theta$ from equation~\eqref{eq:maxwell boundary vector decomposition}, we have
\begin{equation}
	\theta^* = \nu + \epsilon h(m, \zeta) \zeta^*\,,
\end{equation}
where the second term, proportional to $\zeta^*$, gives no contribution in the expression $\t\gs{^i^j} \theta^*_i \p_j \check b$ because of $\zeta^*(\check b^\o0) = 0$, as just shown.
The relevant equations to determine the normal derivatives of the fields are thus
\begin{align}
	n \p_0 \check b - \nu( \check b )
		+ \epsilon \zeta(\check c^\o1)
		+ \epsilon \check b^\o0 j \varepsilon^{ijk} \zeta_i (\p_j \t h{_k_l}) \zeta^{*l}
		= 0\,,
	\\
	n \p_0 \check a^\o1
		+ m(\check a^\o1)
		- \zeta^*(\check c^\o1)
		+ j \varepsilon^{ijk} \zeta_i^* \p_j( \check b^\o0 \t h{_k_l} ) \zeta^{*l}
		= 0\,,
	\\
	\zeta(\check a^\o1) + \zeta^*(\check b^\o1) + m(\check c^\o1)
		= 0\,.
\end{align}
From this, it follows that on the mirror surface $\Sigma'$ it holds that
\begin{align}
	\label{eq:maxwell mirror normal b}
	\nu(\check b)
	&= n \t\p{_0} \check b + O(\Omega)\,,
	\\
	\label{eq:maxwell mirror normal a}
	m(\check a^\o1)
	&= - i n \omega \check a^\o1 + O(\Omega)\,,
	\\
	\label{eq:maxwell mirror normal c}
	m(\check c^\o1)
	&= - i n \omega \check c^\o1 + O(\Omega)\,.
\end{align}
The first equation is immediate, since tangential derivatives of $\check c^\o1$ and all derivatives of the perturbed metric are of order $\omegaG$, as can be seen from \eqref{eq:Maxwell reflection abc} and \eqref{eq:maxwell emitted c}.
For the second equation, we can again neglect the term $\zeta^*(\check c^\o1)$ but must analyse the last term, where the main contribution comes from the derivative of $\check b^\o0$ in the direction of $m$:
\begin{equation}
\begin{split}
	j \t \varepsilon{^i^j^k} \t*\zeta{^*_i} \t\p{_j}(\check b^\o0 \t h{_k_l}) \t \zeta{^*^l} / \omega
		&= j \t \varepsilon{^i^j^k} \t*\zeta{^*_i} \t m{_j} \t h{_k_l} \t \zeta{^*^l} m(\check b^\o0)/ \omega + O(\Omega)\\
		&= h(\zeta^*, \zeta^*) m(\check b^\o0)/ \omega + O(\Omega)\\
		&= -in h(\zeta, \zeta)^* \check b^\o0 + O(\Omega)\\
		&= 2in \check a^\o1 + O(\Omega),
\end{split}
\end{equation}
where we have used \eqref{eq:polarisation cross products}, $m(\check b^\o0) = - i n \omega \check b$, which is the unperturbed form of \eqref{eq:maxwell mirror normal b}, and \eqref{eq:maxwell check a1}.
Finally, to arrive at the last equation, one can use \eqref{eq:Maxwell reflection abc} obtain
\begin{equation}
	m(\check c^\o1) = \zeta^*(a^{\o1 *}) + \zeta(b^{\o1 *})\,.
\end{equation}
Using the fact that the incident field satisfies
\begin{equation}
	\zeta(a^\o1) + \zeta^*(b^\o1) + m(c^\o1) = 0\,,
\end{equation}
as well as the explicit form of $c^\o1$ in \eqref{eq:maxwell emitted c}, and also the reflection condition \eqref{eq:Maxwell reflection abc}, one obtains
\begin{equation}
	m(\check c^\o1)/\omega
		= - m(c^\o1)^*/\omega
		= - in c^\o1 + O(\Omega)
		= - in \check c^\o1 + O(\Omega)\,,
\end{equation}
which establishes \eqref{eq:maxwell mirror normal c}.

\subsection{The Reflected Wave}
\label{s:Maxwell wave equation reflection}

The unperturbed reflected wave is evidently given by
\begin{equation}
	\check Z^\o0
		= - \zeta^* e^{i \check k.x + 2 i n \omega \ell}\,,
\end{equation}
where $\t{\check k}{_\mu} = \omega(-1, - n m)$.
Plugging this into the general wave equation \eqref{eq:maxwell wave equation}, we find
\begin{equation}
\begin{split}
	\Delta(\t{\check Z}{^i}) - n^2 \t*\p{_0^2} \t{\check Z}{^i}
		=&{} - i \epsilon n \omega \left[
			2 \t\Gamma{^{\o1}^i_j_k} \t m{^j} \t \zeta{^*^k}
			- n j \t\varepsilon{^i^j^k} \t m{_j}\t{\dot h}{_k_l} \t\zeta{^*^l}
		\right] e^{i \check k.x + 2 n \omega \ell}\\
		&- \epsilon \left[
			2 \t R{^{\o1}^i_j} \t \zeta{^*^j}
			+ n j \t\varepsilon{^i^j^k} \t{\dot \Gamma}{_k_j_l} \t \zeta{^*^l}
		\right]e^{i \check k.x + 2 n \omega \ell}\,.
\end{split}
\end{equation}
The wave equations for $\check a$, $\check b$ and $\check c$ are now obtained by contracting with the basis vectors $\zeta^*$, $\zeta$ and $m$, respectively.
\begin{align}
	\label{eq:Maxwell a reflected wave eq}
	\begin{split}
		\wop_{\gamma^\o0} \check a^\o1
			=&{} - i n \omegaG \omega ( n + \mn) A(\zeta^*, \zeta^*) \sin(u) \check b^\o0
			\\&
			+ \omegaG^2 \left[
				A(\zeta^*, \zeta^*)
				+ n j \t\varepsilon{^i^j^k} \t*\zeta{^*_i} \t\gamma{_k_j_l} \t\zeta{^*^l}
			\right] \cos(u) \check b^\o0 \,,
	\end{split}
	\\
	\label{eq:Maxwell b reflected wave eq}
	\begin{split}
		\wop_{\gamma^\o0} \check b^\o1
			=&{} - n^2 \omega^2 A(m, m) \cos(u) \check b^\o0
			\\&
			- i n \omegaG \omega \left[
				2 \t \gamma{_i_j_k} \t\zeta{^i} \t m{^j} \t\zeta{^*^k}
				- n A(\zeta, \zeta^*)
			\right] \sin(u) \check b^\o0
			\\&
			+ \omegaG^2 \left[
				A(\zeta, \zeta^*)
				+ n j \t\varepsilon{^i^j^k} \t*\zeta{_i} \t\gamma{_k_j_l} \t\zeta{^*^l}
			\right] \cos(u) \check b^\o0
			\,,
	\end{split}
	\\
	\label{eq:Maxwell c reflected wave eq}
	\begin{split}
		\wop_{\gamma^\o0} \check c^\o1
			=&{} - i n \omegaG \omega (\kh.\zeta^*) A(m,m) \sin(u) \check b^\o0
			\\&
			+ \omegaG^2 \left[
				A(m, \zeta^*) + n j \t\varepsilon{^i^j^k} \t m{_i} \t\gamma{_k_j_l} \t \zeta{^*^l}
			\right] \cos(u) \check b^\o0\,.
	\end{split}
\end{align}
The components $\check a^\o1$, $\check b^\o1$ and $\check c^\o1$ can now be determined using the general formulae of \Cref{s:reflection general}. As done there, we momentarily suppress the GW\ phase offset $\chi$ for notational simplicity and reinstate it only in the final result of the computed field.

\paragraph{The $\check a$ Component}

For the component $\check a^\o1$, the value of $\check \alpha^\pm$ can be read from \eqref{eq:Maxwell a reflected wave eq}, and $\check \beta^\pm_1$ is determined by $- (b^\o1)^*$ according to \eqref{eq:Maxwell reflection abc}, which is given explicitly in \eqref{eq:maxwell emitted b}. Similarly, $\check \beta^\pm_2$ is determined from \eqref{eq:maxwell mirror normal a}. The such obtained non-vanishing parameters are
\begin{align}
	\check \alpha^\pm/e^{i n \omega \ell}
		&= \mp \half n \Omega ( n + \mn) A(\zeta^*, \zeta^*)
		+ O(\Omega^2)\,,
	\\
	\check \beta^\pm_{1,2}/e^{i n \omega \ell}
		&= \quarter A(\zeta^*, \zeta^*) + O(\Omega)
		+ O(\Omega \omegaG \ell)\,.
\end{align}
\checkedtogether{18.5.21}

\paragraph{The $\check b$ Component}

For the component $\check b^\o1$, the parameter $\check \alpha^\pm$ is determined from \eqref{eq:Maxwell b reflected wave eq}, $\check \beta^\pm_1$ is determined from  $- (a^\o1)^*$ according to \eqref{eq:Maxwell reflection abc}, which is given explicitly in \eqref{eq:maxwell emitted a}, and $\check \beta^\pm_2$ can be obtained from \eqref{eq:maxwell mirror normal b}.
These coefficients almost coincide with those for the reflected scalar wave given in \eqref{19V21.1}-\eqref{19V21.3}.
Here we merely state the deviations from the values given there:
\begin{align}
	\delta \check \alpha^\pm/e^{i n \omega \ell}
		&= \mp n \Omega \left(
			\t\gamma{_i_j_k} \t\zeta{^i} \t m{^j} \t\zeta{^*^k} - \half n A(\zeta^*, \zeta)
		\right)
		+ O(\Omega^2)\,,
	\\
	\begin{split}
		\delta \check \beta^\pm_{1,2}/e^{i n \omega \ell}
			&= - \half \left(
				\t\gamma{_i_j_k} \t\zeta{^i} \t m{^j} \t\zeta{^*^k} + \half n A(\zeta^*, \zeta)
			\right) \frac{1 - e^{\pm i \varpi}}{n - \mn}
			\\&\quad
			+ \quarter A(\zeta^*, \zeta) e^{\pm i \varpi}
			+ \half \varsigma^* e^{\pm i \varpi}
			+ O(\Omega) + O(\Omega \omegaG \ell)\,,
	\end{split}
	\\
	\check \gamma^\pm_{1,2}/e^{in\omega l}
		&= - \half[\varsigma^* - j (\alpha_c  \mp i \alpha_s)]e^{\pm i n \omegaG \ell}
		+ O(\Omega) + O(\Omega \omegaG \ell)\,,
	\\
	\check \delta_{1,2}/e^{in\omega l}
		&= \half j \alpha_0
		+ O(\Omega) + O(\Omega \omegaG \ell)\,.
\end{align}
\checkedtogether{19.5.21 and 25.5.21}

\paragraph{The $\check c$ Component}

For the last component, $\check c^\o1$, the parameter $\check \alpha^\pm$ can be inferred from \eqref{eq:Maxwell c reflected wave eq}, $\check \beta^\pm_1$ is determined from the incident field $c^\o1$ in accordance with \eqref{eq:Maxwell reflection abc}, which is given explicitly in \eqref{eq:maxwell emitted c}, and the parameter $\check \beta^\pm_2$ is determined from \eqref{eq:maxwell mirror normal c}.
To the considered level of approximation, the non-vanishing parameters are thus found to be
\begin{align}
	\check \alpha^\pm/e^{i n \omega \ell}
		&= \mp \half n \Omega (\kh.\zeta^*) A(m,m)
		+ O(\Omega^2)\,,
	\\
	\check \beta^\pm_{1,2}/e^{i n \omega \ell}
		&= \quarter (\kh.\zeta^*) A(m,m) \frac{1 - e^{\pm i \varpi}}{n - \mn}
		+ O(\Omega) + O(\Omega \omegaG \ell)\,,
\end{align}
where $\varpi$ is defined in \eqref{eq:varpi def}.
\checkedtogether{18.5.21}

\paragraph{Result}

One checks that the parameters just computed satisfy the conditions~\eqref{eq:counter-propagation reflection} (decorated with the $\pm$ superscripts), so that the components of the reflected electromagnetic field are readily found using \eqref{eq:scalar template reflection}:
\begin{align}
	\check a^\o1
		={}& - \half A(\zeta^*, \zeta^*) \cos(u) \check b^\o0
		+ O(\Omega) + O(\Omega \omegaG \ell)\,,
		\\
	\begin{split}
		\check b^\o1
			={}& \check \phi^\o1
			- \half A(\zeta^*, \zeta) \cos(u_1 + \varpi)\check b^\o0
			\\&
			- \left(
				\t\gamma{_i_j_k} \t\zeta{^i} \t m{^j} \t\zeta{^*^k} - \half n A(\zeta^*, \zeta)
			\right) \frac{\cos(u) - \cos(u_1)}{n + \mn} \check b^\o0
			\\&
			+ \left(
				\t\gamma{_i_j_k} \t\zeta{^i} \t m{^j} \t\zeta{^*^k} + \half n A(\zeta^*, \zeta)
			\right) \frac{\cos(u_1) - \cos(u_1 + \varpi)}{n - \mn} \check b^\o0
			\\&
			- \varsigma^* [\cos(u_1 + \varpi) - \cos(- \omegaG t - \omegaG n(m.x - 2 \ell) + \chi)]
			\\&
			- j \alpha_c \cos(- \omegaG t - \omegaG n(m.x - 2 \ell) + \chi) \check b^\o0
			\\&
			- j \alpha_s \sin(- \omegaG t - \omegaG n(m.x - 2 \ell) + \chi) \check b^\o0
			- j \alpha_0 \check b^\o0
			+ O(\Omega) + O(\Omega \omegaG \ell)\,,
	\end{split}
		\\
	\begin{split}
		\check c^\o1
			={}& - \half (\kh.\zeta^*) A(m,m) \left[
				\frac{\cos(u) - \cos(u_1)}{n + \mn}
				+ \frac{\cos(u_1) - \cos(u_1 + \varpi)}{n - \mn}
			\right] \check b^\o0
			\\&
			+ O(\Omega) + O(\Omega \omegaG \ell)\,.
	\end{split}
\end{align}
\checkedtogether{25.5.21}

\paragraph{Returning Field}
Finally, we evaluate the reflected field at the spatial coordinate origin $\t x{^i} = 0$, where the beam splitter is positioned.
Let us use the notation
\begin{align}
	u^R
		&:= u_1 |_\Sigma
		= u + (n + \mn) \omegaG \ell\,,
		\\
		u^E
		&:= (u_1 + \varpi) |_\Sigma
		= u + 2 n \omegaG \ell\,,
\end{align}
to denote the values of the GW\ phase at the (retarded) times of light reflection and emission, respectively.
In order to make contact with the notation of \cite{Mieling:2021a}, given a vector $\tilde k$ we set
\begin{equation}
	\label{eq:def integrated Christoffel}
	\t\varGamma{^i_j}(\tilde k,u_2, u_1)
		= \frac{1}{\gamma^\o0(\tilde k, \kappa)} \int_{u_1}^{u_2} \t \Gamma{^i_j_l}(u) \t {\tilde k}{^l} \,\dd u\,,
\end{equation}
so that
\begin{align}
	\t\varGamma{^i_j}(k,u_2,u_1)
		&= - \t \gamma{^i_j_k} \t m{^k} \frac{\cos(u_2) - \cos(u_1)}{n - \mn}\,,
	\\
	\t\varGamma{^i_j}(\check k,u_2,u_1)
		&= + \t \gamma{^i_j_k} \t m{^k} \frac{\cos(u_2) - \cos(u_1)}{n + \mn}\,.
\end{align}
Using the function $\alpha(t)$ defined in \eqref{eq;maxwell emitted polarisation perturbation}, we have
\begin{align}
	\label{eq:maxwell result a}
	\check a^\o1|_0
		={}&\! - \half h(\zeta^*, \zeta^*) \check b^\o0
		+ O(\Omega) + O(\Omega \omegaG \ell)\,,
		\\
	\label{eq:maxwell result b}
	\begin{split}
		\check b^\o1|_0
			={}& \check \phi^\o1
			- \half h(\zeta^*, \zeta)|_{u = u^E} \, \check b^\o0
			- j \alpha(t - 2 n \ell) \, \check b^\o0
			\\&
			- \t\zeta{_i} \t \varGamma{^i_j}(\check k, u, u^R) \t\zeta{^*^j} \check b^\o0
			- \t*\zeta{_i} \t \varGamma{^i_j}(k, u^R, u^E) \t\zeta{^*^j} \check b^\o0
			\\&
			+ \quarter n^2 \omegaG \omega \left(
				\frac{h(m,m)}{\gamma^\o0(\check k, \kappa)}\bigg|^{u}_{u^R}
				+ \frac{h(m,m)}{\gamma^\o0(k, \kappa)}\bigg|^{u^R}_{u^E}
			\right) \check b^\o0
			+ O(\Omega) + O(\Omega \omegaG \ell)\,,
	\end{split}
	\\
	\label{eq:maxwell result c}
	\begin{split}
		\check c^\o1|_0
			={}& \half n \omegaG \omega (\kh.\zeta^*) \left(
				\frac{h(m,m)}{\gamma^\o0(\check k, \kappa)}\bigg|^{u}_{u^R}
				+ \frac{h(m,m)}{\gamma^\o0(k, \kappa)}\bigg|^{u^R}_{u^E}
			\right) \check b^\o0
			+ O(\Omega) + O(\Omega \omegaG \ell)\,.
	\end{split}
\end{align}
\checkedtogether{25.5.21} 

%% file: optics_subfiles/maxwell_interferometry.tex
\section{Michelson Interferometers}
\label{s:interferometry}

\begin{figure}[ht]
	\centering
	\includegraphics[width=0.5\textwidth]{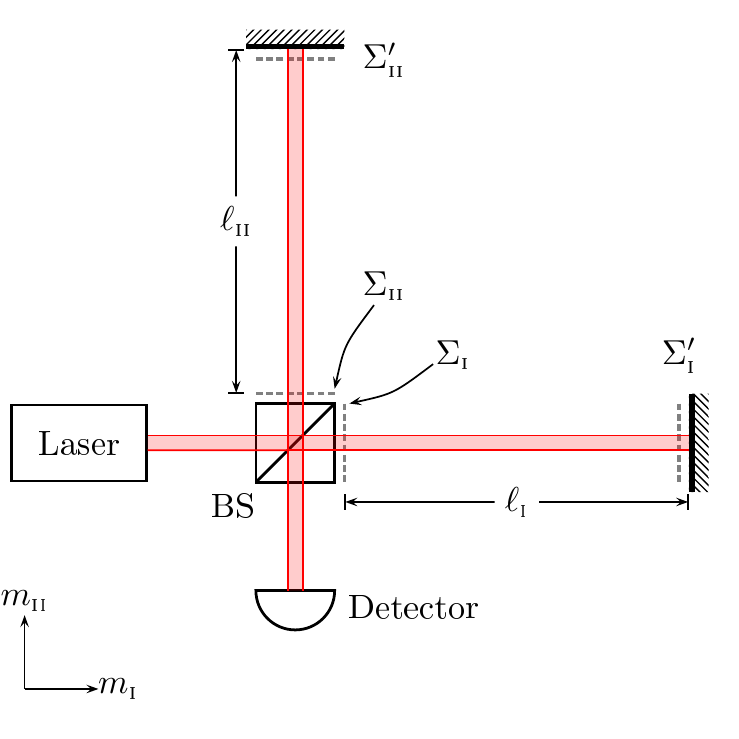}
	\caption{Schematic drawing of a Michelson interferometer.}
	\label{fig:Michelson}
\end{figure}

We are now in a position to describe the gravitational wave response of a Michelson interferometer, as sketched in \Cref{fig:Michelson}.

We use the  approach of Ref.~\cite{Mieling:2021a} to describe such interferometers.
In Minkowski spacetime, i.e.\ the unperturbed geometry, let $m_\cI$ and $m_\cII$ denote the unit vectors pointing along the two interferometer arms (away from the beam splitter), defining the two emission surfaces $\Sigma_\cI =  \{ m_\cI . x = 0 \}$ and $\Sigma_\cII = \{ m_\cII . x = 0 \}$.
The laser emits coherent light with wave vector $k = \omega(m_\cI - \dd t)$ and polarisation $\zeta$.
Part of the ray passes through the beam splitter unchanged except for a reduction in amplitude, and thus defines boundary data on $\Sigma_\cI$, corresponding to normal emission with frequency $\omega$ and polarisation $\zeta$.
Similarly, the deflected part of the beam gives rise to boundary data on $\Sigma_\cII$, corresponding again to normal emission with frequency $\omega$, but with the polarisation vector $\eta = R_\text{BS}^\o0 \zeta$ where $R_\text{BS}^\o0$ is the orthogonal matrix interchanging $m_\cI$ and $m_\cII$ while leaving their orthogonal complements unchanged (see below for details).
There is also a phase shift at the beam splitter, which depends upon the details of the apparatus~\cite[Section~2.4]{Bond_2016}. Since overall phases are irrelevant for our analysis, it suffices to include the relative phase shift of $\pi$ in the Michelson interferometer, which we incorporate in the output field below; see \eqref{s:interferometer output field}.

This description of the beam splitter can be carried over to the perturbed geometry with minimal modifications.
To this end, we prescribe boundary data for Maxwell’s equations on the same surfaces $\Sigma_\cI$ and $\Sigma_\cII$ (whose unit normal one-forms $\nu_\cI$ and $\nu_\cII$ are proportional to $m_\cI$ and $m_\cII$, respectively) as in the Minkowski case.
We assume again normal emission from both surfaces $\Sigma_\cI$ and $\Sigma_\cII$, and the polarisation on $\Sigma_\cI$ is assumed to be of the form given in \eqref{eq:Maxwell boundary z} (as dictated by the assumption of normal emission of plane waves), with the polarisation on $\Sigma_\cII$ (sent into the second arm) related to the one on $\Sigma_\cI$ by the orthogonal map $R_\text{BS}$ which now interchanges $\nu_\cI$ and $\nu_\cII$ while leaving their orthogonal complement untouched.

A simple renaming of the relevant data in the solutions derived above provides then the new solutions, as needed to determine the resulting field at the interferometer output.

As such, there are various readout schemes used in gravitational wave detection. For simplicity, we focus here on the DC readout scheme currently used in the LIGO and Virgo detectors. In this scheme, the arm lengths are chosen to be unequal ($\Delta \ell = \ell_\cI - \ell_\cII \neq 0$) such that even in the absence of a gravitational wave there is a constant output signal registered by the 
photodiode, and the gravitational wave then causes a small fluctuation of this signal, which we determine below.

To describe the output of the interferometer, we collect the results obtained so far to describe the field returning to the beam splitter as a function of the emitted field.
We then use our model of the beam splitter to describe how the output of the laser is transferred into the two interferometer arms,
 and to describe how the returning rays are transferred to the detector.

\subsection{The Returning Field}
 \label{s9VII21.1}

For ease of further reference, we summarise our calculations so far. We have solved Maxwell’s equations in the GW\ metric
\begin{align}
	\t g{_\mu_\nu} &= \t \eta{_\mu_\nu} + \epsilon \t h{_\mu_\nu}
	\equiv \t \eta{_\mu_\nu} + \epsilon \t A{_\mu_\nu} \cos(u)\,,
	\\
\shortintertext{where}
	u &= \kappa.x + \chi
	\equiv \omegaG (-t + \kh.x) + \chi
	\,,
\end{align}
with the electromagnetic field prescribed on the emission surface
\begin{equation}
	\Sigma = \{m.x = 0\}\,,
\end{equation}
to be emitted normally and to have unit norm at the coordinate origin, so that
\begin{align}
	Z_\text{in}|_\Sigma
		&= e^{- i \omega t} \left[
			\zeta (1 + \epsilon j \tilde a - \half \epsilon h(\zeta^*, \zeta))
			- \half \epsilon \zeta^*\, h(\zeta, \zeta)
		\right]\,,
 \label{10VII21.1}
\end{align}
where $\zeta$ is the unperturbed polarisation vector, free to choose subject to the conditions \eqref{eq:polarisation conditions unperturbed}, the asterisk indicates $j$-complex conjugation,
and $\tilde a$ is the function
\begin{equation}
	\label{eq: tilde a recall}
	\tilde a
		= \alpha(t)
		+ j \varsigma [ \cos(\kappa.x) - \cos(-\omegaG t + \chi) ]\,,
\end{equation}
where $\alpha$ and $\varsigma$ are defined in \eqref{eq;maxwell emitted polarisation perturbation} and \eqref{eq: varsigma def}, respectively.

To write the returning field in a concise notation, we use the definitions of $H$, $\mathring H$ and $\t \varGamma{^i_j}$ from \eqref{eq:def integrated waveform}, \eqref{eq:def difference in waveform}, and \eqref{eq:def integrated Christoffel}, as well as
\begin{align}
	\t k{_\mu} &= \omega(-1, + n \t m{_i})\,,
	&
	\t {\check k}{_\mu} &= \omega(-1, - n \t m{_i})\,,
\end{align}
together with the unperturbed optical metric
\begin{equation}
	\t \gamma{^{\o0}^\mu^\nu} = \diag(-n^2, 1, 1, 1)\,,
\end{equation}
and set
\begin{align}
	u^E
		&= u + 2 n \omegaG \ell\,,
	\\
	u^R
		&= u + (n + \mn) \omegaG \ell\,,
	\\
	\Delta h
		&= \frac{h(m, m)}{\gamma^\o0(\kappa, \check k)}\bigg|^{u}_{u^R}
		+ \frac{h(m, m)}{\gamma^\o0(\kappa, k)}\bigg|^{u^R}_{u^E}\,.
\end{align}
Then, the returning field at the coordinate origin, given by \eqref{eq:maxwell result a}—\eqref{eq:maxwell result c} and \eqref{eq:scalar result phase perturbation} as well as \eqref{eq:scalar result amplitude perturbation}, can be written as
\begin{align}
	\label{eq:michelson round-trip transformation}
	Z_\text{out}
		&= - \Aa \Zz e^{i \psi}\,,
	\\
\shortintertext{with}
	\label{eq:michelson round-trip phase}
	\psi
		&= - \omega t + 2 n \omega \ell
			+ \epsilon H(\check k, u, u^R)
			+ \epsilon H(k, u^R, u^E)
			\,,
		\\
	\begin{split}
	\label{eq:michelson round-trip transformation amplitude}
		\Aa
			&= 1 - \half \epsilon \omegaG \ell A(m, m) \frac{1 - (m.\kh)^2}{n^2 - (m.\kh)^2} \left[
				(m.\kh) \sin(u^R) - (n + m.\kh) \sin(u^E)
			\right]
			\\&\qquad
			+ \half \epsilon (n^2 - 1) \omegaG^2 \left[
				\frac{\mathring H(\check k, u, u^R)}{\gamma^\o0(\kappa, \check k)}
				+ \frac{\mathring H(k, u^R, u^E)}{\gamma^\o0(\kappa, k)}
			\right]\,,
	\end{split}
	\\
	\begin{split}
	\label{eq:michelson round-trip transformation polarisation}
		\Zz
			&= \zeta^* \bigg[
				1 - \epsilon j \alpha(t - 2 n \ell)
				- \half \epsilon h(\zeta^*, \zeta)\big|_{u^E}
				\\&\qquad
				- \epsilon \t \zeta{_i} [ \t\varGamma{^i_j}(\check k, u, u^R) + \t\varGamma{^i_j}( k, u^R, u^E) ] \t \zeta{^*^j}
				+ \quarter \epsilon n^2 \omegaG \omega \Delta h
			\bigg]
			\\&\quad
			- \half \epsilon \zeta\, h(\zeta^*, \zeta^*)
			+ \half \epsilon m\, n \omegaG \omega (\zeta^*.\kh) \Delta h\,.
	\end{split}
\end{align}
\checkedtogether{25.6.21}

Since the returning field at the origin only involves the function $\tilde a$ evaluated again at the spatial coordinate origin (but at a retarded time) the $\varsigma$ terms in \eqref{eq: tilde a recall}, which were necessary to satisfy the constraint equations, are irrelevant for the interferometer output.

This expression serves as the basis for the description of interferometers, as it describes how light emitted from a surface $\Sigma$ returns after being reflected at a perfect mirror placed a distance $\ell$ away.

\subsection{Non-Polarising Beam Splitters}

To relate the light polarisation vectors sent into the two interferometer arms (which are both derived from the incoming laser beam) and to compute the output laser field from the returning light rays, one must model the polarisation transfer at the beam splitter.
Following the model put forward in \cite{Mieling:2021a}, we assume the polarisation of the field emitted by the laser to have the form
\begin{equation}
	\label{eq:michelson input field}
	\Zz_\text{in}|_{\t x{^i} = 0}
		= \sqrt 2 \zeta (1 + \epsilon j \alpha - \half \epsilon h(\zeta^*, \zeta)) - \half \epsilon h(\zeta, \zeta) \zeta^*\,,
\end{equation}
as already considered in the previous calculation (up to the scaling factor $\sqrt 2$ which is conventional).
Assuming a 50:50 beam splitter, the amplitudes of the two rays transferred to the two arms are reduced by a factor $1/\sqrt 2$. Concerning the polarisation vectors, we assume that the part of the ray which passes straight through the beam splitter maintains its polarisation, while the polarisation sent into the orthogonal arm is given by applying the transformation
\begin{align}
	\label{eq:michelson bs reflection}
	\Zz \mapsto - R_\textsc{bs} \Zz^*\,,
\end{align}
analogously to \cref{eq:Maxwell mirror solution}, where $R_\textsc{bs}$ is the orthogonal matrix
which interchanges the unit conormals $\nu_\cI \propto m_\cI$ and $\nu_\cII \propto m_\cII$ (see \Cref{fig:Michelson})
\begin{equation}
	\t{{R_\textsc{bs}}}{^i_j}
	= \t*\delta{^i_j} - [1 - g(\nu_\cI, \nu_\cII)]^{-1} \t g{^i^k} \t{(\nu_\cII - \nu_\cI)}{_k} \t{(\nu_\cII - \nu_\cI)}{_j}\,.
\end{equation}
As shown in \Cref{app:beam splitter reflection}, see \eqref{eq:beam splitter reflection app}, the action of this operator on the polarisation vector \eqref{eq:Maxwell boundary z} yields
\begin{equation}
\begin{split}
	R_\textsc{bs}[
		&\zeta (1 + \epsilon j \alpha - \half \epsilon h(\zeta^*, \zeta))
		- \half \epsilon \zeta^* \, h(\zeta, \zeta)
		]
	\\&=
		\eta ( 1 + \epsilon j (\alpha + \breve a) - \half h (\eta^*, \eta^*))
		- \half \epsilon \eta^*\, h(\eta, \eta)\,,
\end{split}
\end{equation}
where $\eta$ is the unperturbed polarisation vector of the second arm:
\begin{equation}
	\eta = \zeta - (m_\cII - m_\cI) (m_\cII. \zeta) \,,
\end{equation}
and $\breve a$ is the $j$-imaginary part of the function $h(\zeta^*, \zeta - \eta )$:
\begin{equation}
	\breve a
		= \Im_j[ h(\zeta^*, \zeta - \eta ) ]\,.
\end{equation}

For the returning rays, the behaviour of the fields is similar. The ray which first passed through the beam splitter without deflection is now deflected towards the detector and thus undergoes the transformation \eqref{eq:michelson bs reflection} with its amplitude reduced by $1/\sqrt 2$.
The other beam, which was deflected before completing its round trip, now maintains its polarisation vector when being transferred towards the detector. Similarly to the first ray, its amplitude is reduced by $1/\sqrt 2$, but additionally there is the standard phase shift of $\pi$.

For a more detailed description of the beam splitter transformations, see \cite[Sect.~6]{Mieling:2021a}.

\subsection{Output Field}
\label{s:interferometer output field}

We are now in the position to compute the output field reaching the detector, which is a sum of the fields from the two interferometer arms.

\paragraph{Ray I}

Consider the ray which is deflected only after its round-trip in one of the interferometer arms.
According to our model of the beam splitter, the corresponding field emanating from $\Sigma_\cI$ is thus (see~\eqref{10VII21.1}—\eqref{eq: tilde a recall})
\begin{equation}
	Z_{\cI\,\text{in}}|_{\t x{^i} = 0}
		= + e^{- i \omega t} [
			\zeta (1 + \epsilon \alpha - \half \epsilon h(\zeta^*, \zeta))
			- \half \epsilon \zeta^*\, h(\zeta, \zeta)
		]\,.
\end{equation}
\checkedtogether{25.6.21}
The returning field is given by  \eqref{eq:michelson round-trip transformation}—\eqref{eq:michelson round-trip transformation polarisation}.
To evaluate this at the coordinate origin, we set
\begin{align}
	u^E_{\cI}
		&= u + 2 n \omegaG \ell_{\cI}
 \,,
	&u^R_{\cI}
		= u + (n + \kh.m_\cI) \omegaG \ell_{\cI}
 \,.
\end{align}
Recall that at the coordinate origin we have $u|_{\t x{^i} = 0} = - \omegaG t + \chi$.
Applying the beam splitter transformation \eqref{eq:michelson bs reflection}, reducing the amplitude by $1/\sqrt 2$, and taking the $i$-real part, one obtains the contribution of this ray to the field at the detector output
\begin{align}
	\label{eq:michelson field 1}
	Z_\cI
		&= \tfrac{1}{\sqrt 2} \Aa_\cI \Zz_\cI \cos(\psi_\cI)\,,
	\\
\shortintertext{where}
	\label{eq:michelson phase 1}
	\psi_\cI
		&= - \omega t + 2 n \omega \ell_\cI
			+ \epsilon H(\check k_\cI, u, u^R_\cI)
			+ \epsilon H(k_\cI, u^R_\cI, u^E_\cI)
			\,,
		\\
	\label{eq:michelson amplitude 1}
	\begin{split}
		\Aa_\cI
			&= 1 - \half \epsilon \omegaG \ell_\cI A(m_\cI, m_\cI) \frac{1 - (m_\cI.\kh)^2}{n^2 - (m_\cI.\kh)^2} \left[
				(m_\cI.\kh) \sin(u^R_\cI) - (n + m_\cI.\kh) \sin(u^E_\cI)
			\right]
			\\&\qquad
			+ \half \epsilon (n^2 - 1) \omegaG^2 \left[
				\frac{H(\check k_\cI, u, u^R_\cI)}{\gamma^\o0(\kappa, \check k_\cI)}
				+ \frac{H(k_\cI, u^R_\cI, u^E_\cI)}{\gamma^\o0(\kappa, k_\cI)}
			\right]\,,
	\end{split}
	\\
	\label{eq:michelson polarisation 1}
	\begin{split}
		\Zz_\cI
			&= \eta \bigg[
				1 + \epsilon j(\alpha(t - 2 n \ell_\cI) + \breve a(t))
				- \half \epsilon h(\eta^*, \eta)
				+ \half \epsilon h(\zeta^*, \zeta)\big|^{u}_{u^E_\cI}
				\\&\qquad
				- \epsilon \t*\zeta{^*_i} [ \t\varGamma{^i_j}(\check k_\cI, u, u^R_\cI) + \t\varGamma{^i_j}( k_\cI, u^R_\cI, u^E_\cI) ] \t \zeta{^j}
				+ \quarter \epsilon n^2 \omegaG \omega \Delta h_\cI
			\bigg]
			\\&\quad
			- \half \epsilon \eta^*\, h(\eta, \eta)
			+ \half \epsilon m_\cII\, n \omegaG \omega (\kh.\zeta) \Delta h_\cI\,,
	\end{split}
\end{align}
and
\begin{equation}
	\Delta h_\cI
		:= \frac{h(m_\cI, m_\cI)}{\gamma^\o0(\kappa, \check k_\cI)}\bigg|^{u}_{u^R_\cI}
		+ \frac{h(m_\cI, m_\cI)}{\gamma^\o0(\kappa, k_\cI)}\bigg|^{u^R_\cI}_{u^E_\cI}\,.
\end{equation}

\paragraph{Ray II}

The other ray is deflected at the beam splitter before the round-trip.
The corresponding field emanating from $\Sigma_\cII$ is
\begin{equation}
	Z_{\cII\,\text{in}}|_{\t x{^i} = 0}
		= - e^{- i \omega t} [
			\eta^* (1 - \epsilon j(\alpha + \breve a) - \half \epsilon h(\eta^*, \eta))
			- \half \epsilon \eta\, h(\eta^*, \eta^*)
		]\,,
\end{equation}
and the returning field is directly obtained from \eqref{eq:michelson round-trip transformation}—\eqref{eq:michelson round-trip transformation polarisation} via the substitutions
\begin{align}
	\zeta &\to \eta^*\,,
	&
	\zeta^* &\to \eta\,,
	&
	\alpha &\to - (\alpha + \breve a)
	&
	m &\to m_\cII
	&
	k &\to k_\cII
 \,,
\end{align}
together with a change of the overall sign. This ray passes through the beam splitter with its amplitude reduced by $1/\sqrt 2$ and the phase shifted by $\pi$, but without modification of the polarisation vector.
Finally, taking the $i$-real part to obtain the physical field, the contribution to the output field from this ray is
\begin{align}
	\label{eq:michelson field 2}
	Z_\cII
		&= \tfrac{1}{\sqrt 2} \Aa_\cII \Zz_\cII \cos(\psi_\cII)\,,
	\\
\shortintertext{where}
	\label{eq:michelson phase 2}
	\psi_\cII
		&= - \omega t + 2 n \omega \ell_\cII + \pi
			+ \epsilon H(\check k_\cII, u, u^R_\cII)
			+ \epsilon H(k_\cII, u^R_\cII, u^E_\cII)
			\,,
		\\
	\label{eq:michelson amplitude 2}
	\begin{split}
		\Aa_\cII
			&= 1 - \half \epsilon \omegaG \ell_\cII A(m_\cII, m_\cII) \frac{1 - (m_\cII.\kh)^2}{n^2 - (m_\cII.\kh)^2} \left[
				(m_\cII.\kh) \sin(u^R_\cII) - (n + m_\cII.\kh) \sin(u^E_\cII)
			\right]
			\\&\qquad
			+ \half \epsilon (n^2 - 1) \omegaG^2 \left[
				\frac{H(\check k_\cII, u, u^R_\cII)}{\gamma^\o0(\kappa, \check k_\cII)}
				+ \frac{H(k_\cII, u^R_\cII, u^E_\cII)}{\gamma^\o0(\kappa, k_\cII)}
			\right]\,,
	\end{split}
	\\
	\label{eq:michelson polarisation 2}
	\begin{split}
		\Zz_\cII
			&= \eta \bigg[
				1 + \epsilon j(\alpha(t - 2 n \ell_\cII) + \breve a(t - 2 n \ell_\cII))
				- \half \epsilon h(\eta^*, \eta)
				\\&\qquad
				- \epsilon \t*\eta{^*_i} [ \t\varGamma{^i_j}(\check k_\cII, u, u^R_\cII) + \t\varGamma{^i_j}( k_\cII, u^R_\cII, u^E_\cII) ] \t \eta{^j}
				+ \quarter \epsilon n^2 \omegaG \omega \Delta h_\cII
			\bigg]
			\\&\quad
			- \half \epsilon \eta^*\, h(\eta, \eta)
			+ \half \epsilon m_\cII\, n \omegaG \omega (\kh.\eta) \Delta h_\cII\,,
	\end{split}
\end{align}
with
\begin{equation}
	\Delta h_\cII
		:= \frac{h(m_\cII, m_\cII)}{\gamma^\o0(\kappa, \check k_\cII)}\bigg|^{u}_{u^R_\cII}
		+ \frac{h(m_\cII, m_\cII)}{\gamma^\o0(\kappa, k_\cII)}\bigg|^{u^R_\cII}_{u^E_\cII}\,,
\end{equation}
and
\begin{align}
	u^E_{\cII}
		&= u + 2 n \omegaG \ell_{\cII}
 \,,
	&u^R_{\cII}
		= u + (n + \kh.m_\cII) \omegaG \ell_{\cII}
 \,.
\end{align}

\subsection{Output Power}
\label{s:interferometry signal general}

Adding the results of the two previous sections, the physical field at the output of the interferometer is found to be
\begin{equation}
	Z_\text{out}
		= Z_\cI + Z_\cII
		\equiv \tfrac{1}{\sqrt 2} \Aa_\cI \Zz_\cI \cos(\psi_\cI)
		+ \tfrac{1}{\sqrt 2} \Aa_\cII \Zz_\cII \cos(\psi_\cII)\,,
\end{equation}
whose energy density is proportional to the norm
\begin{equation}
	\t T{_0_0} \propto g(Z_\cI^* + Z_\cII^*, Z_\cI + Z_\cII)\,,
\end{equation}
compare~\Cref{appendix:energy}.
The exact proportionality factor is irrelevant for our purposes, as we are concerned only with the ratio of the output power relative to the input power. Following the standard time-averaging  procedure, which we denote by $\langle \cdot \rangle$, we find the observable output power, normalised to the input power, to be
\begin{equation}
	P = \langle g(Z_\text{out}^*, Z_\text{out}) \rangle / \langle g(Z_\text{in}^*, Z_\text{in}) \rangle\,.
\end{equation}
As we have normalised the input field \eqref{eq:michelson input field} to $\langle g(Z_\text{in}^*, Z_\text{in}) \rangle = 1$, we get
\begin{equation}
	\begin{split}
		P
			&= \langle g(Z^*_\cI + Z^*_\cII, Z_\cI + Z_\cII) \rangle \\
			&= \quarter \Aa_\cI^2 g(\Zz_\cI^*, \Zz_\cI)
			+ \quarter \Aa_\cII^2 g(\Zz_\cII^*, \Zz_\cII)
			+ \quarter \Aa_\cI \Aa_\cII [ g(\Zz_\cI^*, \Zz_\cII) + g(\Zz_\cI,
               \Zz_\cII^*) ] \cos(\psi_\cI - \psi_\cII) \,.
	\end{split}
\end{equation}
This is directly measured in the commonly used DC~readout scheme of gravitational wave detectors.
In the unperturbed case, the normalised output power takes the simple form
\begin{equation}
	P^\o0
		= \half - \half \cos(2n  \omega \Delta \ell)
		= \sin^2(n \omega \Delta \ell)\,,
\end{equation}
which vanishes for $\Delta \ell \equiv \ell_\cI - \ell_\cII = 0$.
In the perturbed case, a direct calculation based on \cref{eq:michelson polarisation 1,eq:michelson polarisation 2} shows that
\begin{align}
	g(\Zz_\cI^*, \Zz_\cI)
		&= 1 + \epsilon n^2 \omegaG \omega \Delta h_\cI\,,
		\\
	g(\Zz_\cII^*, \Zz_\cII)
		&= 1 + \epsilon n^2 \omegaG \omega \Delta h_\cII\,,
		\\
	2 \Re_j g(\Zz_\cI^*, \Zz_\cII)
		&= 2 + \epsilon n^2 \omegaG \omega [\Delta h_\cI + \Delta h_\cII]\,.
\end{align}
Using the explicit form of the phase and amplitude perturbations given by \cref{eq:michelson phase 1,eq:michelson amplitude 1,eq:michelson phase 2,eq:michelson amplitude 2}, one finds that the perturbed output power can be written in the form
\begin{equation}
	P
		= P^\o0
		+ \epsilon \delta_\psi P
		+ \epsilon \delta_\Aa P
		+ \epsilon \delta_n P
		+ \epsilon \delta_K P
		+ O(\epsilon^2)\,,
\end{equation}
where
\begin{align}
	\begin{split}
		\delta_\psi P
			={}& \half \sin(2 n \omega \Delta \ell)
			\\&
			\times \left(
				H(\check k_\cI, u, u^R_\cI) + H(k_\cI, u^R_\cI, u^E_\cI)
				- H(\check k_\cII, u, u^R_\cII) - H(k_\cII, u^R_\cII, u^E_\cII)
			\right)\,,
	\end{split}
	\\
	\begin{split}
		\delta_\Aa P
			={}& - \half \sin^2(n \omega \Delta \ell)
			\\&
			\times \bigg(
				\omegaG \ell_\cI A(m_\cI, m_\cI) \frac{1 - (m_\cI.\kh)^2}{n^2 - (m_\cI.\kh)^2} \left[
					(m_\cI.\kh) \sin(u^R_\cI) - (n + m_\cI.\kh) \sin(u^E_\cI)
				\right]
				\\&\qquad
				+ \omegaG \ell_\cII A(m_\cII, m_\cII) \frac{1 - (m_\cII.\kh)^2}{n^2 - (m_\cII.\kh)^2} \left[
					(m_\cII.\kh) \sin(u^R_\cII) - (n + m_\cII.\kh) \sin(u^E_\cII)
				\right]
			\bigg)\,,
	\end{split}
	\\
	\begin{split}
		\delta_n P
			={}& \half (n^2 - 1) \omegaG^2 \sin^2(n \omega \Delta \ell)
			\\&
			\times \bigg(
				\frac{\mathring H(\check k_\cI, u, u^R_\cI)}{\gamma^\o0(\kappa, \check k_\cI)}
				+ \frac{\mathring H(k_\cI, u^R_\cI, u^E_\cI)}{\gamma^\o0(\kappa, k_\cI)}
				+ \frac{\mathring H(\check k_\cII, u, u^R_\cII)}{\gamma^\o0(\kappa, \check k_\cII)}
				+ \frac{\mathring H(k_\cII, u^R_\cII, u^E_\cII)}{\gamma^\o0(\kappa, k_\cII)}
			\bigg)\,,
	\end{split}
	\\
	\begin{split}
		\delta_K P
			={}& \half n^2 \omegaG \omega \sin^2(n \omega \Delta \ell)
			\\&
			\times \bigg(
				\frac{h(m_\cI, m_\cI)}{\gamma^\o0(\kappa, \check k_\cI)}\bigg|^{u}_{u^R_\cI}
				+ \frac{h(m_\cI, m_\cI)}{\gamma^\o0(\kappa, k_\cI)}\bigg|^{u^R_\cI}_{u^E_\cI}
				+ \frac{h(m_\cII, m_\cII)}{\gamma^\o0(\kappa, \check k_\cII)}\bigg|^{u}_{u^R_\cII}
				+ \frac{h(m_\cII, m_\cII)}{\gamma^\o0(\kappa, k_\cII)}\bigg|^{u^R_\cII}_{u^E_\cII}
			\bigg)\,.
	\end{split}
\end{align}
This confirms the results of~\cite{Mieling:2021a} and generalises them beyond the vacuum case $n = 1$: the terms $\delta_\psi P$, $\delta_\Aa P$ and $\delta_K P$ are trivial modifications of the corresponding expressions for vacuum, while the term $\delta_n P$ has no counterpart in the vacuum case.

Notably, both in vacuum and in dielectrics, the perturbation of the polarisation vectors produce no (linear) perturbation of the output intensity. This is because $\epsilon$-perturbations of these vectors produce only $\epsilon^2$ terms in their inner products with the unperturbed vectors.

As in~\cite{Mieling:2021a}, $\delta_\psi P$ can be attributed to the GW\ perturbation of the optical phase, and $\delta_K P$ corresponds to a perturbation of the EM\ frequency, which rescales the unperturbed signal (as can be seen from the $\sin^2(\omega \Delta \ell)$ term).
While in vacuum the perturbation of the amplitude produces only one correction similar to $\delta_\Aa P$, in the case of $n \neq 1$ there is a further term $\delta_n P$ arising from the difference in propagation speeds of the electromagnetic and gravitational waves.

\checked{9.7.21}{stp}

\subsection{Low Frequency Limit}
\label{s:interferometry signals LFL}

\begin{figure}[H]
	\centering
	\begin{subfigure}[c]{0.28\columnwidth}
		\includegraphics[width=\columnwidth]{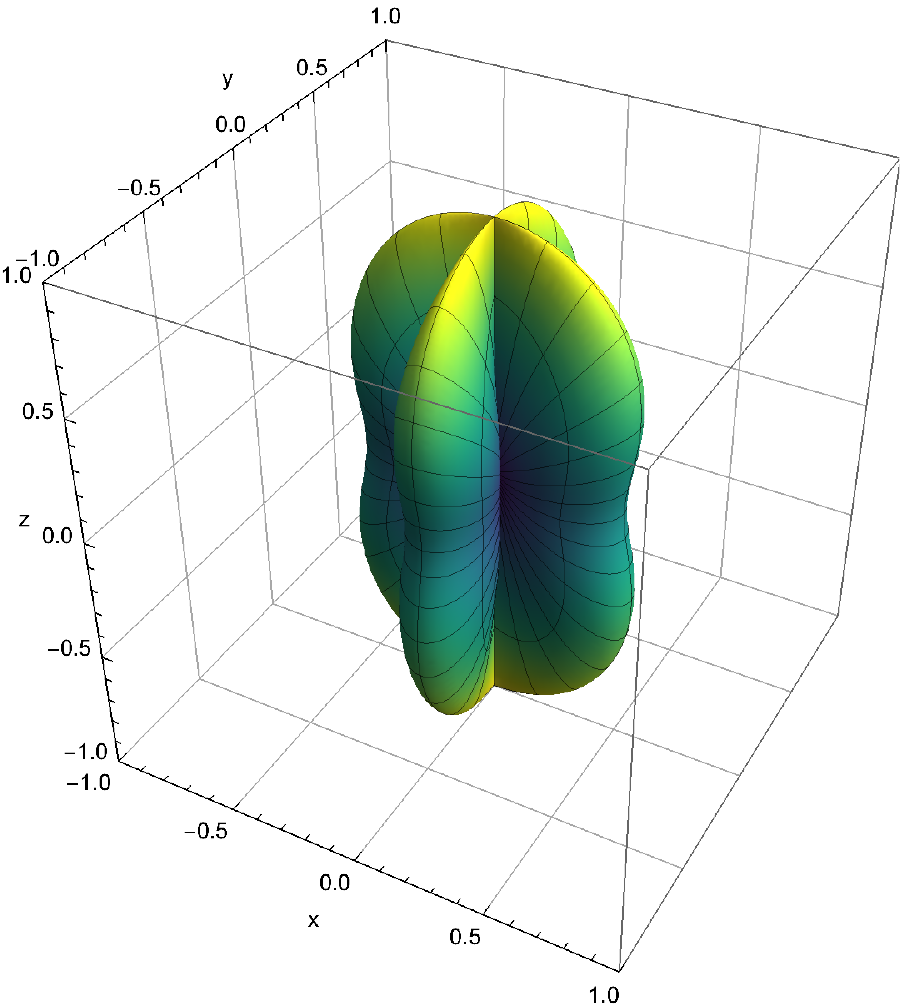}
		\subcaption{Pattern function $F_+$}
		\label{fig: F+ vacuum}
	\end{subfigure}
	\hspace{0.05\columnwidth}
	\begin{subfigure}[c]{0.28\columnwidth}
		\includegraphics[width=\columnwidth]{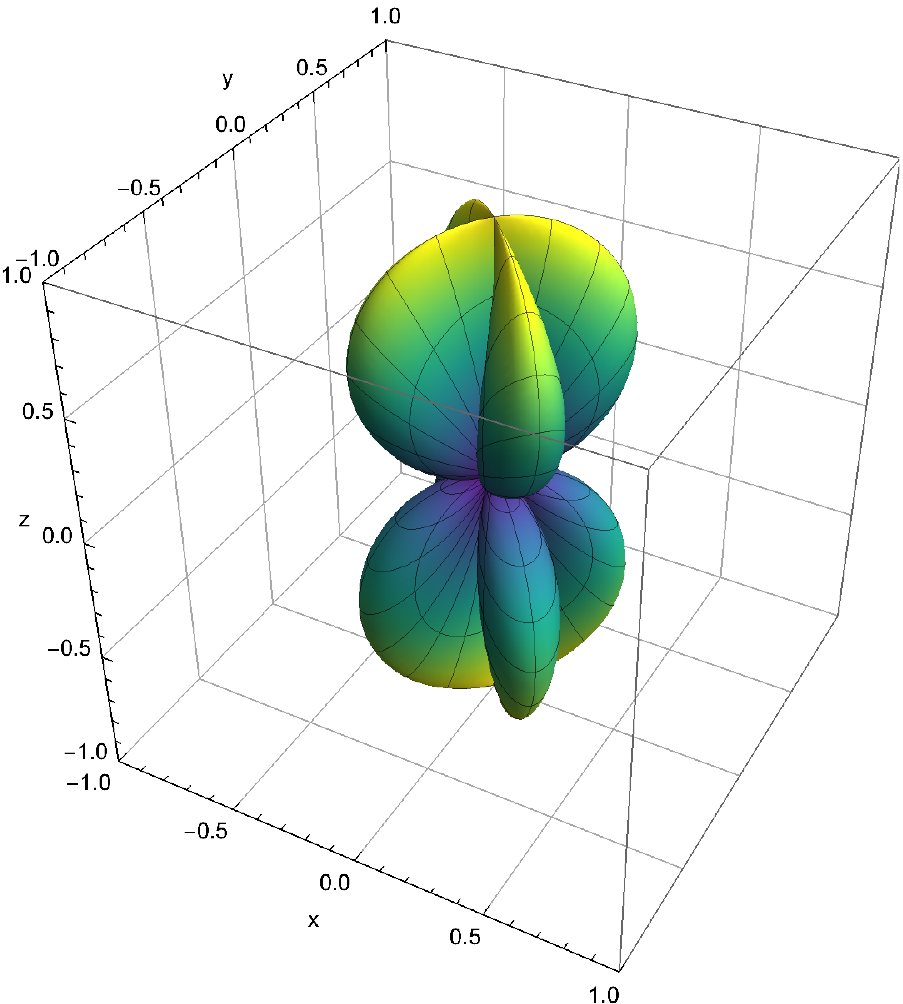}
		\subcaption{Pattern function $F_\times$}
		\label{fig: Fx vacuum}
	\end{subfigure}
	\hspace{0.05\columnwidth}
	\begin{subfigure}[c]{0.28\columnwidth}
		\includegraphics[width=\columnwidth]{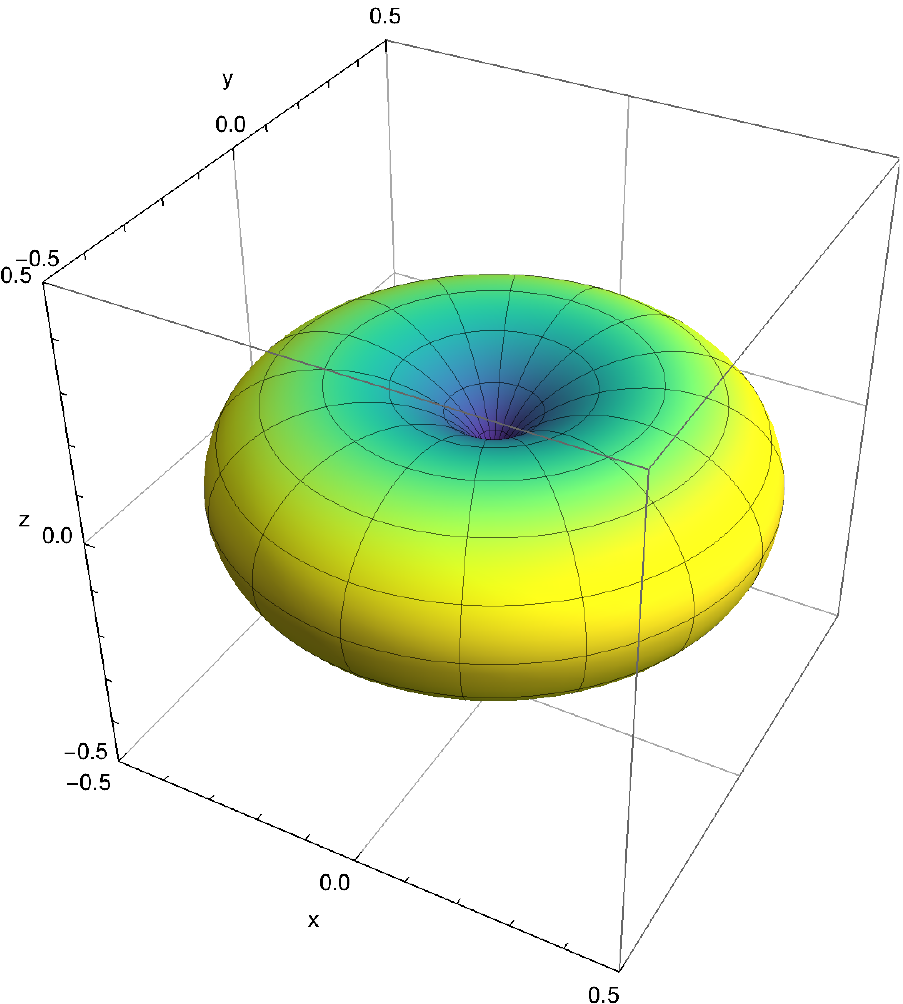}
		\subcaption{Pattern function $f_+$}
		\label{fig:pattern beyond eikonal vacuum}
		\label{fig: f+ vacuum}
	\end{subfigure}
	\caption{Detector response pattern functions for vacuum ($n = 1$). The left two panels (a) and (b) show the normalised eikonal response, while the rightmost panel (c) shows the corrections beyond the eikonal.}
	\label{fig:pattern functions vacuum}
\end{figure}

For current interferometers, the arm lengths are short compared to the gravitational wavelength, so that it suffices to expand to leading order in $\omegaG \ell_\cI$ and $\omegaG \ell_\cII$.
Moreover, the difference in the arm lengths is typically very small: $\Delta \ell / \ell \ll 1$. Neglecting relative errors of this size, we replace $\ell_\cI$ and $\ell_\cII$ by $\ell$ everywhere except in the multiplicative factors $\sin^2(n \omega \Delta \ell)$ and $\sin(n \omega \Delta \ell)$.
Using these approximations, we obtain
\begin{align}
	\delta_\psi P
		&\approx \half n \omega \ell \sin(2 n \omega \Delta \ell) [
			h(m_\cI, m_\cI) - h(m_\cII, m_\cII)
		] \,,
	\\
	\delta_\Aa P
		&\approx - \half n \omegaG \ell \sin^2(n \omega \Delta \ell) \left[
			h'(m_\cI, m_\cI) \frac{1 - (\kh.m_\cI)^2}{n^2 - (\kh.m_\cI)^2}
			+ h'(m_\cII, m_\cII) \frac{1 - (\kh.m_\cII)^2}{n^2 - (\kh.m_\cII)^2}
		\right]\,,
	\\
	\delta_n P
		&\approx - \half n \omegaG \ell (n^2 - 1) \sin^2(n \omega \Delta \ell) \left[
			\frac{h'(m_\cI, m_\cI)}{n^2 - (\kh.m_\cI)^2}
			+ \frac{h'(m_\cII, m_\cII)}{n^2 - (\kh.m_\cII)^2}
		\right]\,,
	\\
	\delta_K P
		&\approx n \omegaG \ell \sin^2(n \omega \Delta \ell) \left[
			h'(m_\cI, m_\cI) + h'(m_\cII, m_\cII)
		\right]\,,
\end{align}
where
\begin{equation}
	h'(m,m) = \p_u h(m,m) = - A(m,m) \sin(u)\,.
\end{equation}
In the low frequency limit, the perturbations “beyond the eikonal”, $\delta_\Aa P$, $\delta_n P$ and $\delta_K P$, have a similar form and can thus be summarised as
\begin{equation}
	\begin{split}
		\delta_{b\psi} P
			:=&\; \delta_\Aa P + \delta_n P + \delta_K P
		\\
		\approx &\; \half n \omegaG \ell \sin^2(\omega \Delta \ell) [
			h'(m_\cI, m_\cI) + h'(m_\cII, m_\cII)
			]\,,
	\end{split}
\end{equation}
Factoring out trigonometric functions depending on the GW phase and the interferometer arm lengths, and splitting the GW polarisation into the two polarisation states as $A = \alpha_+ A_+ + \alpha_\times A_\times$, where $\alpha_+$ and $\alpha_\times$ denote the amplitudes of the respective polarisation modes, see \cite{Mieling:2021a}, the detector response in the low frequency limit takes the form
\begin{align}
	\label{eq:michelson detector response lfl}
	\delta_\psi P &= n \omega \ell \cos(u) \sin(2 n \omega \Delta \ell) ( \alpha_+ F_+ + \alpha_\times F_\times )\,,
	\\
	\delta_{b\psi} P &= n \omegaG \ell \sin(u) \sin(n \omega \Delta \ell)^2 ( \alpha_+ f_+ + \alpha_\times f_\times )\,,
\end{align}
with the detector response pattern functions
\begin{align}
	F_\lambda
		&= + \half [A_\lambda(m_\cI, m_\cI) - A_\lambda(m_\cII, m_\cII)]\,,
	\\
	f_\lambda
		&= - \half [A_\lambda(m_\cI, m_\cI) + A_\lambda(m_\cII, m_\cII)]\,,
\end{align}
where $\lambda$ takes the values $+$ and $\times$.
Parametrising the GW propagation direction and polarisation using Euler angles $\Phi, \Theta, \Psi$ as in \cite{Mieling:2021a} with the polarisation angle $\Psi$ set to zero as usual, we find
\begin{align}
	F_+
		&= \half (1 + \cos^2 \Theta) \cos 2\Phi\,,
	&
	F_\times
		&= - \cos \Theta \sin 2\Phi\,,
	\\
	f_+
		&= \half \sin^2 \Theta  \cos 2 \Phi\,,
	&
	f_\times
		&= 0\,,
\end{align}
as in \cite{Mieling:2021a} (which was concerned with vacuum only).
The function $f_\times$ vanishes identically, and the remaining functions $F_+$, $F_\times$ and $f_+$ are plotted (in absolute values) in \Cref{fig:pattern functions vacuum}.

In the low-frequency limit, the presence of a dielectric thus results merely in a rescaling of the detector response by the refractive index $n$, as can be seen from \eqref{eq:michelson detector response lfl}.

\subsection{Comparison with Fibre Optics}
\label{s25VI21.1}

Finally, we compare our results (applying to plane waves in infinitely extended dielectrics) with those of \cite{Mieling:2021b} for guided waves in step-index optical fibres.
As the final expressions given there are restricted to one-way propagation of light in the low frequency limit, we consider the same setting here.

For one-way light propagation, the phase perturbation in the current setup reads
\begin{equation}
	\delta\psi
		= - \half n A(m,m) \frac{\sin(u) - \sin(u + (n - m.\kh) \omegaG m.x)}{\Omega(n - m.\kh)}
\end{equation}
where $\Omega = \omegaG /\omega$ and $u = \omegaG (\t\kh{_i} \t x{^i} - t) + \chi$, cf.~\eqref{eq:scalar field emitted}.
Evaluating this at a distance $\ell$ from the point of emission, at $\t x{^i} = \ell \t m{^i}$, the low-frequency limit $\omegaG \ell \ll 1$ yields
\begin{equation}
	\delta\psi
		= \half n A(m,m) \cos(- \omegaG t + \chi) \omega \ell + O(\omegaG \ell)\,,
\end{equation}
which is to be compared with the phase perturbation in optical fibres \cite[Eq.~(6.11)]{Mieling:2021b}
\begin{equation}
	\delta\psi_\text{fibre}
		= \half c_1 A(m,m) \cos(- \omegaG t + \chi) + O(\omegaG \ell)
\label{9VII21.1}
\end{equation}
(with the phase offset $\chi$  added as needed), where we have adapted the overall sign of the phase for consistency with the sign convention used here.
The multiplicative coefficient $c_1$ in \eqref{9VII21.1}
had to be determined numerically as a function of the core and cladding refractive indices $n_1$, $n_2$, and also of the core radius $\rho$, but was generally found to be close to the effective refractive index $c_1 \approx \bar n \approx n_1$ \cite[Sect.~8]{Mieling:2021b}.
Hence, we find good agreement of the phase perturbations acquired by plane waves in infinitely extended dielectrics and Bessel modes of azimuthal mode index $m = 1$ in step-index optical fibres.

Concerning perturbations of the amplitude and polarisation, we note that such effects arise in the current setup only at next-to-leading order in the ratio $\omegaG/\omega$, which was neglected in \cite{Mieling:2021b}. Nonetheless, perturbations of this kind were obtained for optical fibres \cite[Sect.~7]{Mieling:2021b}, but they were found to be suppressed relatively to the phase perturbation by a factor $(1 - n_2/n_1)^2$ which vanishes in the case considered here.
So although the results for the amplitude and polarisation perturbations are not directly comparable, they agree to the extent that such effects in media are smaller than the phase perturbation by a factor of $\omegaG/\omega$.
In any case a more accurate comparison of such effects can only be obtained by repeating the calculations here using Bessel waves in lieu of plane waves.

%% file: optics_subfiles/discussion.tex
\section{Discussion}

We have computed the perturbation of monochromatic plane wave solutions of Maxwell’s equations due to a plane gravitational wave and determined the
resulting signal in Michelson interferometers.

While the problem is simpler in the commonly used geometric optics approximation, the validity of such an approximation was not clear. Our analysis of the full Maxwell equations confirms its validity, putting the approximation on firm grounds. The unperturbed electromagnetic field, together with its first order perturbation, can indeed be approximated to current experimental accuracy by using a form which separates amplitude and phase, with the amplitude function analytic in the frequency ratio $\omegaG/\omega$ within the setup considered here.

By computing both phase and amplitude perturbations for arbitrary incidence angles of the gravitational wave, our result generalises previous ones which have either allowed for arbitrary incidence angles but have neglected amplitude perturbations \cite{Montanari1998} or, conversely, described amplitude or polarisation perturbations only for specific alignments \cite{Cooperstock1993,Calura1999,Lobato2021}.

Moreover, by including a dielectric refractive index, we were able to generalise our previous result \cite{Mieling:2021a} (which relied on the geometric optics approximation), and compare with analogous calculations for fibre optics \cite{Mieling:2021b}. In both cases, we find agreement of the essential results.

While we admit that the calculations presented here are unlikely to have a direct impact on signal analysis and interpretation of current gravitational wave detectors, we stress that previous models based on geometric optics relied on the tacit assumption that the true solution of the Maxwell equations in this setup admits a convergent eikonal expansion. Our analysis here describes a model where this is indeed the case. 

%% file: optics_subfiles/maxwell_wave_eq.tex
\section{The Wave Equation for the Electromagnetic Field}
\label{s:Maxwell wave equation}

Starting from the field equations \eqref{eq:maxwell 3+1} in the form
\begin{align}
	n \t\p{_0} \t Z{^i}
	+ j \t\varepsilon{^i^j^k} \t\nabla{_j} \t Z{_k} &= 0\,,
	&
	\t\nabla{_i} \t Z{^i} &= 0\,,
\end{align}
we now derive a wave equation for the electromagnetic field $\t Z{^i}$.
Throughout this section indices are raised and lowered with the \emph{perturbed} spatial metric $\t \gs{_i_j}$ by default. To derive an equation of second order, we apply $n \t\p{_0}$ to the first equation, use $\t\p{_0} \t\nabla{_j} \t Z{_k} = \t\p{_j} \t\p{_0} \t Z{_k} - \t\p{_0} \t\Gamma{^l_j_k} \t Z{_l}$, and thus obtain
\begin{equation}
	n^2 \t{\ddot Z}{^i}
	+ j \t\varepsilon{^i^j^k} \t\nabla{_j}(n \t\p{_0} \t Z{_k}) = 0\,,
\end{equation}
since the connection is torsion-free. Here and elsewhere, an overset dot indicates differentiation with respect to the time coordinate $t \equiv x^0$.
Next, lowering the index of the first-order evolution equation, one obtains
\begin{equation}
	n \t{\dot Z}{_k}
	= - j \t\varepsilon{_k^l^m}\t\nabla{_l} \t Z{_m} + n \t{\dot \gs}{_k_l} \t Z{^l}\,,
\end{equation}
so that the second-order equation can be written as
\begin{equation}
	n^2 \t{\ddot Z}{^i}
	+ (\t\gs{^i^l} \t\gs{^j^m} - \t\gs{^i^m} \t\gs{^j^l}) \t\nabla{_j} \t\nabla{_l} \t Z{_m}
	+ j \t\varepsilon{^i^j^k} \t\nabla{_j}( n \t{\dot \gs}{_k_l} \t Z{^l})
	= 0 \,.
\end{equation}
Using $\t\nabla{_j} \t Z{^j} = 0$, the second term can be simplified to yield
\begin{equation}
	(\t\gs{^i^l} \t\gs{^j^m} - \t\gs{^i^m} \t\gs{^j^l}) \t\nabla{_j} \t\nabla{_l} \t Z{_m}
	= \t\gs{^i^l} \t\nabla{_j} \t\nabla{_l} \t Z{^j} - \t\gs{^i^l} \t\nabla{_l} \t\nabla{_j} \t Z{^j} - \Delta \t Z{^i}
	= - \Delta \t Z{^i} + \t R{^i_j} \t Z{^j}\,,
\end{equation}
where $\Delta = \t\gs{^i^j} \t\nabla{_i} \t\nabla{_j}$ is the spatial Laplacian operator (here acting on a vector field) and $\t R{^i_j}$ is the spatial Ricci tensor.
This leads to the equation
\begin{equation}
	n^2 \t{\ddot Z}{^i} - \Delta \t Z{^i}
	+ \t R{^i_j} Z^j
	+ j \t\varepsilon{^i^j^k} \t\nabla{_j}( n \t{\dot\gs}{_k_l} \t Z{^l})
	= 0\,.
\end{equation}
Using $\t\varepsilon{^i^j^k}\t\nabla{_j} \t\alpha{_k} = \t\varepsilon{^i^j^k}\t\p{_j} \t\alpha{_k}$ and $\t\p{_i} \t\gs{_j_k} = \t\Gamma{_j_k_i} + \t\Gamma{_k_i_j}$, as well as the symmetry of the Christoffel symbols, one finally obtains
\begin{equation}
	n^2 \t{\ddot Z}{^i} - \Delta \t Z{^i}
	+ \t R{^i_j} \t Z{^j}
	+ j \t\varepsilon{^i^j^k}[ n \t{\dot \Gamma}{_k_j_l} \t Z{^l} + n \t{\dot\gs}{_k_l} \t\p{_j} \t Z{^l} ]
	= 0\,.
\end{equation}
A direct calculation shows that the components of the Laplacian acting on the vector field $Z$ are related to the Laplacian acting on the component functions $Z^i$ by
\begin{equation}
	\label{eq:maxwell wave equation A}
	\t{(\Delta Z)}{^i}
	= \Delta(\t Z{^i})
	+ \epsilon \t\delta{^j^k} \t\p{_j} \t\Gamma{^{\o1}^i_k_l} \t Z{^l}
	+ 2 \epsilon \t\delta{^j^k} \t\Gamma{^{\o1}^i_j_l} \t\p{_k} \t Z{^l}
	- \epsilon \t\delta{^j^k} \t\Gamma{^{\o1}^l_j_k} \t\p{_l} \t Z{^i}
	+ O(\epsilon^2)\,.
\end{equation}
For the particular spatial metric \eqref{eq:spatial metric}, the Christoffel symbols take the form
\begin{equation}
	\t \Gamma{^{\o1}_i_j_k}
	= - \half \epsilon [
		\t\kappa{_j} \t A{_k_i}
		+ \t\kappa{_k} \t A{_i_j}
		- \t\kappa{_i} \t A{_j_k}
	] \sin(\kappa.x + \chi) \,,
\end{equation}
so that the last term in \eqref{eq:maxwell wave equation A} is seen to vanish. Consequently, one has
\begin{equation}
	\t\delta{^j^k} \t\p{_j} \t\Gamma{^{\o1}_i_k_l} = - \half \omegaG^2 \t\gs{^{\o1}_i_l} = - \t R{^{\o1}_i_l}\,,
\end{equation}
where $\t R{_i_j}$ is the spatial Ricci tensor with components
\begin{equation}
	\t R{^{\o1}_i_j} = \half \omegaG^2 \t\gs{^{\o1}_i_j}\,.
\end{equation}
Combining these intermediate results, the wave equation takes the form
\begin{equation}
	n^2 \t{\ddot Z}{^i} - \Delta(\t Z{^i})
	+ 2 \epsilon \t R{^{\o1}^i_j} \t Z{^j}
	- 2 \epsilon \t\delta{^j^k} \t\Gamma{^{\o1}^i_j_l} \t\p{_k} \t Z{^l}
	+ j \epsilon \t\varepsilon{^i^j^k}[ n \t{\dot \Gamma}{^{\o1}_k_j_l} \t Z{^l} + n \t{\dot\gs}{^{\o1}_k_l} \t\p{_j} \t Z{^l} ]
	= 0\,.
\end{equation}
In all terms involving derivatives of the metric (i.e.\ its time derivative, Christoffel symbols or the Ricci tensor), we may use the unperturbed expression for the electromagnetic field.

%% file: optics_subfiles/appendix-emission-covariant.tex
\section{Covariant Formulation of the Emission Condition}
\label{app:covariant emission equation}

Here, we discuss a covariant formulation of the emission condition \eqref{eq:Maxwell emission assumption}, which states that
\begin{equation}
	\t Z{^i} - j \t\varepsilon{^i^j^k} \t\nu{_j} \t Z{_k} = 0\,,
\end{equation}
where $\nu$ is the conormal of the emission surface $\Sigma$.

Using a timelike, future-pointing unit normal field $u$, every two-form $F$ admits the covariant representation
\begin{equation}
	\label{eq:app decomposition F}
	\t F{_\mu_\nu} = \t u{_\mu} \t E{_\nu} - \t u{_\nu} \t E{_\mu} + \t\epsilon{_\mu_\nu_\rho} \t B{^\rho} \,,
\end{equation}
with a vector $B$ and a covector $E$ satisfying $\t E{_\mu}\t u{^\mu} = 0 = \t B{^\mu} \t u{_\mu}$. Therein, the “spatial epsilon tensor” $\t \epsilon{_\mu_\nu_\rho}$ is obtained from the epsilon tensor (with $\t\epsilon{_0_1_2_3} = \sqrt{-\det \t g{_\mu_\nu}}$) via the contraction $\t \epsilon{_\mu_\nu_\rho} := \t u{^\sigma} \t \epsilon{_\sigma_\mu_\nu_\rho}$. The projections $\t E{_i} := \t P{^\mu_i} \t E{_\mu}$ and $\t B{^i} := \t P{^i_\mu} \t B{^\mu}$ with ${\t P{^\mu_\nu} := \t*\delta{^\mu_\nu} + \t u{^\mu} \t u{_\nu}}$ yield the 3-vectors that reproduce the  representation \eqref{eq:decomposition DEBH} of the field strength tensor in coordinates with $\t u{_\mu}=(-1/\sqrt{-\t g{^0^0}},0,0,0)$. Analogously, the displacement tensor may be written as
\begin{equation}
	\label{eq:app decomposition Fb}
	\t\Fb{^\mu^\nu} = \t u{^\mu} \t D{^\nu} - \t u{^\nu} \t D{^\mu} + \t \epsilon{^\mu^\nu^\rho} \t H{_\rho}\,.
\end{equation}
For the dual field strength tensor $\t {*F}{^\mu^\nu} := \half \t \epsilon{^\mu^\nu^\alpha^\beta} \t F{_\alpha_\beta}$ one obtains
\begin{equation}
	\t{*F}{^\mu^\nu} = - \t u{^\mu} \t B{^\nu} + \t u{^\nu} \t B{^\mu} + \t \epsilon{^\mu^\nu^\rho} \t E{_\rho}\,.
\end{equation}
Assuming now a linear isotropic dielectric with $\t D{^\mu}= \permittivity \t g{^\mu^\nu} \t E{_\nu}$, $\t B{^\mu}= \permeability \t g{^\mu^\nu} \t H{_\nu}$ and $n := \sqrt{\permittivity\permeability}$, we consider the $j$-complex combination
\begin{equation}
	\label{eq:Maxwell complex tensor}
	\t{\mathcal F}{^\mu^\nu}
		:= \permeability \t \Fb{^\mu^\nu} - j n \,\t{*F}{^\mu^\nu}
		= \t u{^\mu} \t Z{^\nu} - \t u{^\nu} \t Z{^\mu} -\frac{j}{n} \t \epsilon{^\mu^\nu^\rho} \t Z{_\rho}\,,
\end{equation}
where
\begin{align}
	\t Z{^\mu} &:= \permeability\, \t D{^\mu} + jn \t B{^\mu}\,,
	&
	\t Z{_\mu} &:= \t g{_\mu_\nu} \t Z{^\nu} = n^2 \t E{_\nu} + j \permeability n \t H{_\nu}\,.
\end{align}
Note that the electromagnetic field in a general medium can be expressed similarly using two independent complex vectors, cf.~\cite{Kryuchkov2017}.

In accordance with geometric optics, cf.\ e.g.\ \cite{Mieling:2021a}, we consider light of frequency $\omega$ (as measured by observers with 4-velocity $u$) emitted orthogonally to the surface $\Sigma$, with normal $\nu$, which we assume orthogonal to $u$. The corresponding wave vector then reads
\begin{equation}
	\t k{_\mu} = \omega (\t u{_\mu} + n \t \nu{_\mu})\,.
\end{equation}
In such a setting one imposes on $\Sigma$ the conditions
\begin{align}
	\t\Fb{^\mu^\nu} \t k{_\nu} &= 0\,,
	&
	\t{*F}{^\mu^\nu} \t k{_\nu} &= 0\,.
\end{align}
In our complex notation, this can be summarized as
\begin{equation}\label{eq:covariant boundary condition}
	\t {\mathcal F}{^\mu^\nu} \t k{_\nu} = 0\,,
\end{equation}
which can be rewritten as
\begin{equation}
	\label{eq:boundary convariant 1}
	\t Z{^\mu} + n \t u{^\mu} \t Z{^\nu} \t \nu{_\nu} - j \t \epsilon{^\mu^\nu^\rho} \t \nu{_\nu}Z_{\rho} = 0\,.
\end{equation}
Contracting this equation with $\t u{_\mu}$ yields $\t \nu{_\mu} \t Z{^\mu} = 0$, so that \eqref{eq:boundary convariant 1} simplifies to
\begin{equation}\label{eq:covariant boundary condition linear dielectric}
	\t Z{^\mu} - j \t \epsilon{^\mu^\nu^\rho} \t \nu{_\nu} \t Z{_\rho} = 0\,,
\end{equation}
as used in the main body of the paper.

To calculate the invariants of the electromagnetic field we recall the definition of the dual of the excitation tensor $\t \Fb{^\alpha^\beta}$:
\begin{equation}
	\t {*\Fb}{_\mu_\nu}
		:= \half \t \epsilon{_\mu_\nu_\alpha_\beta} \t \Fb{^\alpha^\beta}
		= - \t u{_\mu} \t H{_\nu} + \t u{_\nu} \t H{_\mu} + \t \epsilon{_\mu_\nu_\rho} \t D{^\rho}\,.
\end{equation}
To relate $\t {*\Fb}{_\mu_\nu}$ to $\t {*F}{^\rho^\sigma}$, let $\t \gamma{_\mu_\nu}$ denote the inverse to the contravariant optical metric $\t \gamma{^\mu^\nu}$, which evaluates to
\begin{equation}
	\label{eq:optical metric covariant}
	\t \gamma{_\mu_\nu} = \t g{_\mu_\nu} + (1 - n^{-2}) \t u{_\mu} \t u{_\nu}\,.
\end{equation}
One then finds
\begin{equation}
	\permeability \t {*\!\Fb}{_\mu_\nu}
		= \frac{\det(\t \gamma{^\alpha^\beta})}{\det(\t g{^\alpha^\beta})} \t \gamma{_\mu_\rho} \t \gamma{_\nu_\sigma} \t {*\!F}{^\rho^\sigma}
		= n^2 \t \gamma{_\mu_\rho} \t \gamma{_\nu_\sigma} \t {*\!F}{^\rho^\sigma}\,.
\end{equation}
In the last identity we used the fact that ${\det(\t \gamma{^\alpha^\beta})}/{\det(\t g{^\alpha^\beta})}$ is a scalar, so that it may be computed in the local inertial system of the medium, where $\t g{^\alpha^\beta} = \t \eta{^\alpha^\beta}$ and $\t \gamma{^\alpha^\beta} = \diag(-n^2, 1,1,1)$.
Using \eqref{eq:optical metric maxwell}, contraction of $\t{\mathcal F}{^\rho^\sigma}$ with the optical metric then yields
\begin{equation}
	\t {\bar{\mathcal F}}{_\mu_\nu}
		:= \t \gamma{_\mu_\rho} \t \gamma{_\nu_\sigma} \t {\mathcal F}{^\rho^\sigma}
		= \t F{_\mu_\nu} - \frac j n \permeability\, \t {*\!\Fb}{_\mu_\nu}
		= n^{-2} ( \t u{_\mu} \t Z{_\nu} - \t u{_\nu} \t Z{_\mu}-j n \t \epsilon{_\mu_\nu_\rho} \t Z{^\rho} )\,.
\end{equation}
Thus setting both invariants $F_{\mu\nu}\Fb^{\mu\nu}$ and $F_{\mu\nu}*F^{\mu\nu}$ of the electromagnetic field to zero is seen to be  equivalent to the complex equation
\begin{equation}
	\label{eq:maxwell invariants complex notation}
	\t {\mathcal F}{^\mu^\nu} \t {\bar{\mathcal F}}{_\mu_\nu}
		\equiv 2 \permeability \, \t F{_\mu_\nu} \t \Fb{^\mu^\nu} - 2 j n \t F{_\mu_\nu} \t {*\!F}{^\mu^\nu}
		\equiv - \frac 4 {n^2} \t Z{^\nu} \t Z{_\nu}=0 \,,
\end{equation}
which is already implied by \eqref{eq:covariant boundary condition linear dielectric}.

%% file: optics_subfiles/appendix-constraint.tex
\section{The Boundary-Constraint Equation}
 \label{ss9VIII21.1}

Maxwell’s equations written in terms of the tensor field \eqref{eq:Maxwell complex tensor} read
\begin{equation}
	\label{eq:Maxwell complex tensoreq}
	\t\nabla{_\mu} \t{\mathcal F}{^\mu^\nu} = 0 \,,
\end{equation}
or equivalently
\begin{equation}
	\label{eq:Maxwell complex tensoreq2}
	\t\p{_\mu} (\sqrt{- \det g} \t{\mathcal F}{^\mu^\nu} ) = 0 \,.
\end{equation}
Let us choose a coordinate system so that a \emph{non-characteristic hypersurface} $\Sigma$ is given by the equation $\{\t x{^0} = 0\}$; we emphasise that we \emph{do not} assume that $\t x{^0}$ is a time coordinate, only that $\t g{^0^0}|_\Sigma$ has no zeros. The field
\begin{equation}
	\label{eq:Maxwell complex tensoreq3}
	\mathcal C
		:= \t\p{_i} (\sqrt{- \det g} \t{\mathcal F}{^i^0} )
		\equiv  \t\p{_\mu} (\sqrt{- \det g} \t{\mathcal F}{^\mu^0} )\,,
\end{equation}
which involves only derivatives tangential to $\Sigma$,  has to vanish for solutions of \eqref{eq:Maxwell complex tensoreq}. Hence, the equation $\mathcal C = 0$ has the character of a constraint equation when propagating fields away from $\Sigma$.

The complementing part of the constraint equation $\mathcal C = 0$ in \eqref{eq:Maxwell complex tensoreq} are the propagation equations
\begin{equation}
	\label{eq:Maxwell complex tensoreq4}
	\t\p{_0} (\sqrt{- \det g}\t{\mathcal F}{^0^j} )
		=  - \t\p{_i} (\sqrt{- \det g}\t{\mathcal F}{^i^j}) \,.
\end{equation}
Here, \emph{propagation}  does \emph{not} mean propagation in time, but \emph{propagation away from $\Sigma$}.
Every  solution of these propagation equations satisfies
\begin{equation}
	\label{eq:Maxwell complex tensoreq5 }
	\t\p{_0} \mathcal C
	=  \t\p{_0} \t\p{_j} (\sqrt{- \det g}\t{\mathcal F}{^j^0})
	=  \t\p{_j} \t\p{_0} (\sqrt{- \det g}\t{\mathcal F}{^j^0})
	=  \t\p{_j} \t\p{_i}(\sqrt{- \det g}\t{\mathcal F}{^i^j})
	= 0 \,,
\end{equation}
where we have used the antisymmetry of $\t{\mathcal F}{^\mu^\nu}$.
We conclude that if $\mathcal C$ vanishes on $\Sigma$  and if \eqref{eq:Maxwell complex tensoreq4} holds, then $\mathcal C$ vanishes on the domain of definition of the coordinates above.

%% file: optics_subfiles/appendix-energy.tex
\section{The Energy-Momentum Tensor}
\label{appendix:energy}

The field equations can be derived from a variational principle based on the Lagrangian density
\begin{equation}
	\mathfrak L
		= - \quarter \sqrt{-g}\, \t\Fb{^\alpha^\beta} \t F{_\alpha_\beta}\,,
\end{equation}
where $F$ is regarded as the curvature form of an abelian gauge potential $F = \dd A$.
 The overall scaling depends on the choice of units, which is irrelevant here as we assume no external charges or currents.
An energy tensor $\t T{_\mu_\nu}$ is obtained using the standard formula \cite[Eq.~(E.1.26)]{Wald}
\begin{equation}
	\delta \mathfrak L
		= - \tfrac{1}{2 \pi} \sqrt{-g}\, \t T{_\mu_\nu} \t{\delta g}{^\mu^\nu}\,,
\end{equation}
where the 4-velocity $u$ is varied according to $\t{\delta u}{^\mu} = - \half \t u{^\mu} \t u{_\alpha} \t u{_\beta} \t{\delta g}{^\alpha^\beta}$ in order to preserve the normalisation $\t g{_\mu_\nu}\t u{^\mu} \t u{^\nu} = -1$.
Using $\t \gamma{_\mu_\nu}$ as defined in \eqref{eq:optical metric covariant}, this can be written as
\begin{equation}
	4\pi \t T{_\mu_\nu}
		= \t\gamma{_\mu_\alpha} \t\Fb{^\alpha^\beta} \t F{_\nu_\beta}
		- \quarter \t\Fb{^\alpha^\beta} \t F{_\alpha_\beta} \t g{_\mu_\nu}
		+ (n^2 - 1)\t\gamma{_\rho_\alpha} \t\Fb{^\alpha^\beta} \t F{_\sigma_\beta} \t u{^\rho} \t u{^\sigma} \t u{_\mu} \t u{_\nu}
		\,,
\end{equation}
cf.~\cite{Antoci1997}.
Using the decomposition of $F$ and $\Fb$ from \eqref{eq:app decomposition F} and \eqref{eq:app decomposition Fb}, and letting $\t \gs{_\mu_\nu} = \t g{_\mu_\nu} + \t u{_\mu} \t u{_\nu}$ be the “spatial metric” (which can be used to raise and lower indices of spatial vectors), this can be written as
\begin{equation}
	\label{eq:energy tensor real fields}
	4 \pi \t T{_\mu_\nu}
		= \half (E.D + B.H) (\t u{_\mu} \t u{_\nu} + \t \gs{_\mu_\nu})
		+ 2 \t u{_(_\mu} \t \epsilon{_\nu_)^\rho^\sigma} \t E{_\rho} \t H{_\sigma}
		- \t D{_\mu} \t E{_\nu}
		- \t B{_\mu} \t H{_\nu}
		\,.
\end{equation}
This tensor is evidently symmetric and trace-free with respect to the spacetime metric. In terms of the complex field $Z$ it may be expressed as
\begin{equation}
	\label{eq:energy tensor Z}
	8 \pi \permeability n^2
	\t T{_\mu_\nu}
		=
		 Z^*_\rho Z^{\rho} (\t u{_\mu} \t u{_\nu} + \t \gs{_\mu_\nu})
		+ 2 j n^{-1} \t u{_(_\mu} \t \epsilon{_\nu_)^\rho^\sigma} \t Z{_\rho} \t Z{^*_\sigma}
		- \t Z{_\mu} \t Z{^*_\nu}
		- \t Z{^*_\mu} \t Z{_\nu}
		\,,
\end{equation}
so that the energy density is seen to be given by
\begin{equation}
	\label{eq:energy tensor complex norm}
	\t T{_0_0} =
	\frac{1}{8 \pi} (E.D + B.H)
	= \frac{1}{8 \pi \permeability n^2} Z^*.Z\,.
\end{equation}

%% file: optics_subfiles/appendix-reflection-covariant.tex
\section{Compatibility of Reflection with Constraints}
\label{app:reflection constraint consistent}

Consider the reflection conditions
\begin{align}
	\nu \cdot (\check Z - Z^*) &= 0\,,
	&
	\nu \times (\check Z + Z^*) &= 0\,,
\end{align}
where $Z$ is the incident field, $\check Z$ is the reflected field, and $\nu$ is the normal to the reflecting surface.
This can be covariantly rewritten using the complex tensor $\mathcal F$ defined in \eqref{eq:Maxwell complex tensor}:
\begin{equation}
	\t{(\check{\mathcal F} - \mathcal F^*)}{^\alpha^\beta} \nu_\beta = 0\,.
\end{equation}

Let us verify that this prescription is consistent in the following sense: if the incident field $Z$ satisfies the constraint equation which arises from Maxwell’s equations, then so does the reflected field $\check Z$.
Recall that the constraint equation is obtained from the time-evolution equation by contraction with the conormal $\nu_i$. Note that since $\nu_i$ is proportional to $m_i$, we may use the latter instead, which has the advantage that all its derivatives vanish.
By assumption, the incident field $Z$ satisfies the (conjugate) constraint equation
\begin{equation}
	n \p_0 (m_i Z^{*i}) - j \varepsilon^{ijk}\p_j(m_i Z^*_k) = 0\,.
\end{equation}
Using $m_i Z^{*i} = m_i \check Z^i$ and $\varepsilon^{ijk} m_i Z^*_k = - \varepsilon^{ijk} m_i \check Z_k$, we obtain
\begin{equation}
	n \p_0 (m_i \check Z^i) + j \varepsilon^{ijk}\p_j(m_i \check Z_k) = 0\,,
\end{equation}
so the reflected field indeed satisfies the constraint equation. The remaining projections of Maxwell’s equations then determine the normal derivatives of the various components of $Z$, thereby encoding the law of reflection.

In fact, we can formulate a covariant argument which shows that the requirement that partial derivatives of $m$ vanish is inessential. Since $\nu$ is hypersurface-orthogonal, it can be locally written as $\nu = g\, \dd f$, where $f$ and $g$ are smooth functions. By linearity, we may equally formulate the reflection condition with the exact form $df$ instead of $\nu$:
\begin{equation}
	\t{\check{\mathcal F}}{^\alpha^\beta} \t f{_;_\beta}
	- \t{{\mathcal F}}{^*^\alpha^\beta} \t f{_;_\beta}
	= 0\,.
\end{equation}
Taking the divergence, we obtain
\begin{equation}
	\t{\check{\mathcal F}}{^\alpha^\beta_;_\alpha} \t f{_;_\beta}
	- \t{{\mathcal F}}{^*^\alpha^\beta_;_\alpha} \t f{_;_\beta}
	+
	(\t{\check{\mathcal F}}{^\alpha^\beta}
	- \t{{\mathcal F}}{^*^\alpha^\beta}) \t f{_;_\beta_\alpha}
	= 0\,.
\end{equation}
Since the connection used is torsion-free, $\t f{_;_\beta_\alpha}$ is symmetric, so the last term vanishes due to the anti-symmetry of the field tensors.
Thus, $\check{\mathcal F}$ satisfies the constraint equation $\t{\check{\mathcal F}}{^\alpha^\beta_;_\alpha} \t \nu{_\beta}$ if and only if the incident field $\mathcal F$ does.
Note that the presence of sources on the mirror surface has no influence on this argument since the current four-vector is tangent to the mirror surface and thus annihilated by the normal one-form $\nu$.

%% file: optics_subfiles/appendix-polarisation.tex
\section{Polarization Reflection at a Beam Splitter}
\label{app:beam splitter reflection}

We do not attempt to model the precise details of beam splitters in curved spacetime, and use a simplified model which relates the electromagnetic fields at the various surfaces of the beam splitter, pointing towards the laser, detector, and along the two interferometer arms, see \Cref{fig:Michelson}.

Denote by $\Sigma_\cI$ and $\Sigma_\cII$ the surfaces facing the mirrors, as in \Cref{fig:Michelson}.
We describe these surfaces by $\Sigma_\cI = \{m_\cI.x = 0\}$ and $\Sigma_\cII = \{m_\cII.x = 0\}$, where
\begin{equation}
	\label{eq:app reflection orthogonality}
	m_\cI. m_\cII \equiv \t \delta{^i^j} \t {m}{_\cI_i} \t m{_\cII_j} = 0\,,
\end{equation}
so that the unperturbed normals are orthogonal in the background metric.
The normals in the perturbed metric are then
\begin{align}
	\t \nu{_\cI_i}
		&= \t m{_\cI_i} \left( 1 + \half \epsilon h(m_\cI,m_\cI) \right)\,,
	&
	\t \nu{_\cI^i}
		&= \t m{_\cI^i} \left( 1 + \half \epsilon h(m_\cI,m_\cI) \right)
		- \epsilon \t h{^i^j} \t m{_\cI_j}
		\,,
	\\
	\t {\nu}{_\cII_i}
		&= \t {m}{_\cII_i} \left( 1 + \half \epsilon h(m_\cII,m_\cII) \right)\,,
	&
	\t {\nu}{_\cII^i}
		&= \t {m}{_\cII^i} \left( 1 + \half \epsilon h(m_\cII,m_\cII) \right)
		- \epsilon \t h{^i^j} \t {m}{_\cII_j}
		\,,
\end{align}
where indices of $m_\cI$ and $m_\cII$ are raised with the background metric $\t \delta{_i_j}$.
The corresponding reflection operator which interchanges $\nu_\cI$ and $\nu_\cII$ while leaving their orthogonal complement unaltered is
\begin{equation}
	\label{eq:app reflection operator}
	\t R{^i_j}
		= \t*\delta{^i_j} - [ 1 - g(\nu_\cII, \nu_\cI) ]^{-1} \t{(\nu_\cII-\nu_\cI)}{^i} \t{(\nu_\cII-\nu_\cI)}{_j}\,.
\end{equation}
By virtue of \eqref{eq:app reflection orthogonality}, one has $g(\nu_\cII, \nu_\cI) = - \epsilon h(m_\cII, m_\cI) + O(\epsilon^2)$, and thus
\begin{equation}
	\t R{^i_j}
		= \t*\delta{^i_j} - [ 1 - \epsilon h(m_\cII, m_\cI) ] \t{(\nu_\cII-\nu_\cI)}{^i} \t{(\nu_\cII-\nu_\cI)}{_j}
		+ O(\epsilon^2)\,.
\end{equation}
We wish to compute $\t R{^i_j} \t\Zz{^j}$, where
\begin{align}
	\t\Zz{^i}
		&= a \t\zeta{^i} - \half \epsilon h(\zeta, \zeta) \t\zeta{^*^i}\,,
	\\
	\shortintertext{with}
	\label{eq:app reflection a explicit}
	a
		&= 1 + \epsilon j \alpha - \half \epsilon h(\zeta^*, \zeta)\,.
\end{align}
Defining the unperturbed reflected polarization
\begin{equation}
	\label{eq:app reflection unperturbed}
	\t\eta{_i}
		= \t R{^{\o0}_i^j} \t\zeta{_j}
		= \t\zeta{_i} - \t{(m_\cII-m_\cI)}{_i} (m_\cII.\zeta)\,,
\end{equation}
one can expand $\t R{^i_j} \t\Zz{^j}$ in the basis $\eta$, $\eta^*$, $m_\cII$. Since $\Zz$ is orthogonal to $m_\cI$, the reflected vector is orthogonal to $m_\cII$, so that one can write
\begin{equation}
	\label{eq:app reflection expansion}
	\t R{^i_j} \t\Zz{^j}
		= b \t\eta{^i} + \epsilon c \t\eta{^*^i}\,,
\end{equation}
for some constants $b, c$ to be determined.
Since $\Zz$ satisfies $g(\Zz,\Zz) = 0$, the reflected vector also has zero norm. As $b = 1 + O(\epsilon)$ the factor $c$ is determined by
\begin{equation}
	h(\eta, \eta) + 2 c = 0\,,
\end{equation}
so that it remains to determine $b$.
For this, one may use the expansion \eqref{eq:app reflection expansion} to find
\begin{equation}
	b
		= \t*\eta{^*_i} \t R{^i_j} \t\Zz{^j}
		= \t*\zeta{^*_i} \t R{^i_j} \t\Zz{^j} - (m_\cII.\zeta^*) \t*{(m_\cII-m_\cI)}{_i} \t R{^i_j} \t\Zz{^j}\,,
\end{equation}
where we consider the second term first. There, one has $\t {m}{_\cII_i} \t R{^i_j} \t\Zz{^j} \propto \t {\nu}{_\cII_i} \t R{^i_j} \t\Zz{^j} = \t\nu{_\cI_i} \t\Zz{^i} = 0$. For the other term, $\t m{_\cI_i} \t R{^i_j} \t\Zz{^j}$, one may use the explicit form of $\nu_\cI$, $\nu_\cII$ and $z$ to obtain
\begin{equation}
	b
		= \t*\zeta{^*_i} \t R{^i_j} \t\Zz{^j}
		+ a (m_\cII.\zeta^*) (m_\cII.\zeta)[ 1 + \half \epsilon h(m_\cII, m_\cII) - \half \epsilon h(m_\cI,m_\cI) ]
		- \half \epsilon h(\zeta, \zeta) (m_\cII.\zeta^*)^2\,.
\end{equation}
Contracting \eqref{eq:complex basis delta} with $\t m{_\cII_i} \t m{_\cII_j}$ yields $(m_\cII.\zeta^*) (m_\cII.\zeta) = \half$, and using $a = 1 + O(\epsilon)$, this simplifies to
\begin{equation}
	\label{eq:app reflection b intermediate}
	b
		= \t*\zeta{^*_i} \t R{^i_j} \t\Zz{^j}
		+ \half a
		+ \quarter \epsilon h(m_\cII, m_\cII)
		- \quarter \epsilon h(m_\cI, m_\cI)
		- \half \epsilon h(\zeta, \zeta) (m_\cII.\zeta^*)^2\,.
\end{equation}
Consider, now, the first term. Using the explicit expression for the reflection operator, as well as the intermediate results
\begin{align}
	\t*\zeta{^*_i} \t\Zz{^i}
		&= a\,,
		\\
	\t*\zeta{^*_i} \t{(\nu_\cII-\nu_\cI)}{^i}
		&= [1 + \half \epsilon h(m_\cII, m_\cII)] (m_\cII.\zeta^*) - \epsilon h(\zeta^*, m_\cII - m_\cI)\,,
		\\
	\t{(\nu_\cII-\nu_\cI)}{_i} \t\Zz{^i}
		&= [ 1 + \half \epsilon h(m_\cII, m_\cII) ] [ a (m_\cII.\zeta)- \half \epsilon h(\zeta, \zeta)  (m_\cII.\zeta^*)]\,,
\end{align}
and using again $(m_\cII.\zeta^*) (m_\cII.\zeta) = \half$ as well as $a = 1 + O(\epsilon)$, one finds
\begin{equation}
	\t*\zeta{^*_i} \t R{^i_j} \t\Zz{^j}
		= \half a
		+ \half \epsilon h(m_\cII, m_\cI)
		- \half \epsilon h(m_\cII, m_\cII)
		+ \epsilon h(\zeta^*, m_\cII - m_\cI) (m_\cII.\zeta)
		+ \half \epsilon h(\zeta, \zeta) (m_\cII.\zeta^*)^2\,.
\end{equation}
Plugging this into \eqref{eq:app reflection b intermediate} then yields
\begin{equation}
	b
		= a
		+ \epsilon h(\zeta^*, m_\cII - m_\cI)(m_\cII.\zeta)
		- \quarter \epsilon h(m_\cII - m_\cI, m_\cII - m_\cI)\,,
\end{equation}
and inserting the explicit form of $a$ from \eqref{eq:app reflection a explicit}, one arrives at
\begin{equation}
	b
		= 1 + \epsilon j \alpha
		- \half \epsilon h(\zeta^*, \zeta)
		+ \epsilon h(\zeta^*, m_\cII - m_\cI) (m_\cII.\zeta)
		- \quarter \epsilon h(m_\cII-m_\cI, m_\cII-m_\cI)\,.
\end{equation}
Decomposing $b$ analogously as $a$ was decomposed in \eqref{eq:app reflection a explicit} in the form
\begin{equation}
	\begin{split}
		b
			&= 1 + \epsilon j \beta - \half \epsilon h(\eta^*, \eta)
			\\&
			= 1 + \epsilon j \beta
			- \half \epsilon h(\zeta^*, \zeta)
			+ \half \epsilon h(\zeta^*, m_\cII-m_\cI) (m_\cII.\zeta)
			\\&\quad
			+ \half \epsilon h(\zeta, m_\cII-m_\cI) (m_\cII.\zeta^*)
			- \quarter \epsilon h(m_\cII-m_\cI, m_\cII-m_\cI)\,,
	\end{split}
\end{equation}
one arrives at
\begin{equation}
	j \beta
		= j \alpha
		+ \half h(\zeta^*, m_\cII-m_\cI) (m_\cII.\zeta)
		- \half h(\zeta, m_\cII-m_\cI) (m_\cII.\zeta^*)\,,
\end{equation}
or equivalently
\begin{equation}
	\beta = \alpha + \Im_j [ h(\zeta^*, m_\cII-m_\cI) (m_\cII.\zeta) ]\,,
\end{equation}
where $\Im_j$ denotes the $j$-imaginary part.

Summarizing, given a polarization vector of the form
\begin{equation}
	\t\Zz{^i}
		= \t\zeta{^i}[ 1 + j \epsilon \alpha - \half \epsilon h(\zeta^*, \zeta) ]
		- \half \epsilon \t\zeta{^*^i} h(\zeta, \zeta)
		+ O(\epsilon^2)\,,
\end{equation}
the vector obtained by applying the reflection operator \eqref{eq:app reflection operator} is
\begin{equation}
	\label{eq:beam splitter reflection app}
	\t R{^i_j} \t\Zz{^j}
		= \t\eta{^i} \left(
				1
				+ j \epsilon \alpha
				+ j \epsilon \Im_j [ h(\zeta^*, m_\cII-m_\cI) (m_\cII.\zeta) ] - \half \epsilon h(\eta^*, \eta)
			\right)
		- \half \epsilon \t\eta{^*^i} h(\eta, \eta)
		+ O(\epsilon^2)\,,
\end{equation}
where $\eta$ is the unperturbed reflected vector, as defined in \eqref{eq:app reflection unperturbed}.